\documentclass{jfm}

\usepackage{graphicx}
\usepackage{lineno}
\usepackage{multirow}
\usepackage{epstopdf, epsfig}
\usepackage{subfigure}
\usepackage{array}
\usepackage[usenames]{color} 
 \usepackage{amssymb} 
 \usepackage{amsmath} 
 \usepackage[table,usenames,dvipsnames]{xcolor}
 \usepackage{bm}
 \usepackage{mathdots}
 \usepackage[utf8]{inputenc} 
 \usepackage{tikz}
 \usetikzlibrary{shapes}
 \usepackage{pgfplots}
 \usepackage{hyperref}
\hypersetup{
	colorlinks=true, 
	linktoc=all,     
	linkcolor=blue,  
	citecolor=blue,
	urlcolor = blue
}
 \usepackage{todonotes}

 \newcommand{\bluelt}{\raisebox{2pt}{\tikz{\draw[blue,solid,line width=0.9pt](0,0) -- (5mm,0);}}}
 \newcommand{\bluedashdotlt}{\raisebox{2pt}{\tikz{\draw[blue,dash dot,line width=0.9pt](0,0) -- (5mm,0);}}}
 \newcommand{\greendashlt}{\raisebox{2pt}{\tikz{\draw[black!40!green,dashed,line width=0.9pt](0,0) -- (5mm,0);}}} 
 \newcommand{\reddotlt}{\raisebox{2pt}{\tikz{\draw[red,dotted,line width=0.9pt](0,0) -- (5mm,0);}}} 
 \newcommand{\dottedlt}{\raisebox{2pt}{\tikz{\draw[black,dotted,line width=0.9pt](0,0) -- (5mm,0);}}}
 \newcommand{\dashdotlt}{\raisebox{2pt}{\tikz{\draw[black,dash dot,line width=0.9pt](0,0) -- (5mm,0);}}}
 \newcommand{\dashlt}{\raisebox{2pt}{\tikz{\draw[black,dashed,line width=0.9pt](0,0) -- (5mm,0);}}}
 \newcommand{\symbolsquarelt}{\raisebox{0.5pt}{\tikz{\node[draw,scale=0.4,regular polygon, regular polygon sides=4,fill=none](){};}}}
 \newcommand{\symbolsquaresolidline}{\raisebox{0pt}{\tikz{\draw[black,solid,line width = 1.0pt](3.5mm,-0.2mm) rectangle (5.5mm,1.8mm);\draw[-,black,solid,line width = 1.0pt](0.,0.8mm) -- (9mm,0.8mm)}}}    
 \newcommand{\symbolfillsquaresolidline}{\raisebox{0pt}{\tikz{\draw[black,solid,line width = 1.0pt, fill = black](3.5mm,-0.2mm) rectangle (5.5mm,1.8mm);\draw[-,black,solid,line width = 1.0pt](0.,0.8mm) -- (9mm,0.8mm)}}}    
 \newcommand{\symbolfillsquarelt}{\raisebox{0.5pt}{\tikz{\node[draw,scale=0.4,regular polygon, regular polygon sides=4,fill=black](){};}}} 
 \newcommand{\greydashedlt}{\raisebox{2pt}{\tikz{\draw[gray,dashed,line width=0.9pt](0,0) -- (5mm,0);}}}
  
 \newcommand{\solidlt}{\raisebox{2pt}{\tikz{\draw[black,solid,line width=0.9pt](0,0) -- (5mm,0);}}}
 \newcommand{\redlt}{\raisebox{2pt}{\tikz{\draw[red,solid,line width=0.9pt](0,0) -- (5mm,0);}}}
 \newcommand{\bluedottedlt}{\raisebox{2pt}{\tikz{\draw[blue,dotted,line width=0.9pt](0,0) -- (5mm,0);}}} 
 \newcommand{\reddashedlt}{\raisebox{2pt}{\tikz{\draw[red,dashed,line width=0.9pt](0,0) -- (5mm,0);}}} 
 \newcommand{\circlelineblack}{\raisebox{0pt}{\tikz{\draw[black,solid,line width = 1.0pt](4.5mm,0.8 mm) circle (1.mm);\draw[-,black,solid,line width = 1.0pt](0.,0.8mm) -- (9mm,0.8mm)}}}    
 \newcommand{\filledcirclelineblack}{\raisebox{0pt}{\tikz{\draw[black,solid,line width = 1.0pt, fill = black](4.5mm,0.8 mm) circle (1.mm);\draw[-,black,solid,line width = 1.0pt](0.,0.8mm) -- (9mm,0.8mm)}}}    
 
  \definecolor{Gray}{gray}{0.9}

 \newcommand{\DPP}[2]{\dfrac{\partial{#1}}{\partial{#2}}}
 \newcommand{\DCP}[2]{\dfrac{\bm{\text{d}}{#1}}{\text{d}{#2}}}

 \newcommand{\LAP}[1]{\nabla^{2}{#1}}
 \newcommand{\grad}[1]{\nabla{#1}}

 \newcommand{\Div}[1]{\nabla\cdot{#1}}
 \definecolor{myblue}{rgb}{0.858, 0.188, 0.478}
 \long\def\comment#1{}
 \newcommand*{\Scale}[2][4]{\scalebox{#1}{\ensuremath{#2}}}%

\allowdisplaybreaks[1]

\newcommand{\bsig}{\bm{\sigma}}

\newcommand{\bu}{\bm{u}}

\newcommand{\br}{\bm{r}}
\newcommand{\bc}{\mathsfb{c}}

\newcommand{\bF}{\bm{F}}
\newcommand{\bfib}{\bm{F}_{\text{IB}}}

\newcommand{\bn}{\bm{n}}

\newcommand{\T}{\bm{\mathsfb{T}}}

\newcommand{\Vp}{{\mathcal{V}_p}}

\newcommand{\Ip}{{\mathcal{I}_p}}
\newcommand{\Fr}{{\mathcal{F}r}}
\newcommand{\Voro}{Vorono{\"\i} }

\shorttitle{Dense particles in vertical channel flows}
\shortauthor{A. Esteghamatian and T. A. Zaki}

\title{The dynamics of dense particles in vertical channel flows: gravity, lift and particle clusters}

\author{Amir Esteghamatian
\and Tamer A. Zaki\corresp{\email{t.zaki@jhu.edu}}}

\affiliation{Department of Mechanical Engineering, Johns Hopkins University,
	Baltimore, MD 21218, USA}

\begin{document}

\maketitle

\begin{abstract}
The dynamics of dense finite-size particles in vertical channel flows of Newtonian and viscoelastic carrier fluids are examined using particle resolved simulations.  Comparison to neutrally buoyant particles in the same configuration highlights the effect of settling.  
The particle volume fraction is $5\%$, and a gravity field acts counter to the flow direction.
Despite a modest density ratio ($\rho_r = 1.15$), qualitative changes arise due to the relative velocity between the particle and fluid phases. 
While dense particles are homogeneously distributed in the core of the channel, the mean concentration profile peaks in the near-wall region due to a competition between shear- and rotation-induced lift forces.
These forces act in the cross-stream directions, and are analyzed by evaluating conditional averages along individual particle trajectories. 
The correlation between the angular and translational velocities of the particles highlights the significance of the Magnus lift force in both the spanwise and wall-normal directions.
The collective behaviour of the particles is also intriguing.
Using a \Voro analysis, strong clustering is identified in dense particles near the wall, which is shown to alter their streamwise velocities.
This clustering is attributed to the preferential transport of aggregated particles towards the wall. 
The practical implication of the non-uniformity of particle distribution is a significant increase in drag.
When the carrier fluid is viscoelastic, the particle migration is enhanced which leads to larger stresses, thus negating the capacity of viscoelasticity to reduce turbulent drag. 

\end{abstract}

\section{Introduction}
Turbulent particle-laden flows are widely encountered in various environmental and engineering applications.
A vast majority of the literature on particles-turbulence interactions in wall-bounded flows is dedicated to the study of neutrally buoyant particles, where the mean particle slip velocity vanishes throughout the domain except in the near-wall region.
A finite density ratio in the presence of gravity results in settling of particles, and the sustained wake region near the particles substantially alters the nature of particle-turbulence interaction.
In the present work, we investigated the effect of dense finite-size particles in vertical channel flows of Newtonian and viscoelastic carrier fluids using direct numerical simulations.

When the particle size is smaller or on the same order as the viscous dissipation length, turbulence modulation primarily depends on the Stokes number, defined as the ratio of particles to fluid response times \citep[see][for regime classification based on the Stokes number]{elgobashi2006updated}.  
A gravitational acceleration decouples the particles motion from the fluid, increases the Stokes number and, as a result, has an immediate impact on turbulence modulation. 
It is known that small and heavy particles attenuate the turbulence intensity in homogeneous isotropic condition \citep{gore1989effect,elghobashi1993two} as well as in turbulent channel flow \citep{tsuji1984ldv}. 
In the latter case the attenuation is particularly important in the cross-stream direction \citep{kulick1994particle}.
At high particle concentrations, however, the strong inter-phase coupling triggers an alternative turbulence generation mechanism, where the turbulent kinetic energy is entirely generated by particles slip velocities \citep{Capecelatro2018a}.
Particles motion is also influenced by gravity:
In contrast to inertialess particles (or fluid elements) that are trapped inside eddies, dense particles escape the initially-surrounding eddies faster than the eddy decay rate\textemdash a phenomenon known as ``crossing trajectories" which decreases particles dispersion due to the subdued correlated motions \citep{wells1983effects}.

When the particle size lies in the inertial range of the turbulence, its wake interacts with turbulent eddies of different sizes \citep{Cisse2013}, and the region of interaction extends to several particle diameters \citep{Naso2010}.  
\citet{zeng2010wake} studied isolated stationary particles in turbulent channel flow and for a wide range of particle Reynolds numbers  ($\Rey_p\equiv d_p u_\text{slip}/\nu = [42-450]$, where $\nu$, $d_p$ and $u_\text{slip}$ denote fluid kinematic viscosity, particle diameter and slip velocity).
The authors highlighted the dual effect of particles, where the turbulence kinetic energy is 
(i) suppressed due to the damping of the streamwise fluctuations and 
(ii) augmented by vortex shedding and/or wake oscillations.
Towards the lower range of particle Reynolds numbers ($\Rey_p \approx [26-33]$)), \citet{vreman2018turbulent} considered a similar configuration and reported that particles create sinks of turbulence kinetic energy near their surfaces, although the turbulence suppression remains appreciable far from the particles. 
They concluded that the primary reason for turbulence suppression is particle-induced non-uniformity of the mean driving force of the flow, rather than turbulence modulations in the vicinity of the particles.	
At the higher particle Reynolds numbers ($\Rey_p \approx 400$), \citet{kajishima2001turbulence} studied freely-moving finite-size particles and investigated the influence of vortex shedding on turbulence production in a vertical channel. 

At $\Rey_p \approx 136$ and low volume fractions $\varphi \approx 0.5 \%$, \citet{uhlmann2008interface} reported enhanced wall-normal mixing with flattening of the mean velocity profiles.
They attributed the near-wall peak of particles concentrations to migration of particles in the direction opposite to gradients in the turbulence intensity\textemdash a phenomenon known as``turbophoresis" \citep{Caporaloni1975}.   
In a follow-up study, \citep{garcia2012dns} examined the same configuration with a twice longer streamwise length and they confirmed that particle and fluid statistics are not strongly affected by the streamwise extension of the domain, yet extremely long columnar flow structures remain correlated even at streamwise lengths on the order of $16h$, where $h$ is the channel half-width. Settling has also been examined in other canonical configurations including horizontal wall-bounded flows and homogeneous shear. In the former, \citet{Shao2012a} concluded that at high density ratios particles form a sediment layer over the horizontal wall, which contributes to the turbulence intensity by vortex shedding, and \citet{Kidanemariam2013b} observed an apparent lag in the mean particle velocity compared with that of the fluid due to particle accumulation in the low-speed regions.
When gravity is directed perpendicular to the plane of homogeneous shear, \citet{tanaka2017effect} concluded that the Reynolds shear stress is markedly decreased in the wake of particles and, in turn, the amplification of turbulent kinetic energy is reduced. 

Another important implication of gravity and sustained particle wakes is the emergence of particle clusters. 
In quiescent flow conditions and at sufficiently large $\Rey_p$, the reduced drag acting upon trailing particles promotes cluster formation \citep{wu1998dynamics,fortes1987nonlinear}. 
This phenomenon has been extensively studied both experimentally \citep{huisman2016columnar} and numerically \citep{yin2007hindered}, and the onset of clustering is marked by a critical Galileo number which is defined as the ratio between gravitational and viscous forces \citep{Uhlmann2014}.
In the presence of background turbulence, \citep{chouippe2019influence} examined the clustering of finite-size particles in the presence of gravity, and highlighted that the background turbulence hinders cluster formation. 	
\citet{fiabane2012clustering} experimentally examined the clustering of finite-size particles, and concluded that a finite fluid/particle density difference is crucial for cluster formation. 
In a vertical turbulent channel, cluster formation induced by wake attraction is reported for particles with an unsteady wake \citep{kajishima2004influence}, while particle distribution is reportedly more homogeneous than a random distribution when wake oscillations are absent \citep{garcia2012dns}. 
Overall, the influence of gravity on cluster formation is widely associated with the wake attraction mechanism, while the role of lateral forces on particles clustering and collective motion is less explored. 

The literature on particle-turbulence interactions has predominately focused on Newtonian fluids. 
In absence of turbulence, particles in non-Newtonian flows has been studied \citep{Zenit2018}, and the formation of particle-rich vertical columns has been observed in experiments of non-Brownian particles settling in a shear-thinning fluid \citep{bobroff1998nuclear,mora2005structuring}.  Once turbulence enters the picture, the vast literature on viscoelasticity \citep{White2008,samanta2013elasto,Agarwal2014,Choueiri2018} has considered single-phase conditions for three reasons.  Firstly, polymers have been among the most effective drag reduction technologies \citep{Toms1948,Virk1970}.  Secondly, even single-phase viscoelastic flows exhibit puzzling dynamics that defy intuition based on Newtonian conditions. Examples include vorticity propagation as waves in visco-elastic flows \citep{Page2014}, enstrophy amplification in two dimensions due to polymer torque \citep{Page2015,Page2016}, and the initial destabilization of supercritical channel flow by increasing elasticity prior to its stabilization at higher Weissenberg numbers \citep{Lee2017}.  The second reason is the lack of a mathematical framework that can facilitate progress in characterizing the state of polymers in turbulence. This point was addressed recently by a rigorous formulation for evaluating the mean conformation tensor \citep{Hameduddin2019b} and quantifying departures from that mean \citep{Hameduddin2018,Hameduddin2019a}.

Building upon the above studies, it was recently possible to analyze  the interplay of viscoelasticity, particles and turbulence for neutrally buoyant particles \citep{esteghamatian2019dilute,esteghamatian2020viscoelasticity}. 
In a viscoelastic channel flow at $5\%$ particle concentration, particles periodically migrate away and towards the wall, while suppressing the turbulent activity, or reinforcing its hibernation. 
Viscoelasticity increased the formation of particle pairs with an angled alignment with respect to the streamwise direction, while no significant large-scale long-lived structure was identified \citep{esteghamatian2019dilute}.
At $20\%$ particle concentration, a viscoelastic drag increase was reported for elasticity levels above a critical value. 
The enhanced polymer stress was attributed to strong deformations of polymers in the vicinity of particle surface \citep{esteghamatian2020viscoelasticity}.

In this work, we explore the settling effect of particles in vertical channel flow of Newtonian and viscoelastic fluids with direct numerical simulations.
The particles are larger than the Kolmogorov length scale, and eddies at inertial ranges are expectedly affected by particles boundary layer and wake. 
Therefore, the computations resolve the flow at the scale of particles with an immersed boundary method, and the polymer forces are obtained by solving evolution equations for the polymer conformation tensor.
The flow configuration, governing equations and computational setup are described in \S\ref{sec:method}. 
In \S\ref{sec:results}, we examine the impact of particle settling on multiple aspects of the channel flow dynamics including: the mean concentration and velocity profiles (\S\ref{sec:mean}); particle lift forces (\S\ref{sec:lift}); clusters and microstructure  (S\ref{sec:cluster}); and momentum balance and velocity fluctuations (\S\ref{sec:turb}). 
Concluding remarks are provided in \S\ref{sec:conclusion}.

\begin{figure}		
\centering
\includegraphics[width =0.25\textwidth,scale=1]{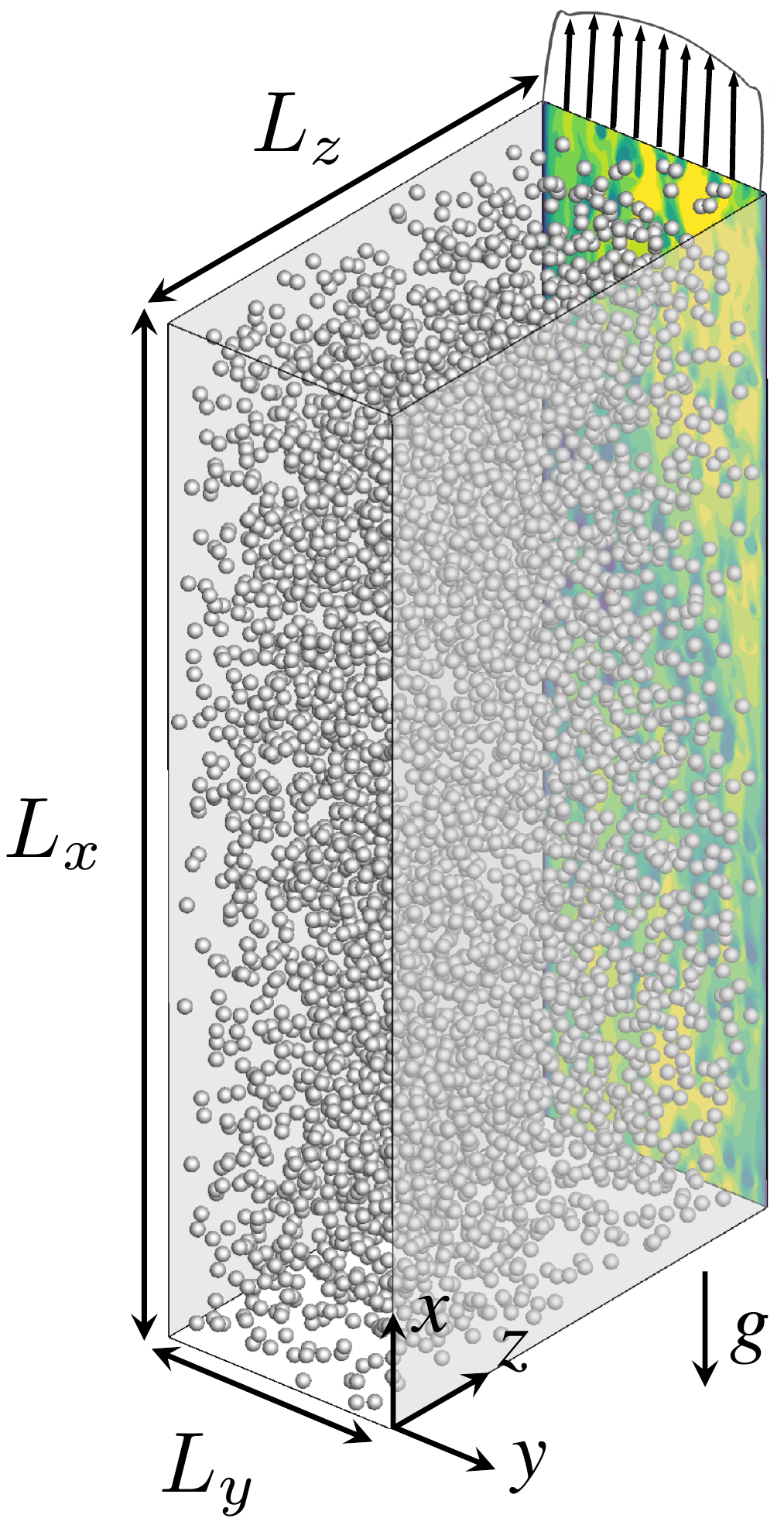}						
\caption{Schematic of the particle-laden simulations. No-slip boundary conditions $\bm{u}_f = 0$ are imposed at $y = \{0, 2\}$, and periodicity is enforced in the $x$ and $z$ directions. }   \label{fig:vis}		
\end{figure}
\section{Governing equations and numerical setup \label{sec:method}}
The computational domain is a plane channel, where the streamwise, wall-normal and spanwise directions are denoted $x$, $y$ and $z$ (figure \ref{fig:vis}). 
The flow is maintained at a constant mass flux in $x$ direction, and a gravitational field $\bm{g}$ acts in the opposite direction.
The bulk velocity, $U_b$, and the channel half-height, $h$, are adopted as characteristic scales. 
The single-phase simulations are characterized by Reynolds number $\Rey = hU_b/\nu$ and Weissenberg number $Wi = \lambda U_b/h$, where $\nu$ and $\lambda$ are the fluid total kinematic viscosity and the polymer relaxation time. 
The particle-laden configuration requires four additional dimensionless numbers:
(i) dimensionless diameter of the particles $d_p/h$;
(ii) density ratio between the particle and fluid phases, $\rho_r \equiv \rho_p/\rho_f$;
(iii) bulk volume fraction $\varphi \equiv N_p \mathcal{V}_p/\mathcal{V}_t$, where $N_p$ is the number of particles, $\mathcal{V}_p$ is the volume of an individual particle, and $\mathcal{V}_t$ is the volume of the computational domain;
(i) Froude number defined as $\Fr =U^2_b/(|\bm{g}|h)$. 
The governing equations for the fluid velocity $\bu_f$, the hydrodynamic pressure $p$, and the polymer conformation tensor $\bc$ in dimensionless form are, 
\begin{align}
& \Div{\bu_f} = 0 \label{eq:mass} \\
& \DPP{\bu_f}{t} + \bu_f \cdot \grad{\bu_f}+\grad{p} - \dfrac{\beta}{Re} \LAP{\bu_f} - \dfrac{1-\beta}{Re} \Div{\T} - \bF = 0 \label{eq:momentum_f} \\
& \DPP{\bc}{t} +\bu_f \cdot \grad{\bc}- \bc \cdot \grad{\bu_f} - (\bc \cdot \grad{\bu}_f)^T - \T = 0. \label{eq:conform} 
\end{align}
In the above equations, $\bF$ is a generic force, and $\beta \equiv {\mu_s}/\left({\mu_s+\mu_p}\right)$ is the ratio of solvent to total viscosity with the latter being comprised of the solvent $\mu_s$ and polymer $\mu_p$ contributions. 
For the present dilute polymer solution, the viscoelastic stress tensor $\T$ is expressed in terms of the conformation tensor using the FENE-P model,
\begin{align}
\T \equiv \dfrac{1}{Wi} \left(\dfrac{\bc}{\psi}-\dfrac{\mathsfbi{I}}{a} \right), \qquad
\psi = 1 - \dfrac{\text{tr}(\bc)}{L^2_\text{max}}, \qquad  
a = 1- \dfrac{3}{L^2_\text{max}}, 
\end{align}
where $\mathsfbi{I}$ is the isotropic tensor and $L_\text{max}$ is the maximum extensibility of the polymer chains.
Particles motion in the presence of a gravitational field is governed by the Newton-Euler equations,
\begin{align}
&	\Vp \DCP{\bu_p}{t} = \oint_{\partial \Vp} \bsig_f \cdot \bn \text{d} \mathcal{A} + \bF_c + \dfrac{1}{\Fr}(\rho_r-1)\dfrac{\bm{g}}{|\bm{g}|}  \label{eq:pvel} \\ 
&	\mathcal{I}_p \DCP{\bm{\Omega}_p}{t} = \oint_{\partial \Vp} \br \times \bsig_f \cdot \bn \text{d} \mathcal{A},
\end{align}
where $\bu_p$, $\bm{\Omega}_p$, $\Vp = \dfrac{\pi d_p^3}{6}$ and $\Ip= \dfrac{\pi d_p^5}{60}$ are the translational and angular velocities and dimensionless volume and moment of inertia of a spherical particle. 
The hydrodynamic stress tensor, $\bm{\sigma}_f$, is integrated over a particle's surface, and $\bn$ denotes the outward unit vector normal to that surface. 
The particle-particle and particle-wall repulsive collision forces, $\bF_c$, are applied in the opposite direction to $\bn$ \citep{Glowinski2001}, and the last term on the right hand side of \eqref{eq:pvel} accounts for the buoyancy force.

A complete description of our numerical algorithm is provided by \citet{esteghamatian2019dilute} and, for completeness, only the main features are described here. 
The flow equations  \eqref{eq:mass} and \eqref{eq:momentum_f} were solved using a fractional step algorithm on a staggered grid with a local volume-flux formulation  \citep{Rosenfeld1991}, and the conformation-tensor equations \eqref{eq:conform} were solved using a third-order accurate Runge-Kutta method. 
The viscous terms were solved implicitly in time using the Crank-Nicolson scheme, while an explicit Adams-Bashforth scheme is adopted for advection and stretching terms in \eqref{eq:momentum_f} and \eqref{eq:conform}. 
Following \citet{Dubief2005}, a semi-implicit approach that ensures the finite extensibility of the polymers was used to discretize the polymer stress term.
In particle-laden cases, a sharp-interface immersed boundary force field \citep{Nicolaou2015} enforces the no-slip boundary condition at the surface of the particles ($\bF = \bfib$ in \eqref{eq:momentum_f}), and the conformation tensor is set to unity inside the particle domain.
The algorithm was extensively validated for simulations of Newtonian and viscoelastic flows \citep{Wang_jcp2019,Lee2017} including particle-laden conditions \citep{esteghamatian2019dilute}.

The set of physical parameters are selected such that they demonstrate the effect of settling on the dynamics of the flow and particles. 
The simulations will contrast dense particles in Newtonian and viscoelastic fluids (designated ``W0P5D" and ``W15P5D") to neutrally buoyant conditions (``W0P5" and ``W15P5") and single-phase flows (``W0" and ``W15").  In the designation, ``W" refers to the Weissenberg number which is either $Wi=0$ for Newtonian or $Wi=15$ for viscoelastic flow; ``P" refers to the presence of a particle phase with bulk particle volume fraction $\varphi  = 5 \%$ and dimensionless particle diameter $d_p/h = 1/9$;  ``D" indicates dense particles.  
A complete list of physical and computational parameters is provided in table \ref{tab:params}. 
In the Newtonian single-phase configuration, the friction Reynolds number $\Rey_\tau \equiv hu_\tau/\nu \approx180$, where $u_\tau \equiv \sqrt{\langle \tau_\text{w} \rangle/\rho}$ denotes the friction velocity and $\langle \tau_\text{w} \rangle$ is the average wall shear stress.
With a particle response time defined by the Stokes drag, $t_p \equiv {d_p}^2 \rho_p/18\mu$, the Stokes number is $St\equiv  h t_p/U_b \sim O(1)$ or $St^+\equiv  t_pu^2_\tau/\nu \sim O(10)$.  Interaction between the fluid and particle phases is therefore expected.

A uniform Cartesian grid was used in all three directions in the particle-laden simulations.
In the single-phase cases, the grid spacing was uniform in the spanwise and streamwise dimensions, and a hyperbolic tangent grid stretching was adopted in the wall-normal direction.
The domain size was $6\times 2 \times 3$ in streamwise, wall-normal, and spanwise directions in single-phase and neutrally-buoyant particle-laden cases, while in dense particle-laden cases the domain size was extended to $8\times 2 \times 4.5$.
The grid size was set to $d_p/ \Delta x=16$ in all particle-laden conditions.
Beyond an initial transient, statistics were collected until convergence. 
Phase-averaged statistics are related to unconditional counterparts with a phase indicator $\chi$ that is zero in the fluid and unity in the particle phase, 
\begin{align}
\langle o\rangle =\langle (1-\chi)o_f\rangle + \langle \chi o_p\rangle  =  (1-\phi)\langle o_f \rangle_f + \phi \langle o_p \rangle_p, \label{eq:stat} 
\end{align}
where $\phi$ denotes the particle volume fraction. For brevity, the subscripts $\{f,p\}$ are hereafter omitted from the averaging symbol. 
Unless otherwise stated, the averaging operation is performed in time and in homogeneous $x$ and $z$ directions.
\begin{table}
\begin{center}
\def~{\hphantom{0}}
\setlength\extrarowheight{2	pt}
\begin{tabular}{c|c|c|c|c|c|c|c|l}	
$Re, \Fr^{-1}$   &   $\beta, L_\text{max}$ & $Wi$ & $\varphi (\%) $  & $\rho_r$&$1/d_p$  & $L_x \times L_y \times L_z$ &$N_x \times N_y \times N_z$ &Case    \\[3pt] \hline
\multirow{3}{*}{}   & \multirow{4}{*}{Newtonian}& \multirow{4}{*}{0} & 0  & -& - & $6 \times 2 \times 3$ & $96 \times 192 \times 80 $& W0  \\ [3pt]	\cline{4-9}					
 & & & 5  & 1 & 9 & $6 \times 2 \times 3$ &  $864 \times 288 \times 432 $& W0P5 \\ [3pt]	\cline{4-9}		
\multirow{1}{*}{2800,} 	&  &  &\cellcolor{Gray} 5 &\cellcolor{Gray} 1.15 & \cellcolor{Gray} 9 &\cellcolor{Gray} $8 \times 2 \times 4.5$  &  \cellcolor{Gray} $1152 \times 288 \times 576 $ & \cellcolor{Gray} W0P5D \\ [3pt] \cline{2-9}			
\multirow{1}{*}{9.81}  & \multirow{2}{*}{$\beta = 0.97$}& \multirow{4}{*}{15} & 0  & -& - & $6 \times 2 \times 3$ & $96 \times 192 \times 80 $& W15  \\ [3pt]	\cline{4-9}					
& \multirow{3}{*}{$L_\text{max} = 70$}& & 5  & 1& 9 & $6 \times 2 \times 3$ & $96 \times 192 \times 80 $& W15P5  \\ [3pt]	\cline{4-9}		
&  &  &\cellcolor{Gray} 5 &\cellcolor{Gray} 1.15&\cellcolor{Gray} 9 &\cellcolor{Gray} $8 \times 2 \times 4.5$  & \cellcolor{Gray} $1152 \times 288 \times 576 $ &\cellcolor{Gray} W15P5D \\ [3pt] \cline{2-9}		\hline
\end{tabular}
\caption{Physical and computational parameters of the simulations. Reynolds number $\Rey \equiv hU_b/\nu$ and Weissenberg number $Wi \equiv \lambda U_b/h$ are based on the channel half-height $h$, the bulk velocity $U_b$, total kinematic viscosity $\nu$ and viscoelastic fluid relaxation time $\lambda$. The domain sizes are $L_{\{x,y,z\}}$ in $\{x,y,z\}$ directions, and the numbers of grid cells are $N_{\{x,y,z\}}$.  }
\label{tab:params}
\end{center}
\end{table}

\section{Results \label{sec:results}}
An instantaneous field from the Newtonian dense-particle-laden case (W0P5D) is shown in figure \ref{fig:vis} which highlights key features that are exclusive to the dense particle cases.
Collective particle motion is evident from their large-scale velocity structures.
Wakes downstream of individual particles highlight the potential impact on the collective motion and cluster formation. 
Particle motion affects the fluid-phase streamwise velocity structures as shown by contours of $u_f$, both at a micro-scale due to the finite slip velocities, and at a global scale due to the collective motion.
Since the flow is dominated by particles, we start by investigating the mechanisms controlling the particles motion. 
Subsequently, a discussion on the impact of gravity and particle wakes on the mean stress balance and fluid fluctuations is provided.

\begin{figure}		
    \centering
    \includegraphics[width =0.6\textwidth,scale=1]{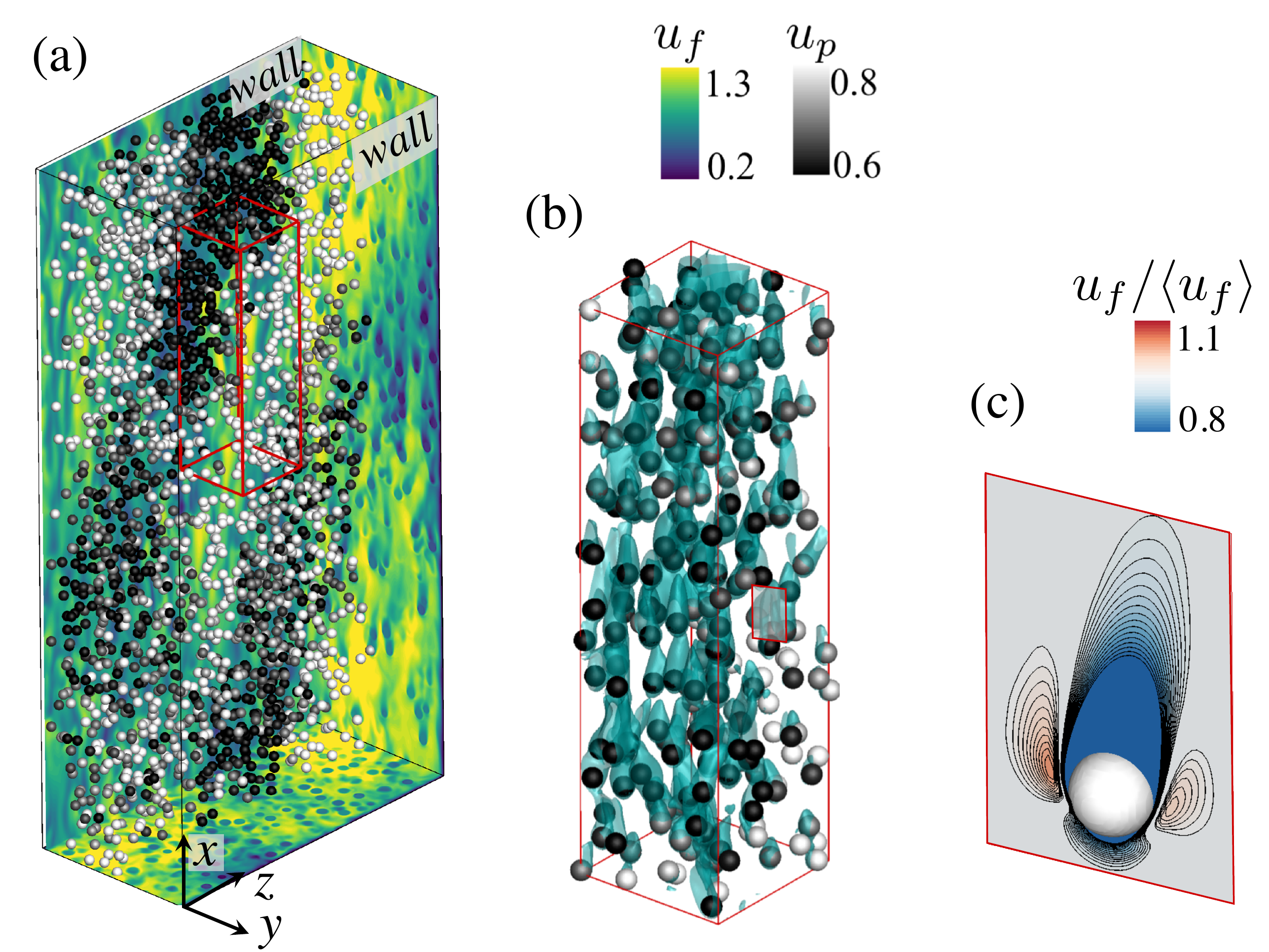}						
    \caption{(a) Instantaneous snapshot of particle positions coloured by their velocity and contours of $u_f$ for case W0P5D. For clarity only particles located at $y<1$ are displayed. (b) A subset of the domain is magnified and reproduced with isosurfaces of $u_f=0.7$. (c) Contours of $u_f/\langle u_f \rangle$ in the $(x,y)$ plane in the vicinity of one particle.  }   
\label{fig:vis}
\end{figure}

\subsection{Particle concentration and mean velocity profiles \label{sec:mean}}

\begin{figure}	
\subfigure[]{\label{fig:up_W0}
\begin{minipage}[b]{0.3\textwidth}
\begin{center}
\includegraphics[width =\textwidth,scale=1]{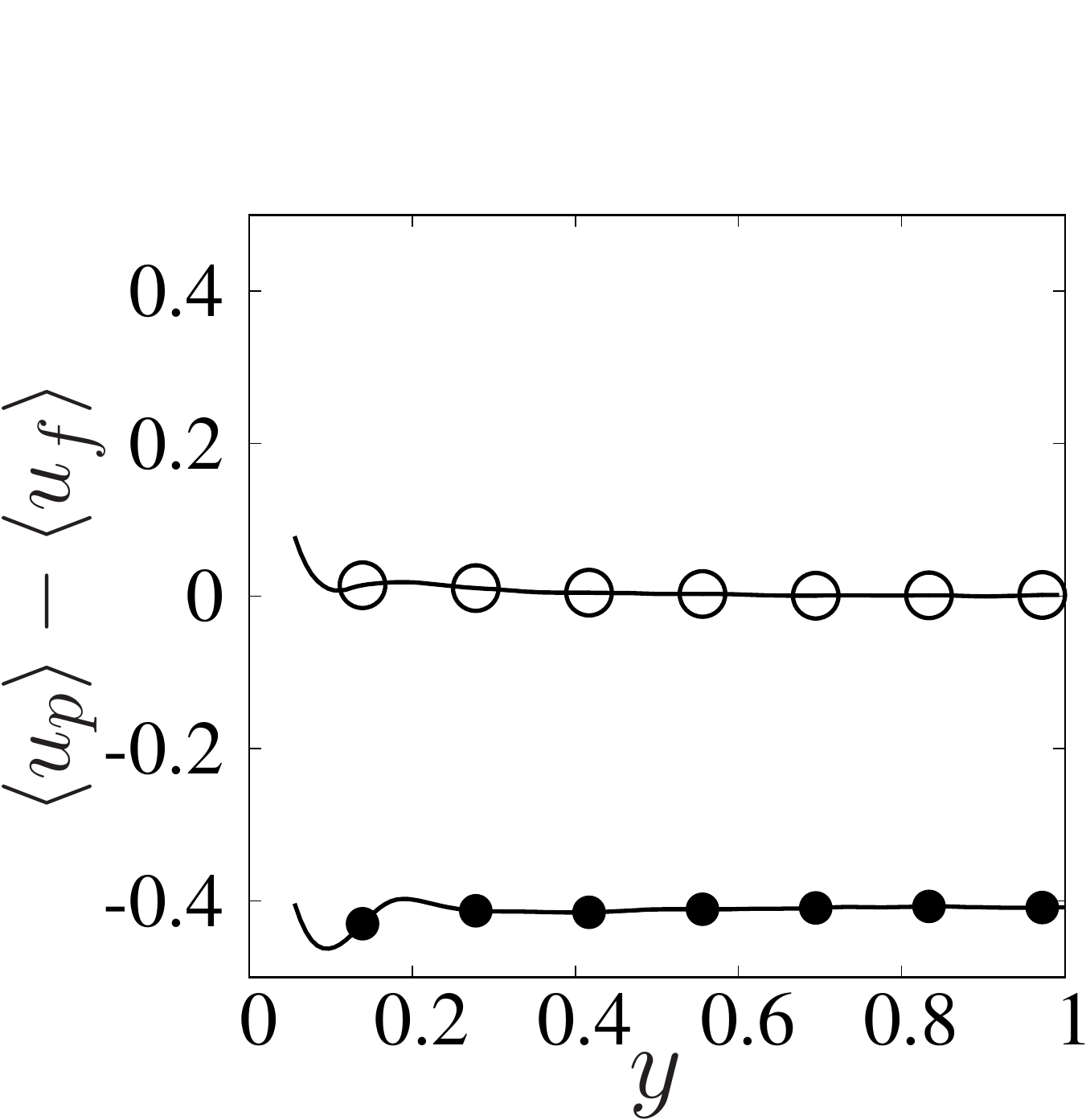}
\end{center}
\end{minipage}}		
\subfigure[]{\label{fig:phi_W0}
\begin{minipage}[b]{0.29\textwidth}
\begin{center}
\includegraphics[width =\textwidth,scale=1]{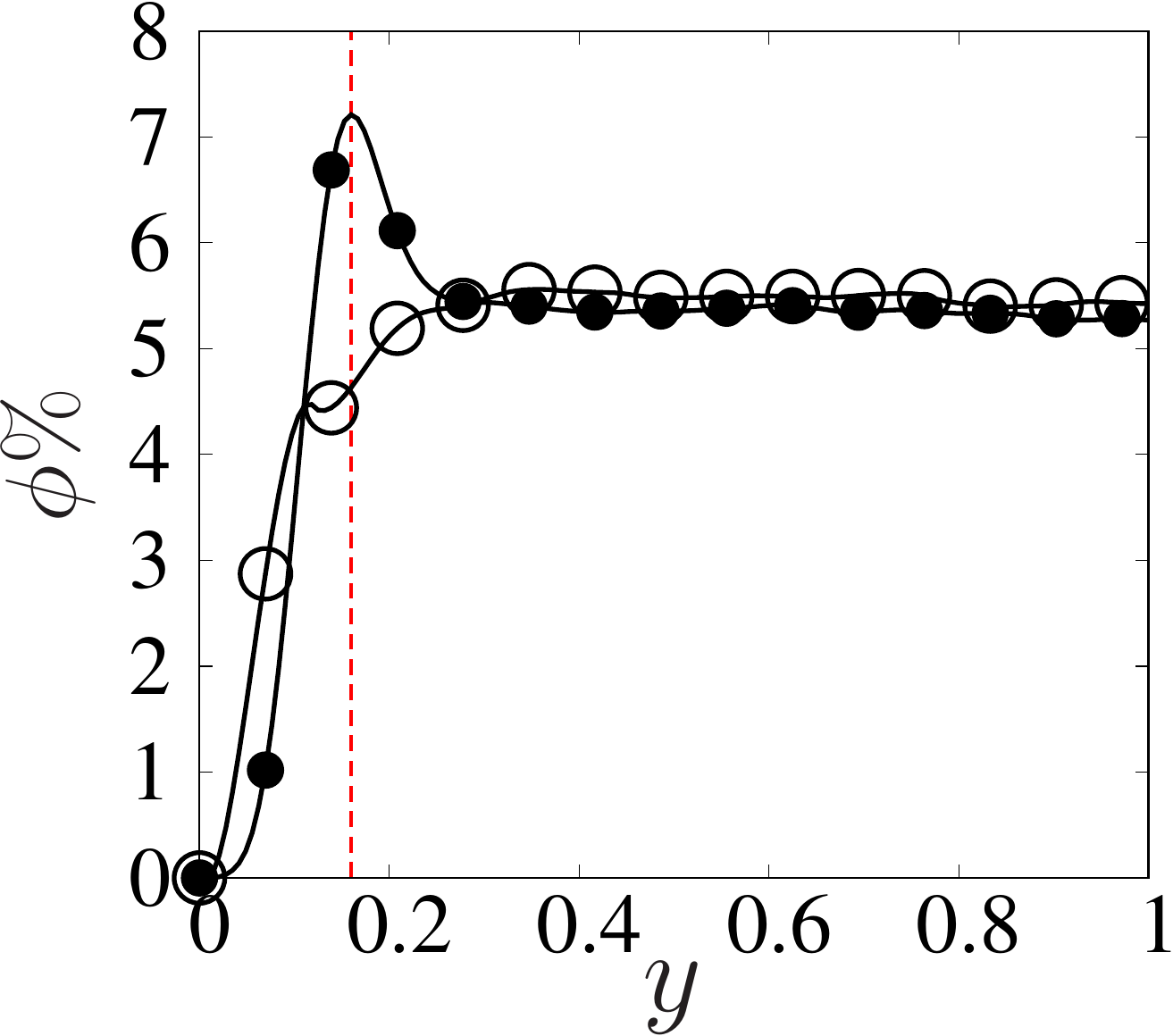}
\end{center}
\end{minipage}}			
\subfigure[]{\label{fig:uf_W0}
\begin{minipage}[b]{0.30\textwidth}
\begin{center}
\includegraphics[width =\textwidth,scale=1]{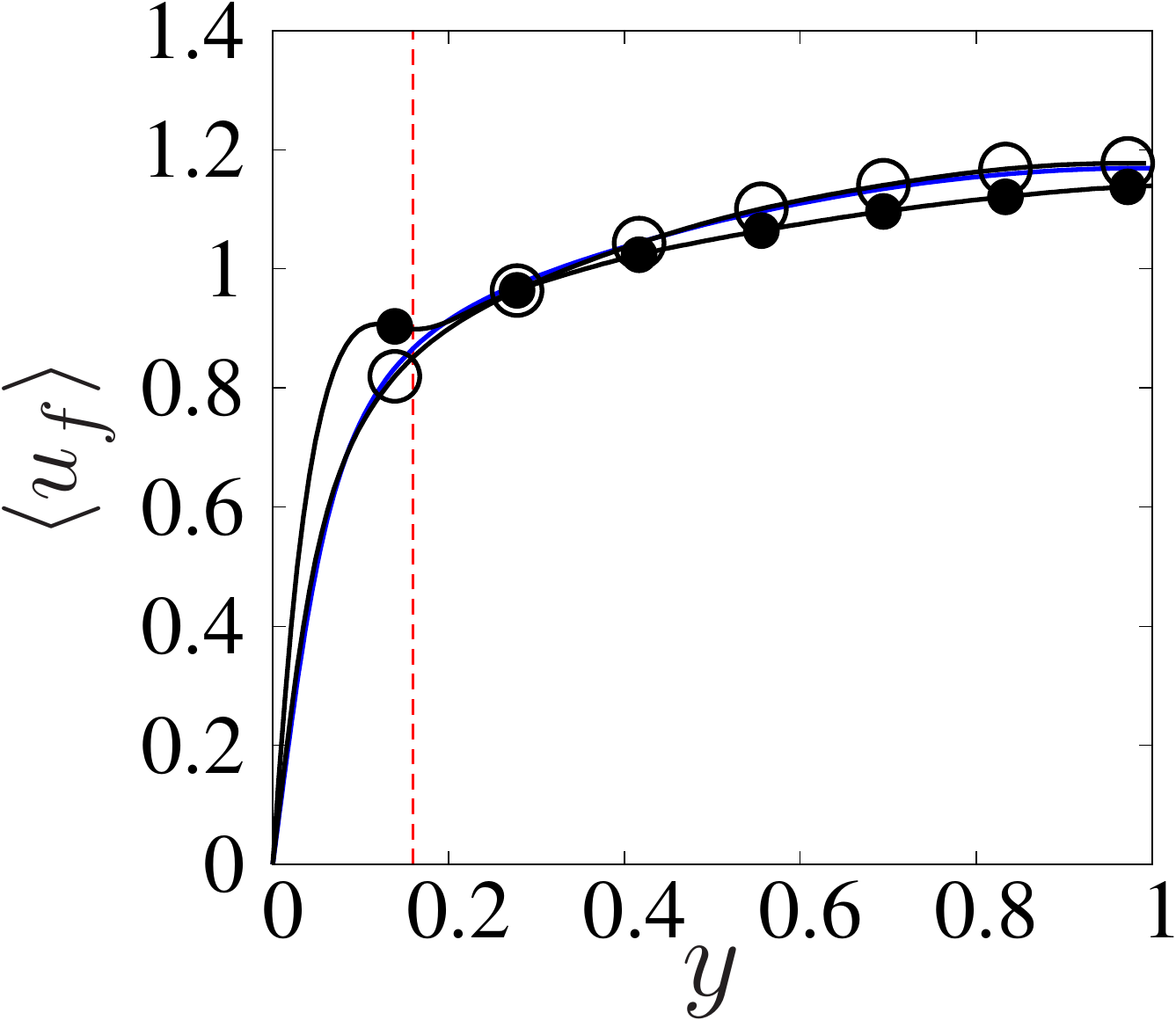}
\end{center}
\end{minipage}}	\\
\subfigure[]{\label{fig:up_W15}
\begin{minipage}[b]{0.3\textwidth}
\begin{center}
\includegraphics[width =\textwidth,scale=1]{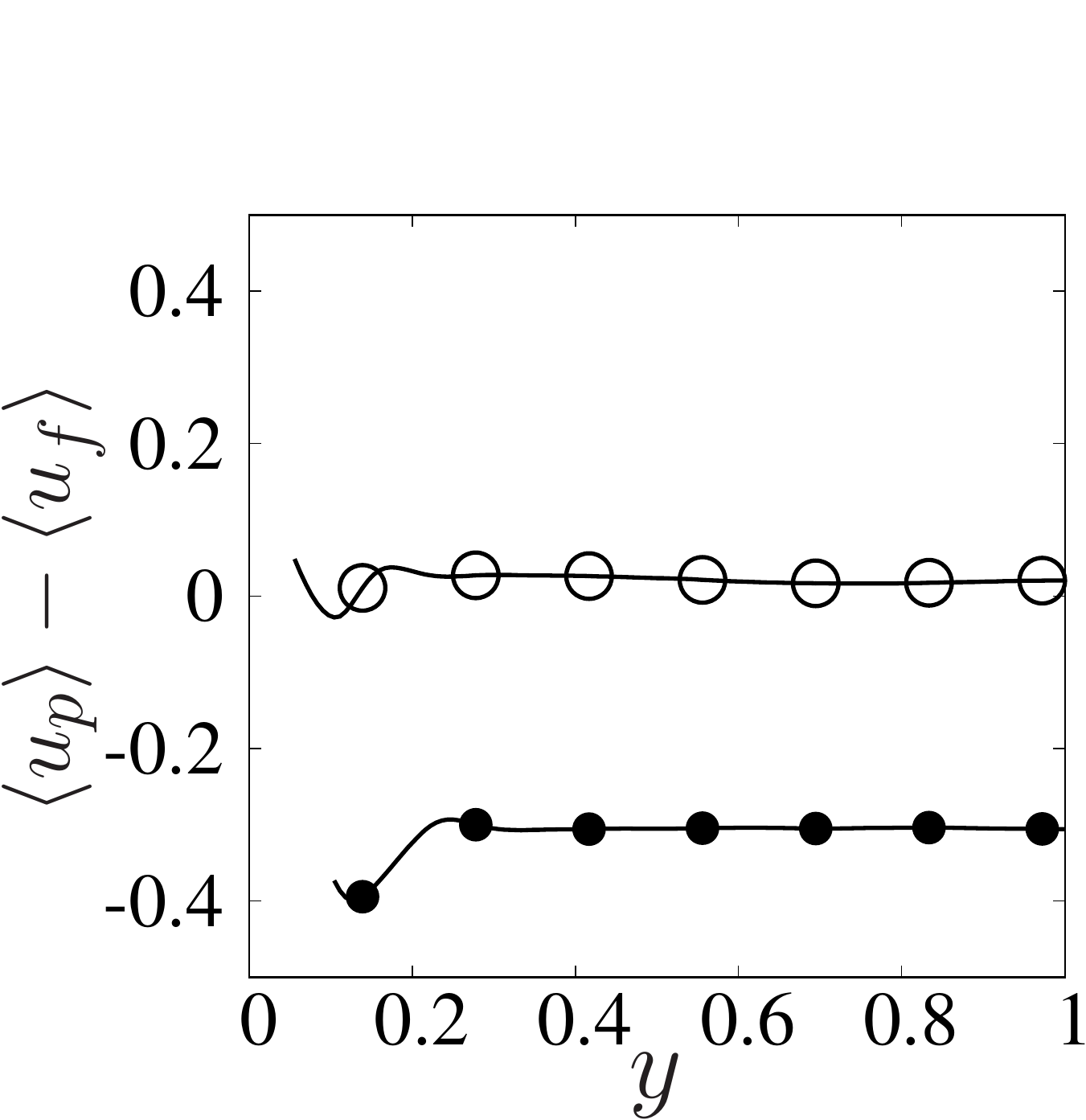}
\end{center}
\end{minipage}}					
\subfigure[]{\label{fig:phi_W15}
\begin{minipage}[b]{0.29\textwidth}
\begin{center}
\includegraphics[width =\textwidth,scale=1]{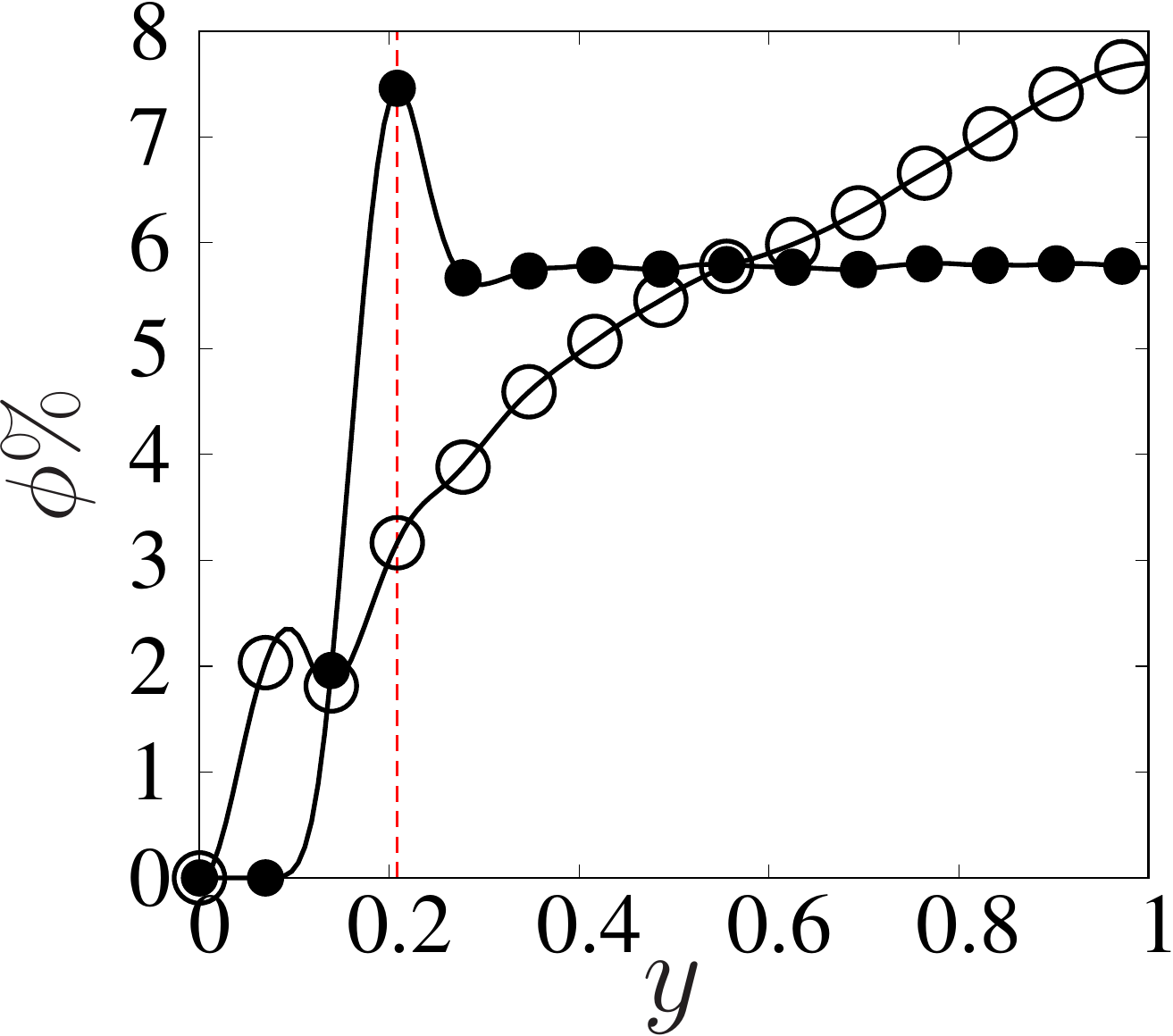}
\end{center}
\end{minipage}}			
\subfigure[]{\label{fig:uf_W15}
\begin{minipage}[b]{0.3\textwidth}
\begin{center}
\includegraphics[width =\textwidth,scale=1]{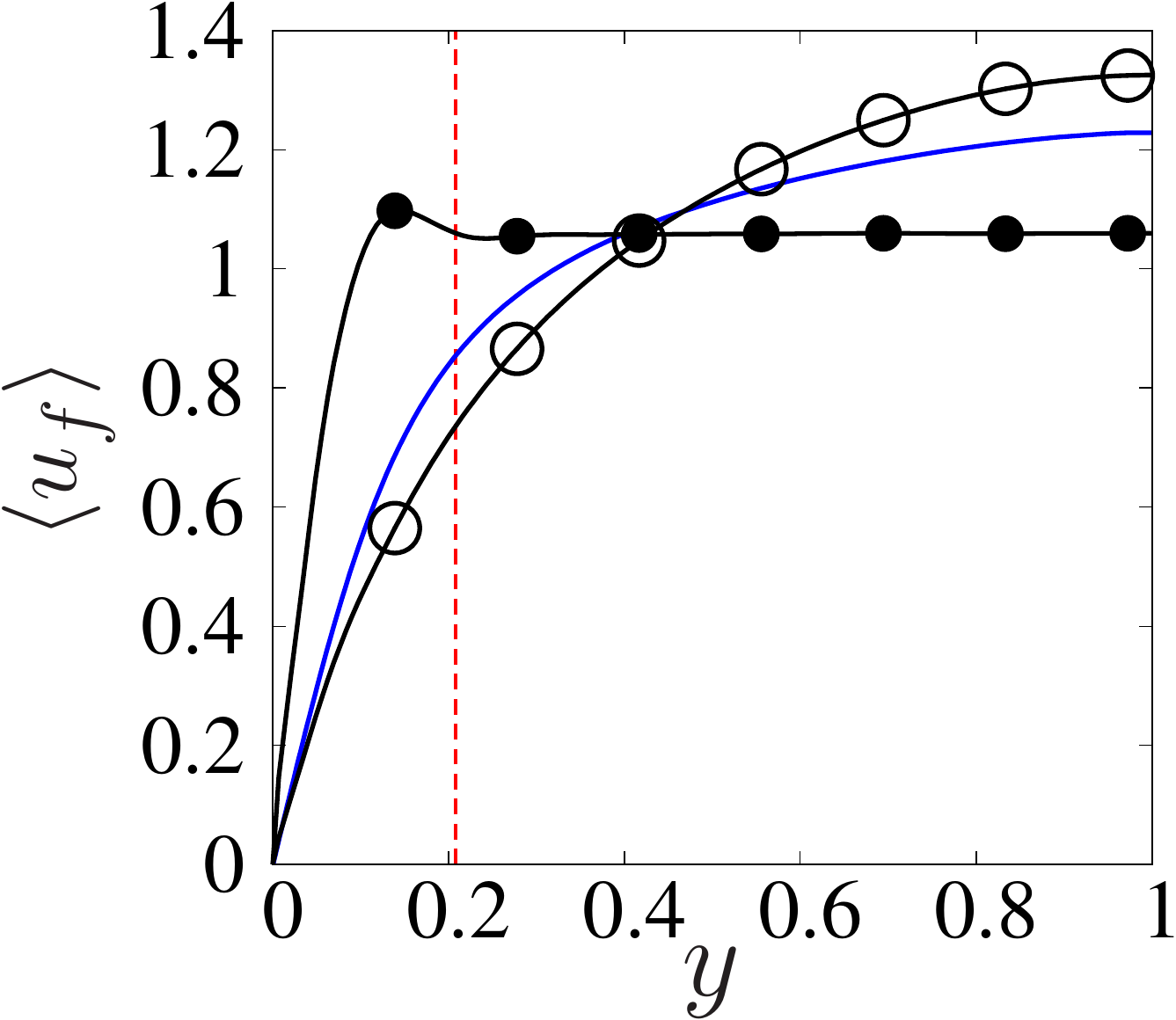}
\end{center}
\end{minipage}}																			
\caption{Profiles of average (a,d) particle velocity relative to the fluid, (b,e) particle concentration and (c,f) fluid velocity. Suspensions with ({\protect\circlelineblack}) neutrally buoyant and ({\protect\filledcirclelineblack}) dense particles; ({\protect\bluelt}) single-phase condition.
The fluid in the top row (a,b,c) is Newtonian and in the bottom row (d,e,f) is viscoelastic.
Vertical dashed lines in (b,c) and (e,f) mark the locations of the peak particles concentration. 
}   \label{fig:phi_uf_up}		
\end{figure}

Multiple physical mechanisms that distinguish dense particle cases from neutrally buoyant counterparts are direct consequences of the relative velocity of the particles with respect to the fluid. 
As shown in figures \ref{fig:up_W0} and \ref{fig:up_W15}, the particles sustain a negative relative velocity and as a result a positive drag that counter-balances the buoyancy force exerted in the negative $x$ direction. 
Since the buoyancy force is constant at all $y$ locations, the particle slip velocity $\langle u_p \rangle - \langle u_f \rangle$ is nearly constant across the bulk of the flow. 
The smaller magnitude of $\langle u_p \rangle - \langle u_f \rangle$ in the viscoelastic case W15P5D compared to the Newtonian fluid W0P5D suggests that particles experience a larger drag coefficient in the former, in agreement with previous studies \citep{chhabra2006bubbles}.

The concentration profiles of neutrally buoyant and dense particles are reported in figure \ref{fig:phi_W0} for Newtonian fluid and in figure \ref{fig:phi_W15} for the viscoelastic case. 
In the latter, neutrally buoyant particles migrate towards the channel center due to imbalance of elastic normal stresses \citep{d2015particle}.
For dense particles, the repulsion away from the wall is enhanced for $y<0.2$, and the absence of mean shear beyond $y=0.2$ eliminates particle migration towards the center. 
The presence of a local maximum near $y=0.2$ for the dense particle cases, both viscoelastic and Newtonian, is remarkable. 
This peak is qualitatively different than the relatively small maximum in $\phi$ that is discernible in the viscoelastic neutrally buoyant case that was previously attributed to the lubrication forces and the asymmetric inter-particle interactions  \citep{esteghamatian2019dilute}.  That mechanisms cannot explain the strong local maximum in dense cases, where the immediate vicinity of the wall ($y < d_p$ in the case W15P5D) is almost void of particles. 
The lift forces acting perpendicular to the motion of the particles are primarily responsible for the observed trends in the particle concentration profile and will be examined in detail.

The strong repulsion from the wall is due to the Saffman lift force \citep{saffman1965lift}, and can be expressed as $F_\text{S} \equiv -C_l \rho_f \sqrt{\nu} d_p^2 (u_p-u_f) \dot{\gamma}/\sqrt{|\dot{\gamma}|}$, where $\dot{\gamma}$ is the shear rate and $C_l$ is a constant coefficient.
The negative sign of $\langle u_p \rangle - \langle u_f \rangle$ in W0P5D and W15P5dD gives rise to a Saffman lift force displacing the particles away from the immediate vicinity of the wall\textemdash an effect that is insignificant in the neutrally buoyant cases due to the inappreciable relative velocity.
Nonetheless, the Saffman lift force alone cannot fully explain the peak in the particle concentration profile.
As will be discussed further in this section, the presence of the peak is related to the coupling of angular and translational velocities of dense particles.

\begin{figure}		
\centering
\subfigure[]{\label{fig:omegaf_W0}
\begin{minipage}[b]{0.4\textwidth}
\begin{center}
\includegraphics[width =\textwidth,scale=1]{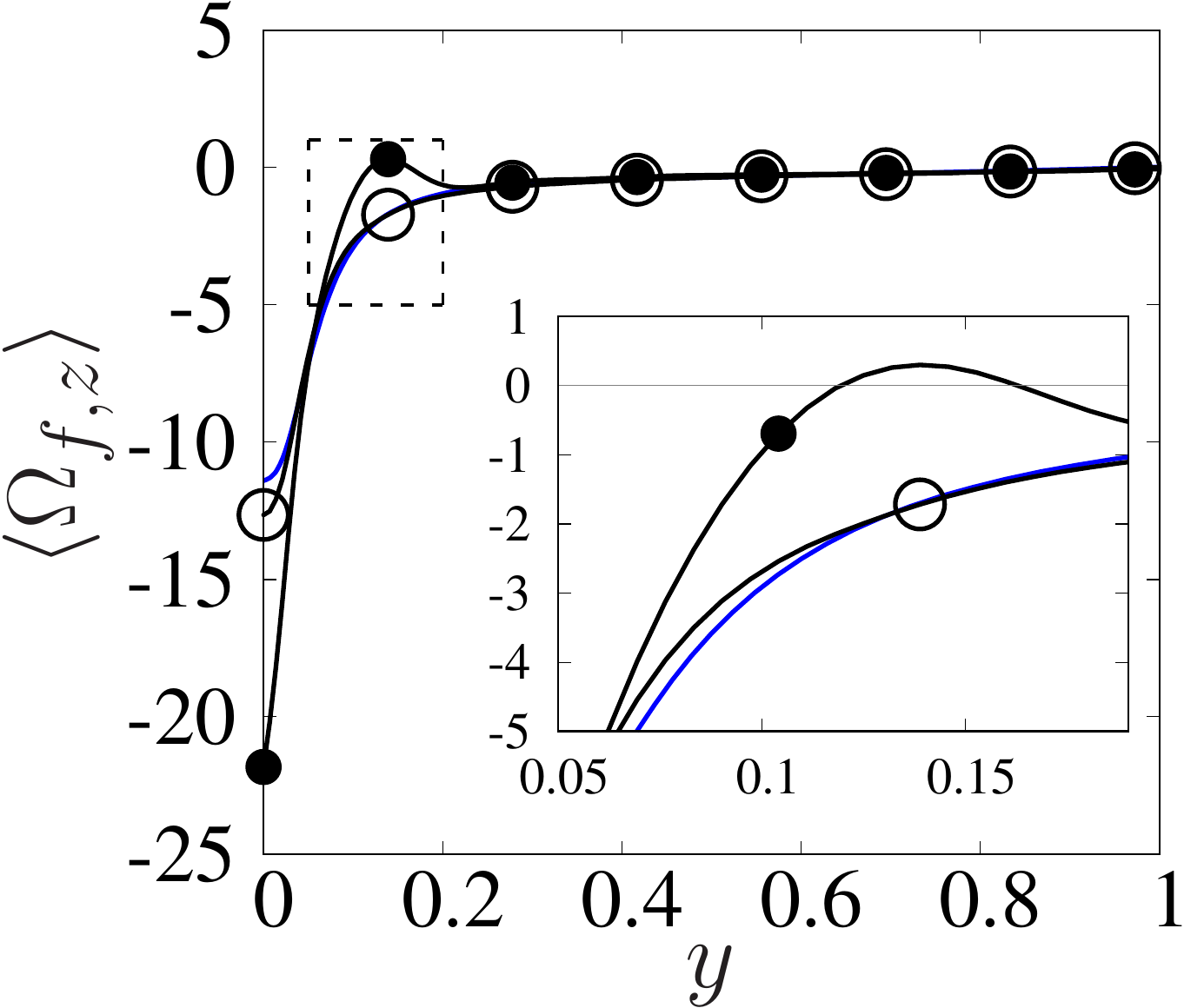}
\end{center}
\end{minipage}}	
\subfigure[]{\label{fig:omegap_W0}
\begin{minipage}[b]{0.4\textwidth}
\begin{center}
\includegraphics[width =\textwidth,scale=1]{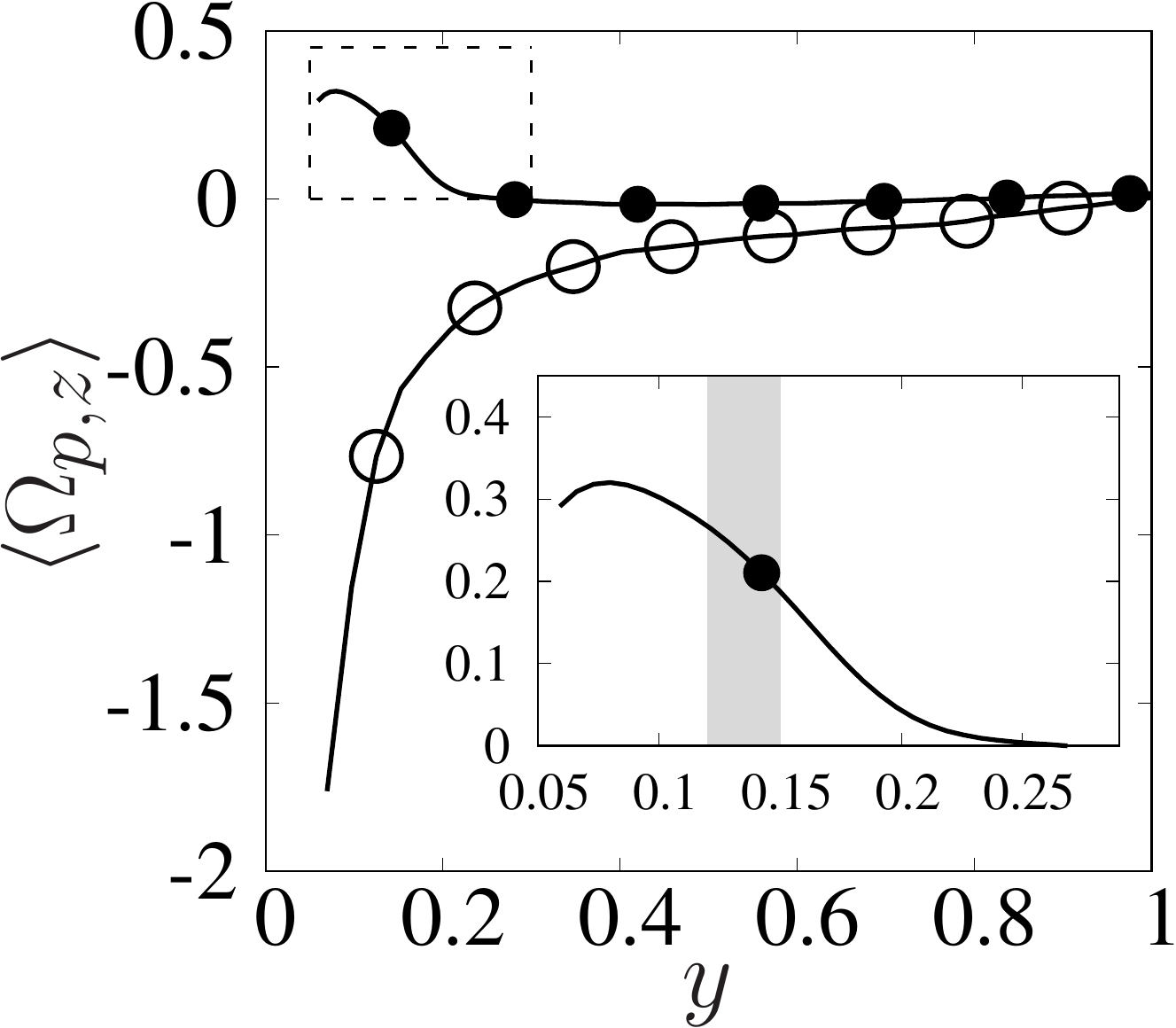}
\end{center}
\end{minipage}}				    
\subfigure[]{\label{fig:omegaf_W15}
\begin{minipage}[b]{0.4\textwidth}
\begin{center}
\includegraphics[width =\textwidth,scale=1]{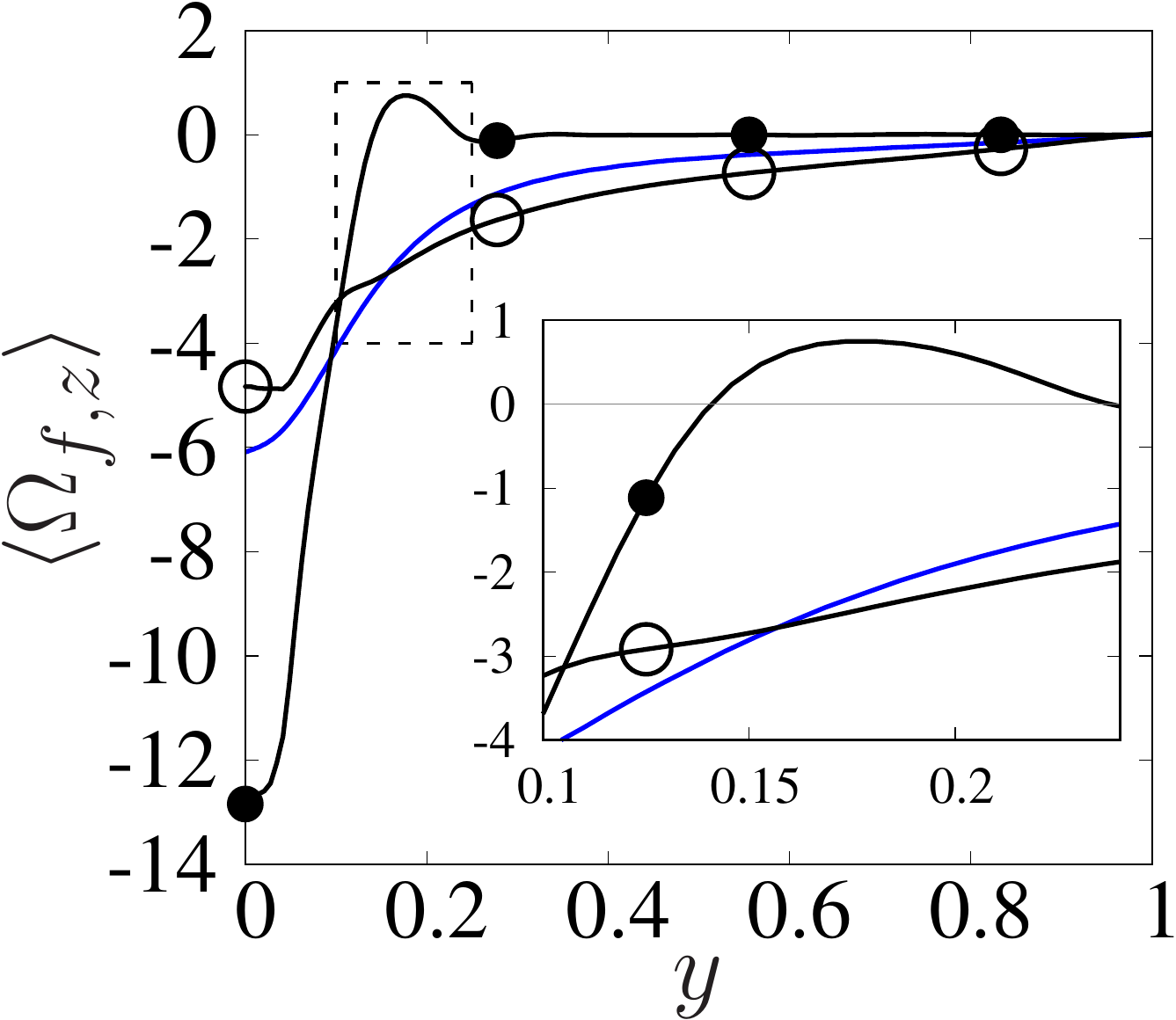}
\end{center}
\end{minipage}}	
\subfigure[]{\label{fig:omegap_W15}
\begin{minipage}[b]{0.4\textwidth}
\begin{center}
\includegraphics[width=\textwidth,scale=1]{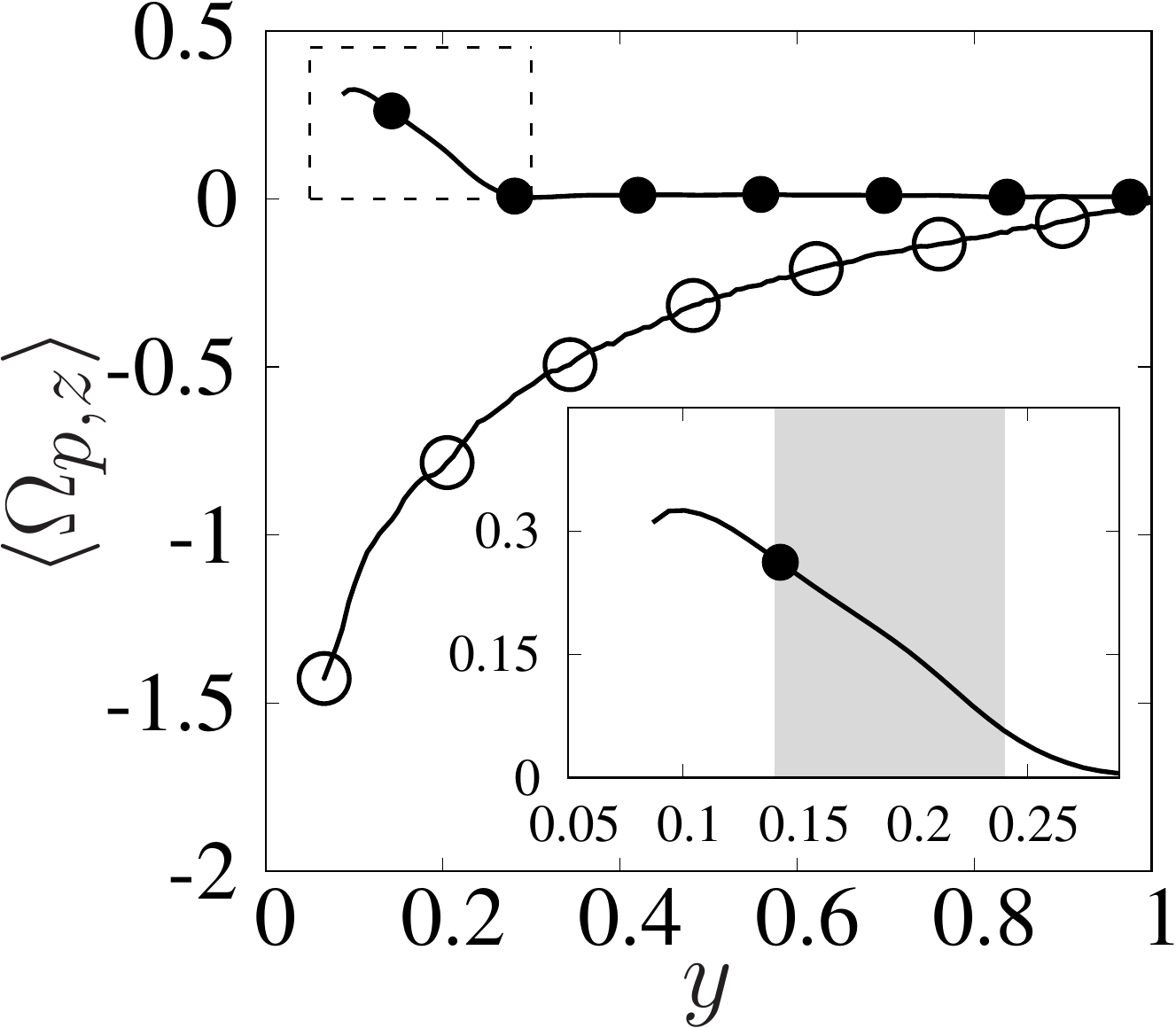}
\end{center}
\end{minipage}}							 																	
\caption{Profiles of average spanwise (a,c) fluid vorticity and (b,d) particle angular velocity.
Suspensions with ({\protect\circlelineblack}) neutrally buoyant and ({\protect\filledcirclelineblack}) dense particles; ({\protect\bluelt}) single-phase condition.
The fluid in the top row (a,b) is Newtonian and in the bottom row (c,d) is viscoelastic.
Marked area is magnified in the inset, and shaded area in the inset of (b,d) indicates the $y$ locations where $\langle \Omega_{f,z}\rangle>0$. }   \label{fig:omega_p_f}		
\end{figure}	

The fluid velocity profiles are significantly different for dense particles than neutrally buoyant, and the change is accentuated in the viscoelastic case. 
Due to the reaction of the particle drag forces on the fluid field, the flow slows down in the channel core.
The mean velocity is nearly flattened in the bulk, in a stark contrast to the W15P5 where the velocity tends to the Poiseuille profile. 
Near the walls ($y\approx \{0.1, 0.15\}$ for W0P5D and W15PD), the fluid speeds up avoiding the resistive forces from the particles in the bulk. 
As a result, in both the viscoelastic and Newtonian cases, a local maximum in fluid velocity is observed between the wall and the peak in the particle concentration.
We now direct our attention on the angular particle velocities, which play an important role in the global dynamics of the suspension (figure \ref{fig:omega_p_f}).
Neutrally buoyant particles sustain a negative spanwise angular velocity $\Omega_{p,z}$ throughout the bottom half of the channel,  similar to the negative mean spanwise vorticity $\Omega_{f,z}$.
In the dense particle cases, $\Omega_{d,z}$ vanishes away from the walls, is positive over a short wall-normal extent, and reaches large negative values in the vicinity of the wall. 
The region of positive mean vorticity occurs below the peak in the particle concentration profile, for instance over $0.12<y<0.15$ when the peak of $\phi$ is at $y=0.16$ in W0P5D. 
That positive vorticity is thus associated with the shear layer generated at the edge of the high concentration of particles moving much slower than the mean flow.
The particles naturally experience positive angular velocity at this location; Interestingly, they preserve this positive velocity closer to the wall where vorticity becomes negative (see the inset of figures \ref{fig:omegap_W0} and \ref{fig:omegap_W15}). 
This observation highlights that the positive angular velocity of the dense particles near the wall cannot only be explained by the local positive fluid vorticity; other physical mechanisms are at play. 

\begin{figure}		
\centering
\subfigure[]{\label{fig:vf_cond_W0}
\begin{minipage}[b]{0.32\textwidth}
\begin{center}
\includegraphics[width =\textwidth,scale=1]{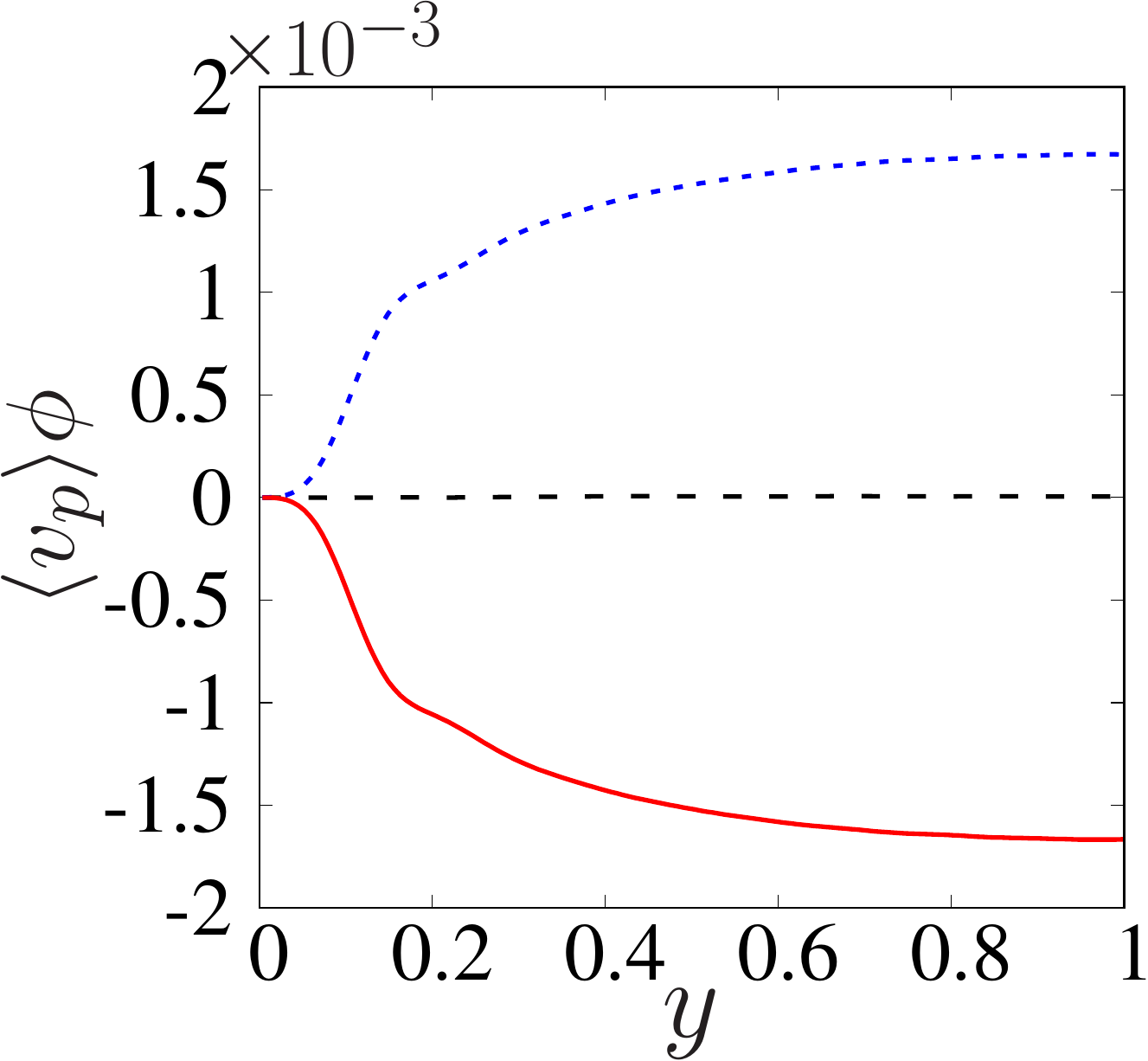}
\end{center}
\end{minipage}}			
\subfigure[]{\label{fig:phi_cond_W0}
\begin{minipage}[b]{0.32\textwidth}
\begin{center}
\includegraphics[width =\textwidth,scale=1]{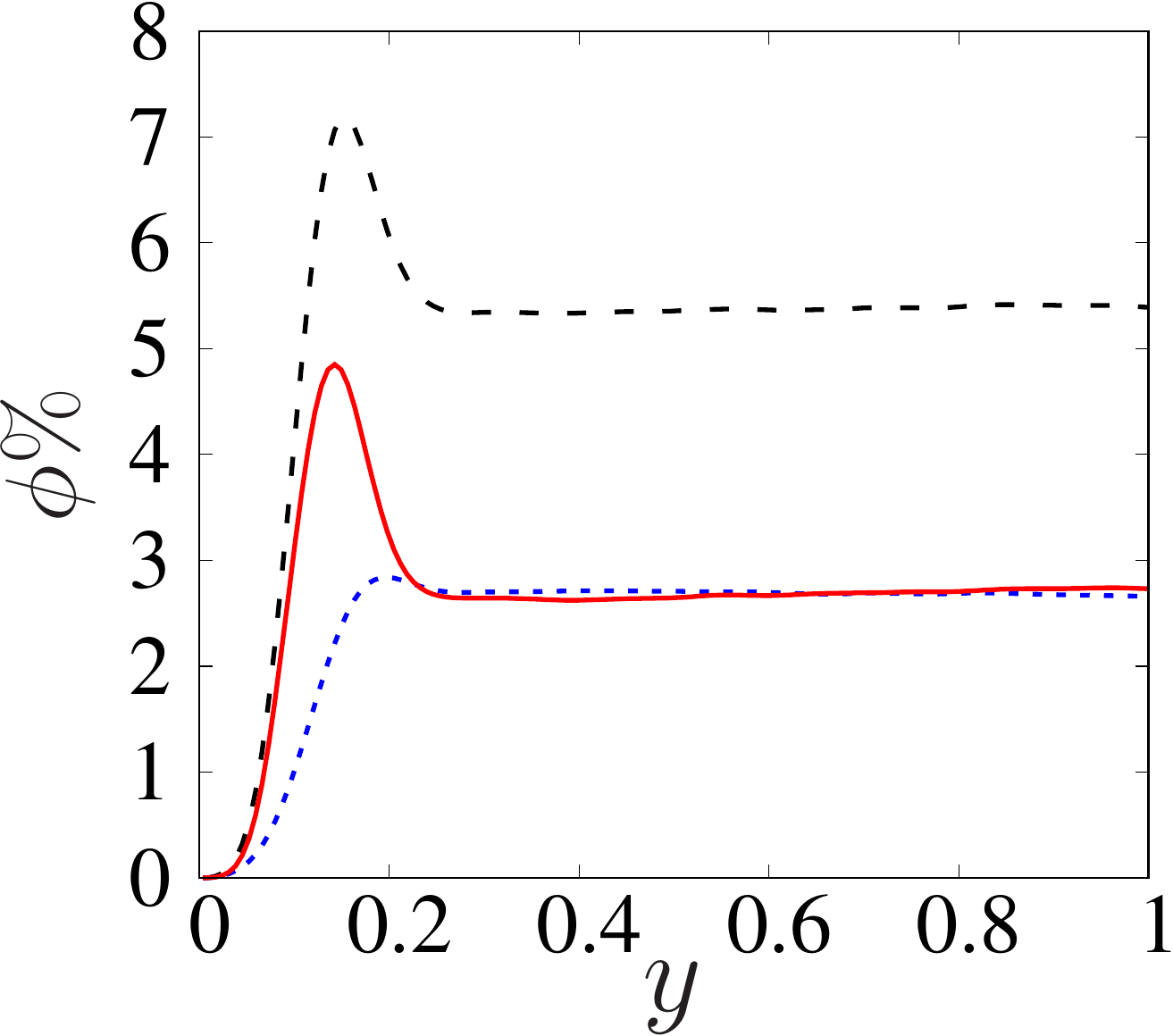}
\end{center}
\end{minipage}}					\\			
\subfigure[]{\label{fig:vf_cond_W15}
\begin{minipage}[b]{0.32\textwidth}
\begin{center}
\includegraphics[width =\textwidth,scale=1]{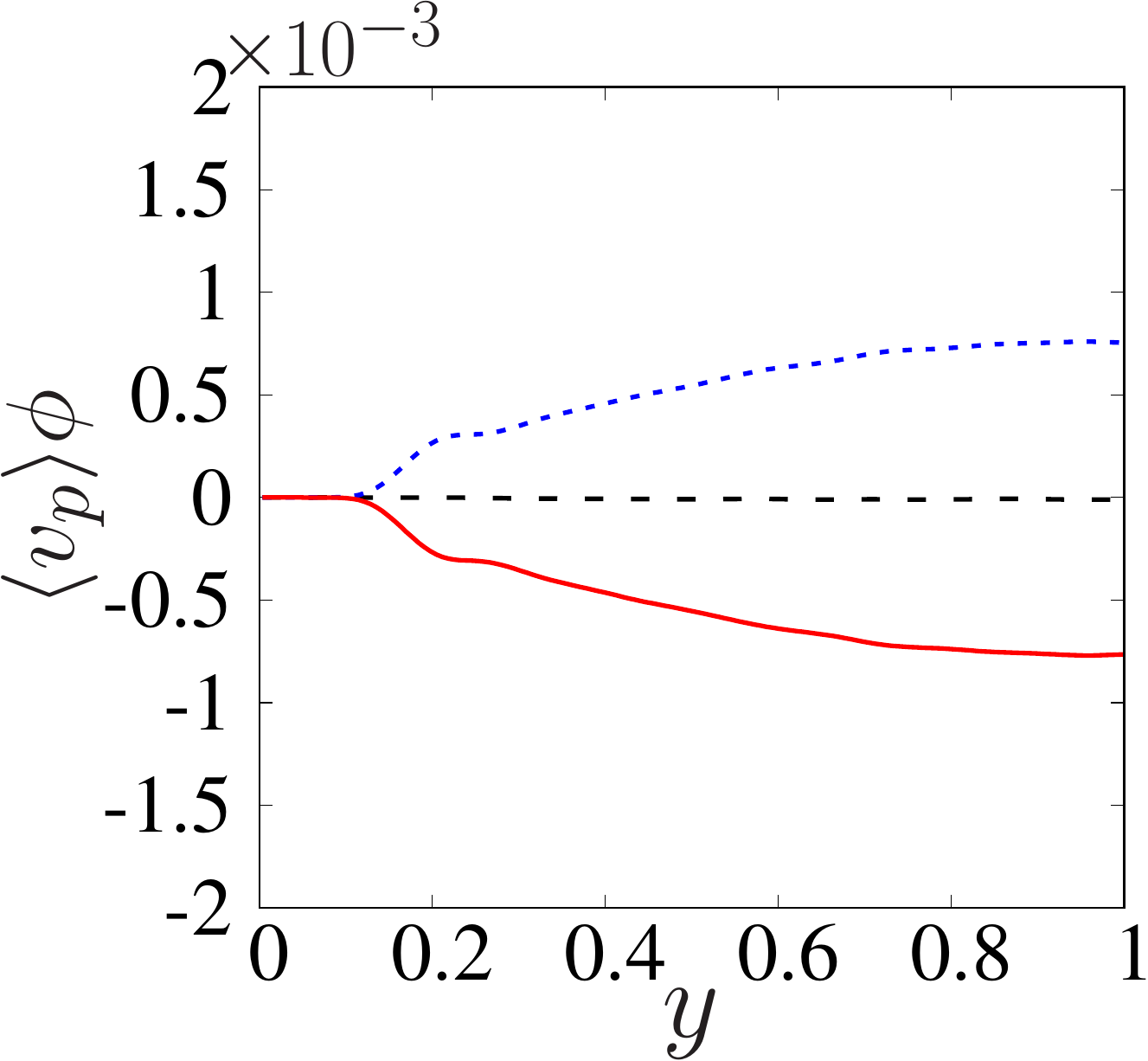}
\end{center}
\end{minipage}}	
\subfigure[]{\label{fig:phi_cond_W15}
\begin{minipage}[b]{0.32\textwidth}
\begin{center}
\includegraphics[width =\textwidth,scale=1]{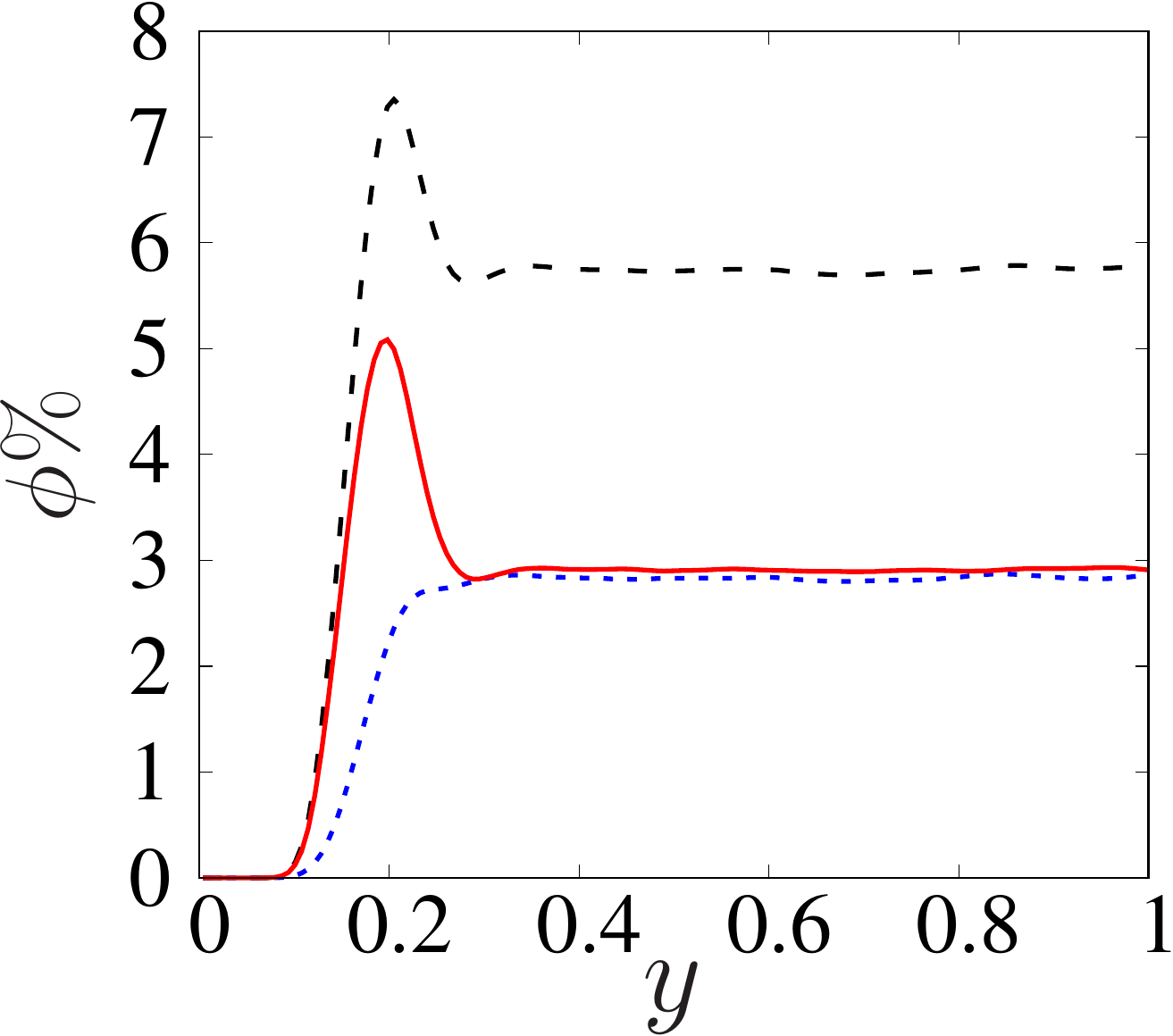}
\end{center}
\end{minipage}}										
\caption{Profiles of conditionally averaged (a,c) wall-normal particle flux and (b,d) particle concentration.
Red solid line ({\protect\redlt}) $\Omega_{p,z} > 0$; 
blue dotted line ({\protect\bluedottedlt}) $\Omega_{p,z} < 0$; 
black dashed line ({\protect\dashlt}) unconditional statistics. 
The fluid in the top row (a,b) is Newtonian and in the bottom row (c,d) is viscoelastic.
}   \label{fig:conditional_eulerian}		
\end{figure}	

To further investigate this matter, conditional statistics of particles with positive and negative spanwise angular velocities are shown in figure \ref{fig:conditional_eulerian}. 
The conditioned wall-normal particle flux demonstrates a clear dependence on its angular velocity (figures \ref{fig:vf_cond_W0} and \ref{fig:vf_cond_W15}): 
particles which rotate in the positive $z$ direction experience downward wall-normal velocities and vice versa. 
As a result, only particles with a positive angular velocity can penetrate into the near-wall region, and therefore the unconditional average particle angular velocity is positive in that region (figures \ref{fig:omegap_W0} and \ref{fig:omegap_W15}). 
Confirming this trend, figures \ref{fig:phi_cond_W0} and \ref{fig:phi_cond_W15} show that the volume fraction of particles with $\Omega_{p,z}>0$ is significantly larger near the wall. 
The coupling of angular and translational velocities also provides an explanation for the unconditional particle concentration profile:
Away from the walls, particles efficiently mix in the wall-normal direction due to the equal positive and negative fluxes $\langle v_{p} \rangle \phi$ associated with positive and negative rotation.
Near the wall, particles with $v_p<0$ and $\Omega_{p,z}>0$ are stabilized by the opposing repulsive force from the wall, which leads to the increased particle concentration in that region. 

The coupling of angular and translational velocities hints to the relevance of rotation-induced lift forces \citep{magnus1853ueber}.
Nonetheless, due to the additive nature of hydrodynamic forces, isolating individual contributions is not a trivial task.
In the following section, ensemble-averages along particle trajectories are investigated to elucidate the hydrodynamic effects that control the motion of particles. 

\subsection{Ensemble-averages along particle trajectories  \label{sec:lift}}
Hydrodynamic forces acting perpendicular to the primary direction of motion of a sphere are broadly divided into shear- and rotation-induced contributions.
In the limit of small and high particle Reynolds numbers, defined as $Re_p \equiv d_p (u_p-u_f) / \nu$, the shear-induced lift is the first-order contribution and the influence of rotation-induced lift is deemed to be negligible. 
For an intermediate range of the particle Reynolds numbers, approximately from 5 to 100, particle's free rotation has a measurable impact on its motion \citep{bagchi2002effect}. 
In our dense particle cases, the average particle Reynolds number is $Re_p \approx \{130, 95\}$ in the Newtonian and viscoelastic conditions, and therefore the rotation-induced lift force is expected to be important. 

\begin{figure}		
\centering
\subfigure[]{\label{fig:traj_sample_xz_D}
\begin{minipage}[b]{0.4\textwidth}
\begin{center}
\includegraphics[width =\textwidth,scale=1]{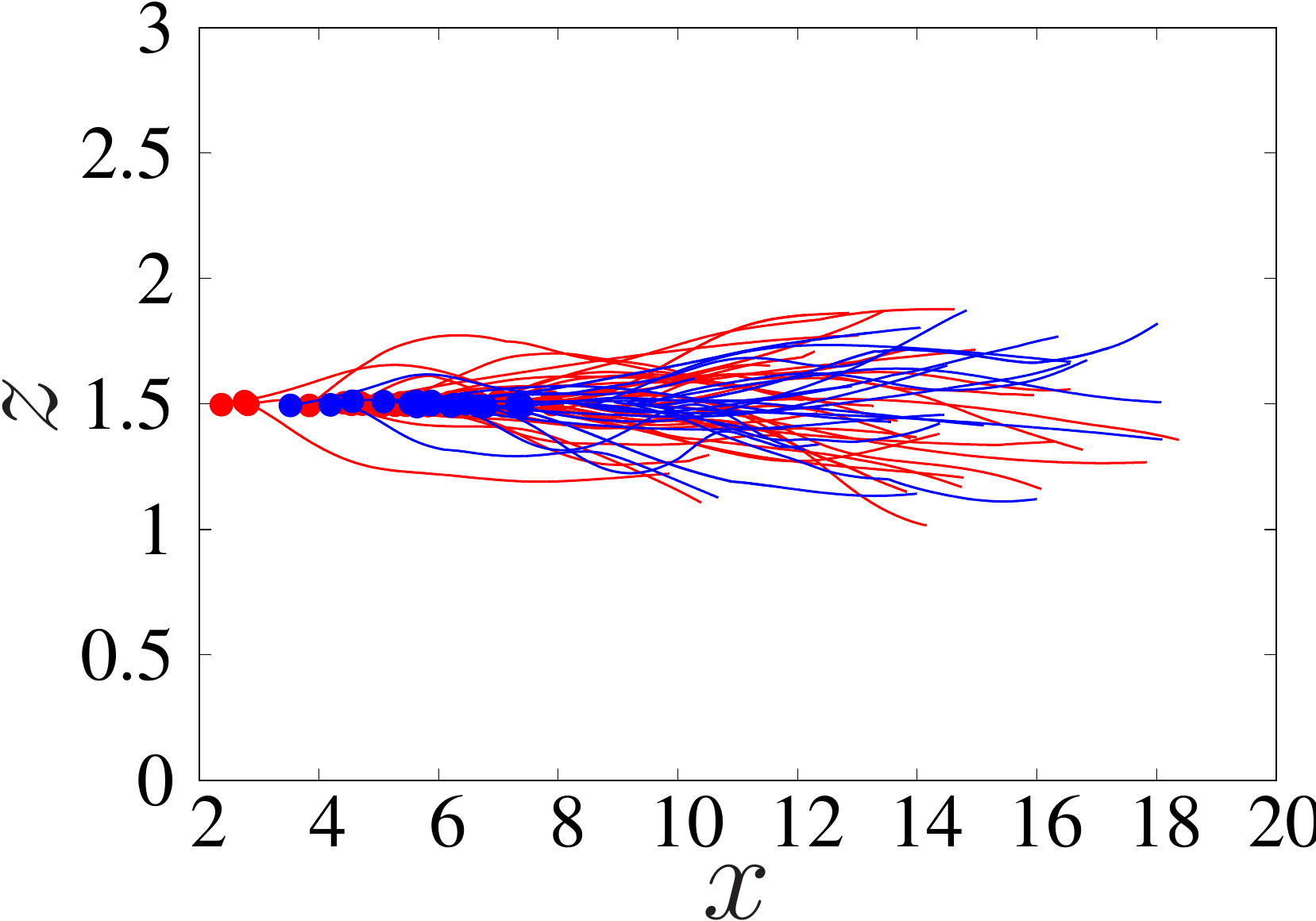}
\end{center}
\end{minipage}}	
\subfigure[]{\label{fig:traj_sample_xz_N}
\begin{minipage}[b]{0.4\textwidth}
\begin{center}
\includegraphics[width =\textwidth,scale=1]{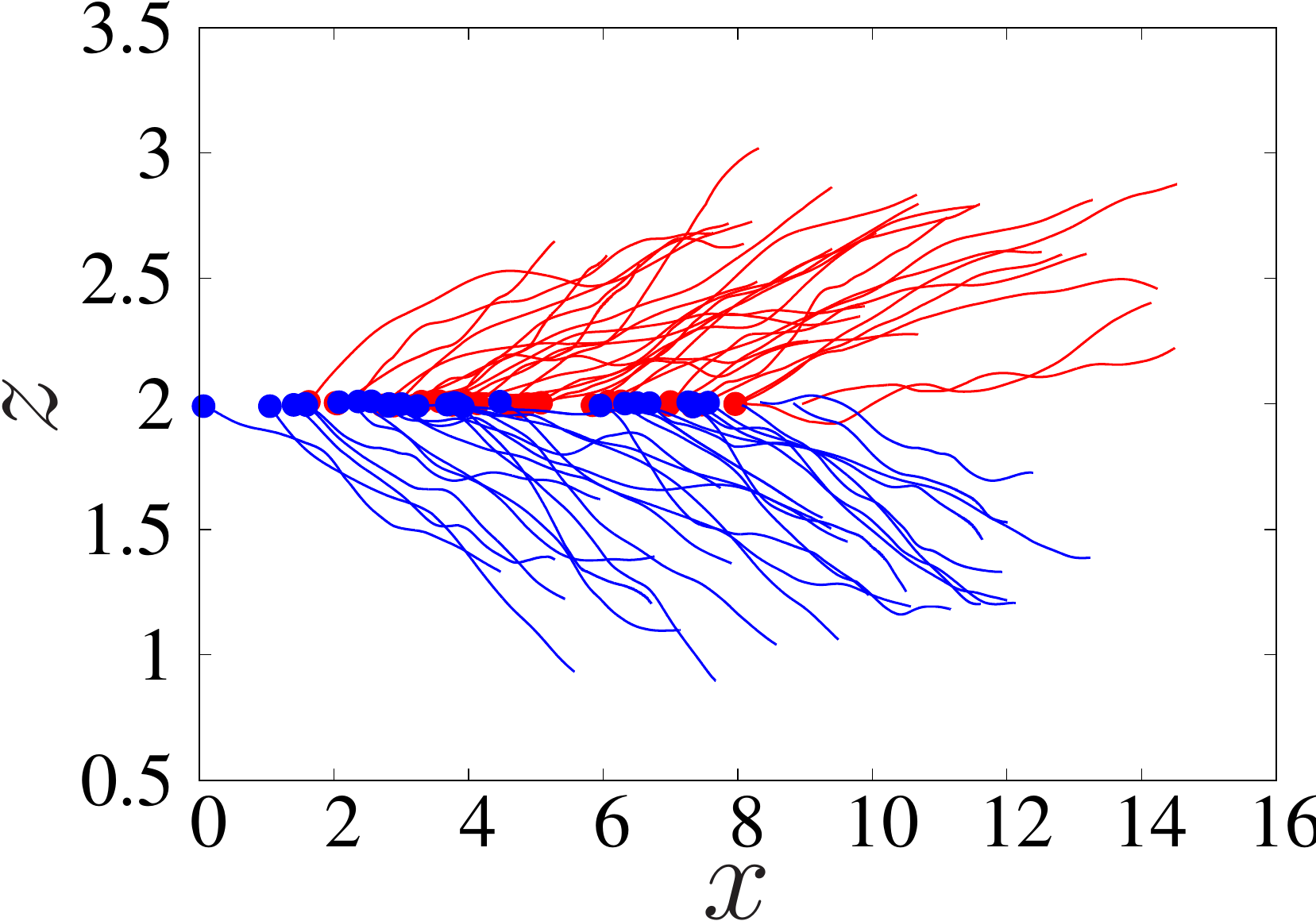}
\end{center}
\end{minipage}}								 																
\caption{Sample particle trajectories projected onto the $xz$ plane for particles with a positive ({\protect\redlt}) or negative ({\protect\bluelt}) wall-normal angular velocity $\Omega_{p,y}$; (a) W0P5 (b) W0P5D.  Particles velocity relative to the fluid is in the negative $x$ direction and the positive $y$ axis points into the page. All particles are located at approximately the same $z$ location at the beginning of the observation time window, during which $\Omega_{p,y}$ does not change sign.}   \label{fig:traj_sample_xz}	
\end{figure}	

We exploit the absence of mean shear in the span to highlight the effect of rotation-induced lift force, by investigating the particle trajectories projected onto the $xz$ plane (figure \ref{fig:traj_sample_xz}). 
Particles with positive and negative $\Omega_{p,y}$ are tracked for approximately 10 time units in the Newtonian cases.
Evidently, the sign of $\Omega_{p,y}$ is a predictor of dense particles' direction of motion in the span, while the neutrally buoyant particles disperse randomly regardless of their angular velocity. 
Similar trends in the trajectories are observed in the viscoelastic cases (not shown for brevity).

\begin{figure}		
\subfigure[]{\label{fig:traj_stat_xz_W0}
\begin{minipage}[b]{0.51\textwidth}
\begin{center}
\includegraphics[width =\textwidth,scale=1]{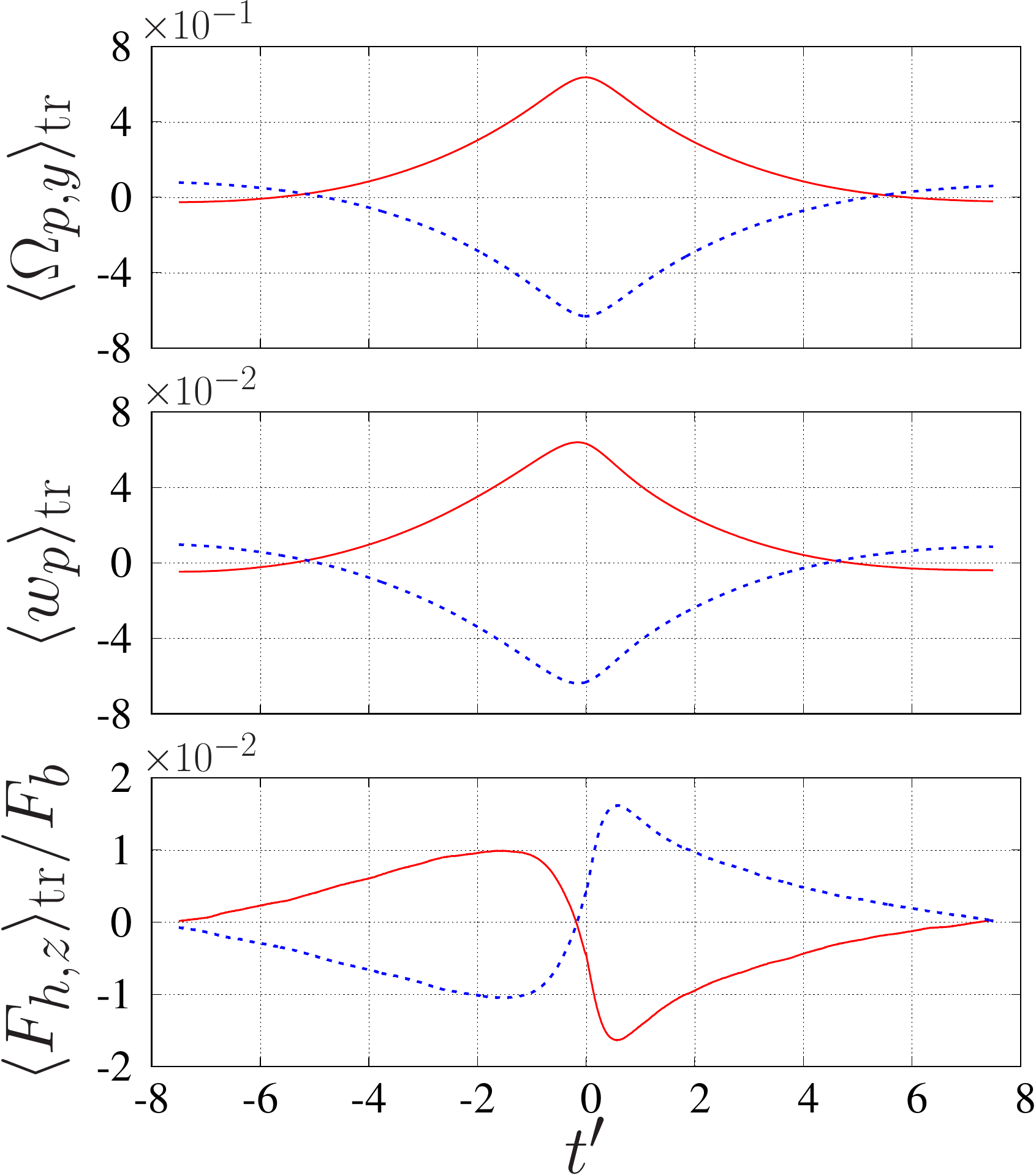}
\end{center}
\end{minipage}}	
\subfigure[]{\label{fig:traj_stat_xz_W15}
\begin{minipage}[b]{0.445\textwidth}
\begin{center}
\includegraphics[width =\textwidth,scale=1,trim={2.19cm 0 0 0},clip]{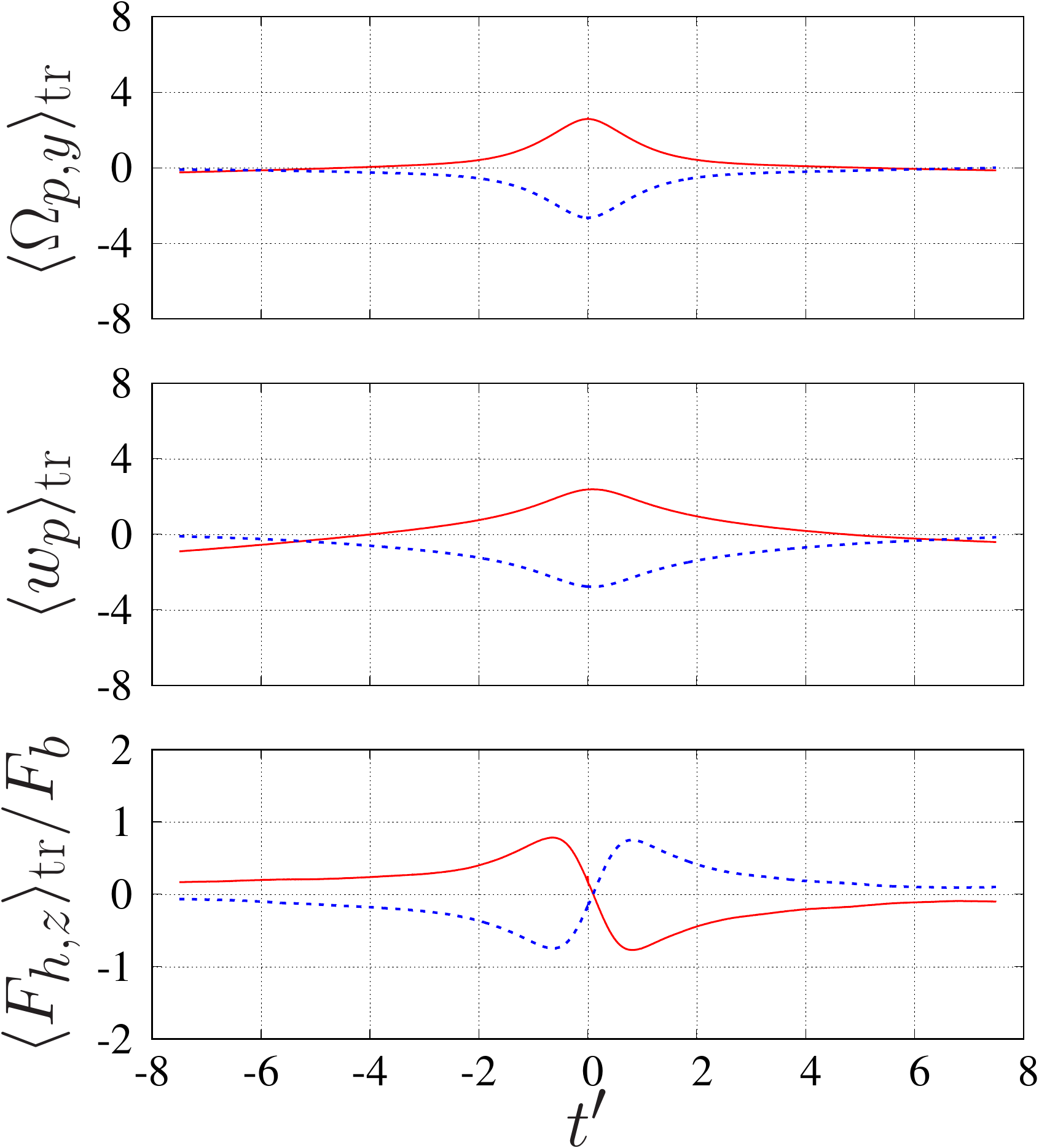}
\end{center}
\end{minipage}}								 																
\caption{ Ensemble-averaged properties along the trajectories for particles with a positive ({\protect\redlt}) or negative ({\protect\bluedottedlt}) wall-normal angular velocity $\Omega_{p,y}$ at $t_0$; (a) W0P5D (b) W15P5D. The shifted time is denoted $ t'\equiv t-t_0$, and the particle properties from top to bottom are wall-normal angular velocity, spanwise translational velocity and spanwise hydrodynamic force exerted on the particle and normalized by the particle buoyancy force $F_b\equiv(\rho_p-\rho_f)V_p|\bm{g}|$. Particles are located within $0.4<y<1$ at $t'=0$.  }   \label{fig:traj_stat_xz}		
\end{figure}

Ensemble-averaged properties along the trajectories of dense particle are plotted for Newtonian and viscoelastic conditions (figure \ref{fig:traj_stat_xz}).  The averaging is conditioned on the sign of $\Omega_{p,y}$ at $t_0$, and the statistics are plotted versus $t' = t-t_0$.
In an inviscid laminar flow, the Magnus lift force is expressed as $\bm{F}_\text{Mag} = C_{lm} d_p^2 \rho_f \bm{\Omega}_p \times (\bm{u}_p-\bm{u}_f)$ \citep{auton1988force}.
Therefore, for the present conditions where particles are moving slower than the mean flow in the streamwise direction, when $\bm{\Omega}_p$ is aligned with the positive $y$ axis the Magnus lift force is in the positive $z$ direction.
In both Newtonian and viscoelastic conditions, figure \ref{fig:traj_stat_xz} shows that the spanwise translational velocity is positively correlated with the wall-normal angular velocity, which confirms the relevance of the Magnus lift forces. 
Interestingly, the peak in the hydrodynamic force, which is induced by the Magnus effect precedes the particles' maximum angular and translational velocities. 
In fact, $\langle F_{h,z} \rangle_\text{tr} $ sharply decreases and changes sign approximately between $-1<t'<1$. 
When a particle begins to gain momentum in $z$ direction due to the Magnus force, the surrounding fluid resists the particle's motion, and the particle experiences a drag force counteracting the Magnus lift. 
Finally, while the importance of the Magnus lift force in particles' motion is evident in both Newtonian and viscoelastic conditions, in the later case has smaller magnitudes of both the translational and angular velocities and of the hydrodynamic force. 

\begin{figure}		
\centering
\includegraphics[width =0.4\textwidth,scale=1]{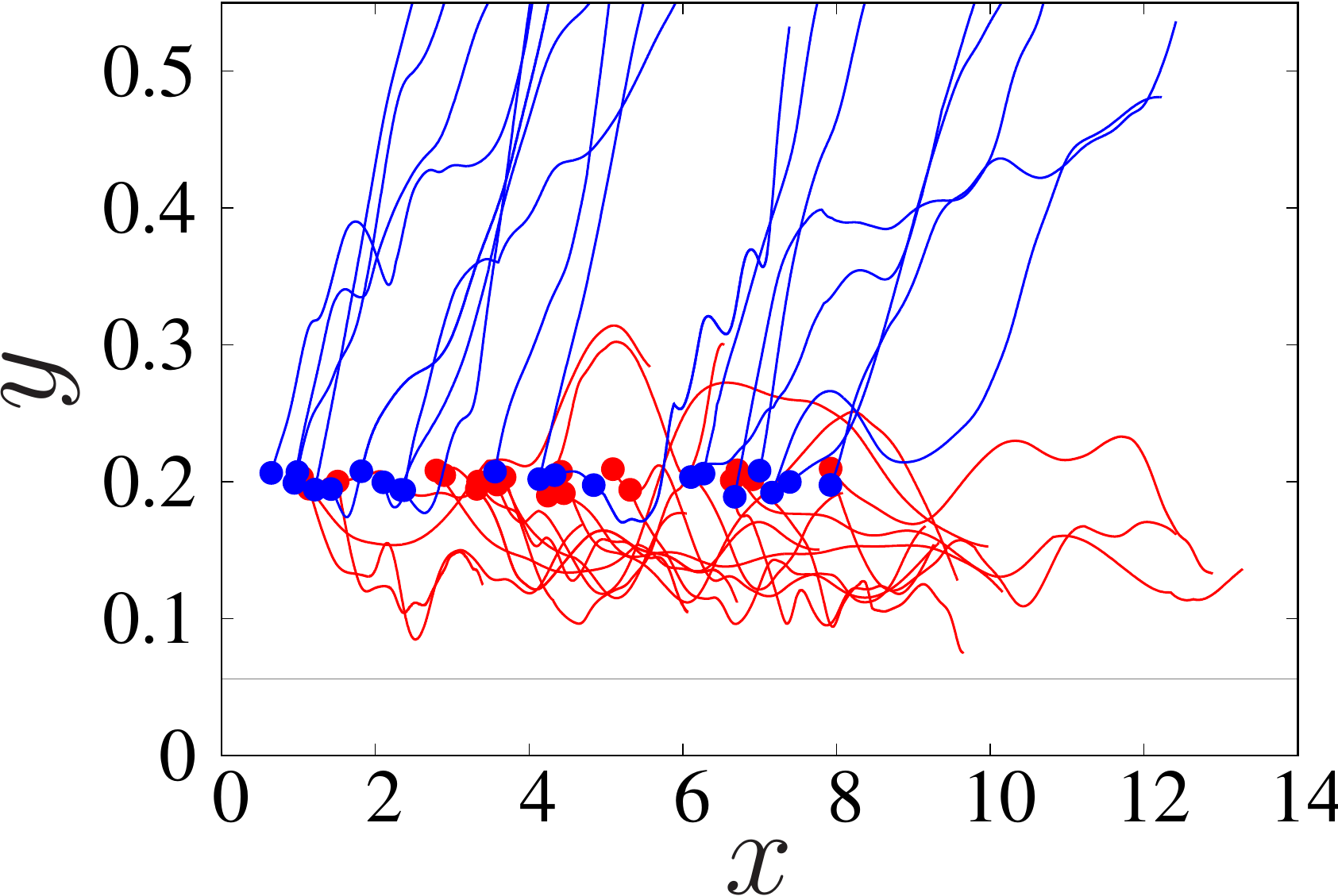}
\caption{Sample particle trajectories projected onto the $xy$ plane for particles with a positive ({\protect\redlt}) or negative ({\protect\bluelt}) spanwise angular velocity $\Omega_{p,z}$ in W0P5D.  Particles velocity relative to the fluid is in the negative $x$ direction and the positive $z$ axis points out of the page. All particles are located at approximately the same $y$ location at the beginning of the observation time window, during which $\Omega_{p,z}$ does not change sign. The solid grey line marks the geometric limit for particles center in $y$ direction.}   \label{fig:traj_sample_xy}		
\end{figure}	

\begin{figure}		
\subfigure[]{\label{fig:traj_stat_xy_W0}
\begin{minipage}[b]{0.51\textwidth}
\begin{center}
\includegraphics[width =\textwidth,scale=1]{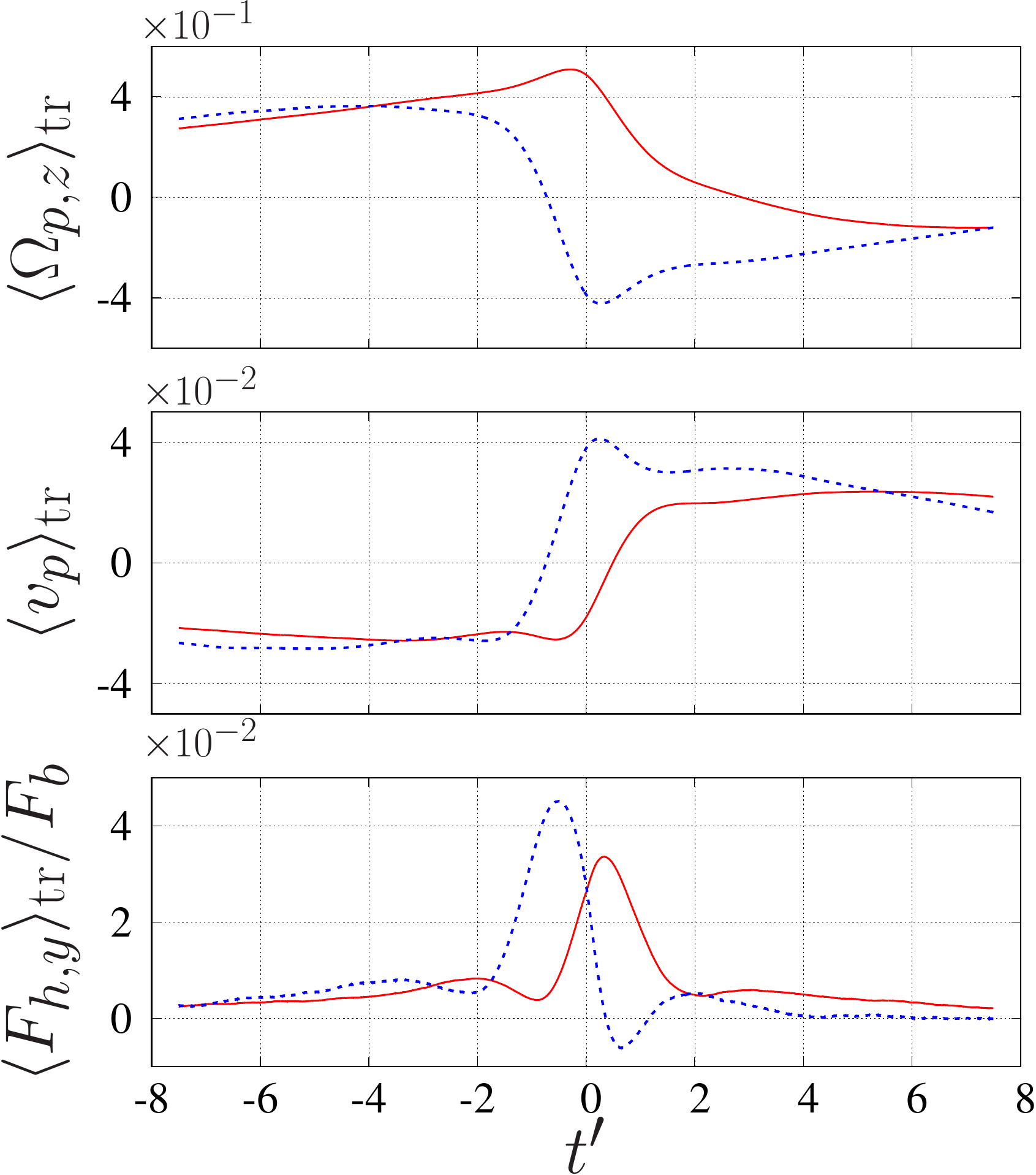}
\end{center}
\end{minipage}}	
\subfigure[]{\label{fig:traj_stat_xy_W15}
\begin{minipage}[b]{0.445\textwidth}
\begin{center}
\includegraphics[width =\textwidth,scale=1,trim={2.19cm 0 0 0},clip]{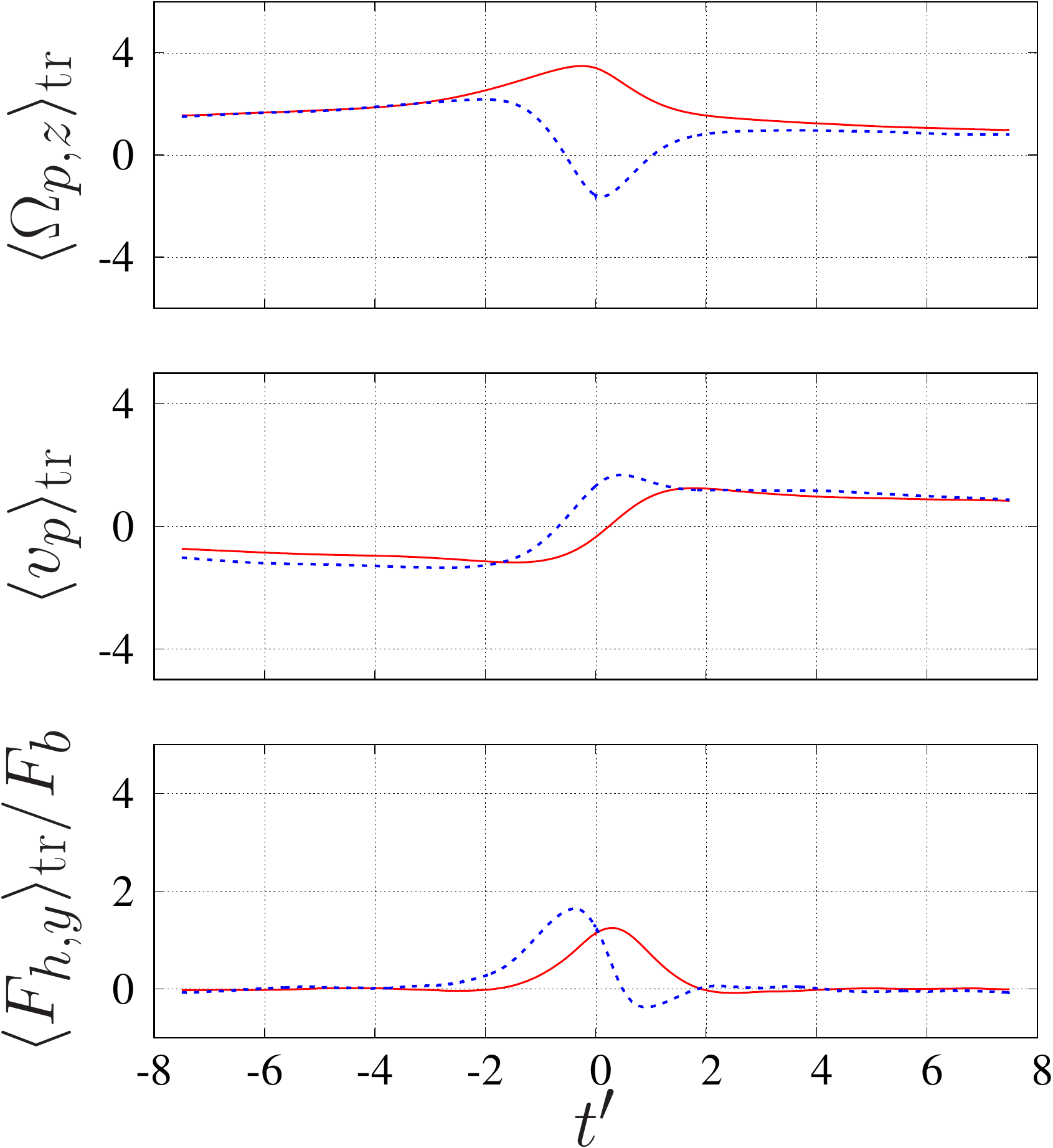}
\end{center}
\end{minipage}}								 																		
\caption{Ensemble-averaged properties along the trajectories for particles with a positive ({\protect\redlt}) or negative ({\protect\bluedottedlt}) spanwise angular velocity $\Omega_{p,z}$ at $t_0$; (a) W0P5D (b) W15P5D. The shifted time is denoted $ t'\equiv t-t_0$, and the particle properties from top to bottom are spanwise angular velocity, wall-normal translational velocity and wall-normal hydrodynamic force exerted on the particle and normalized by the particle buoyancy force $F_b\equiv(\rho_p-\rho_f)V_p|\bm{g}|$. Particles are located within $y<0.4$ at $t'=0$. }   \label{fig:traj_stat_xy}		
\end{figure}	

We now turn to the $xy$ plane, where sample particle trajectories are plotted in figure \ref{fig:traj_sample_xy} for W0P5D; this time the trajectories are conditioned on the spanwise angular velocity of the particles. 
As noted earlier in \S\ref{sec:mean}, a shear-induced lift force repels the particles away from the wall. 
Similar to their spanwise motion, the vertical motion of dense particles is also correlated to their angular velocity.
Particles with $\Omega_{p,z}<0$ migrate towards the center of the channel while particles with positive rotation  penetrate towards the wall.
Nonetheless, due to the strong upward shear-induced lift force, it is very unlikely that the particles directly interact with the wall.

For particles located within $y< 0.4$, ensemble-averages along particle trajectories are shown in figure \ref{fig:traj_stat_xy}. 
It should be noted that due to the rotation-induced lift force, the likelihood of presence of particles with $\Omega_{p,z}<0$ in the near-wall region is low and, as a result, $\Omega_{p,z}$ is on average positive within $y<0.4$ (see figures \ref{fig:omegap_W0} and \ref{fig:omegap_W15}).
Thus, even for those particles with $\Omega_{p,z}<0$, it is more likely that their spanwise angular velocity has changed sign shortly before $t'=0$ (the top row of figure \ref{fig:traj_stat_xy}).
After $t'>0$, however, particles have already reached the vicinity of the wall, and the strong negative fluid vorticity drives their angular velocities to negative values.
Similarly, particles experience negative wall-normal velocity before $t'=0$ and positive values after
(the middle row of figure \ref{fig:traj_stat_xy}). 
At $t'=0$, the effect of the Magnus effect is more evident: particles with positive spanwise angular velocity experience negative wall-normal velocity, and vice versa. 
Investigating the ensemble-averaged hydrodynamic forces is complex, as one should account for simultaneous effects of shear- and rotation-induced lift forces, as well as the resistive drag force opposing any incurred motion (the bottom row of figure \ref{fig:traj_stat_xy}). 
For particles with $\Omega_{p,z}<0$, the shear- and rotation-induced forces are both in the positive $y$ direction, and therefore the particles experience a strong upward force when  $t' < 0$. 
Their upward motion is resisted by drag, resulting in vanishing net force when $t'>0$. 
For particles with $\Omega_{p,z}>0$, the shear- and rotation-induced lift forces are in opposite directions, and the net effect is an upward force with a weaker peak compared to the particles with $\Omega_{p,z}<0$.

\subsection{Clustering, microstructure and particle wake \label{sec:cluster}}

We have seen that the presence of gravity gives rise to sustained slip velocities, lift forces, and significant changes in the mean profiles. 
The statistical averaging, however, masks the impact of gravity on the spatial distribution, or clustering, of particles in the wall-parallel planes.    
In this section, we examine the particles collective motion and potential clustering effects, and relate these macro-scale observations to the micro-scale dynamics, e.g.\,lift forces, micro-structure and particle wake.

\begin{figure}		
\subfigure[]{\label{fig:vor_sample_near_wall}
\begin{minipage}[b]{0.47\textwidth}
\begin{center}
\includegraphics[width =\textwidth,scale=1]{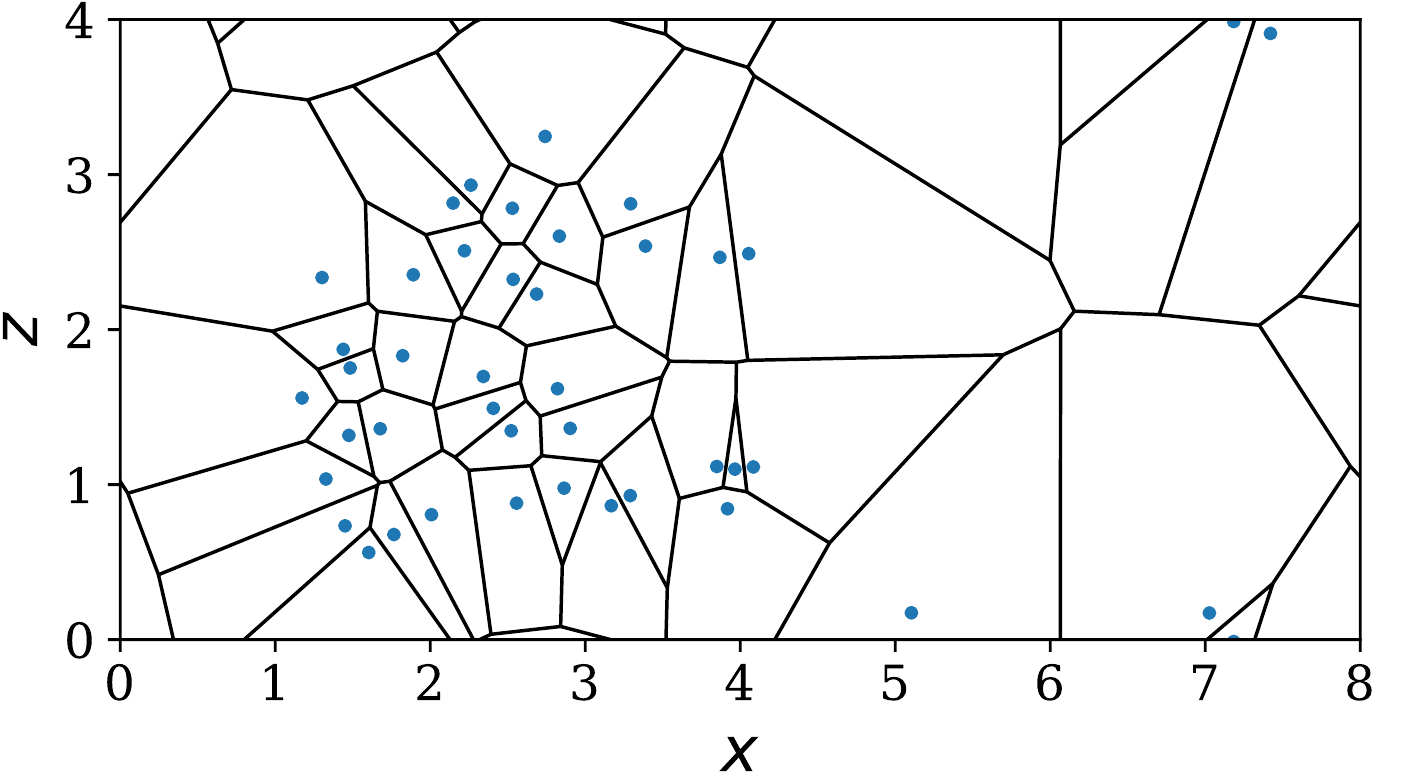}
\end{center}
\end{minipage}}	
\subfigure[]{\label{fig:vor_sample_center}
\begin{minipage}[b]{0.47\textwidth}
\begin{center}
\includegraphics[width =\textwidth,scale=1]{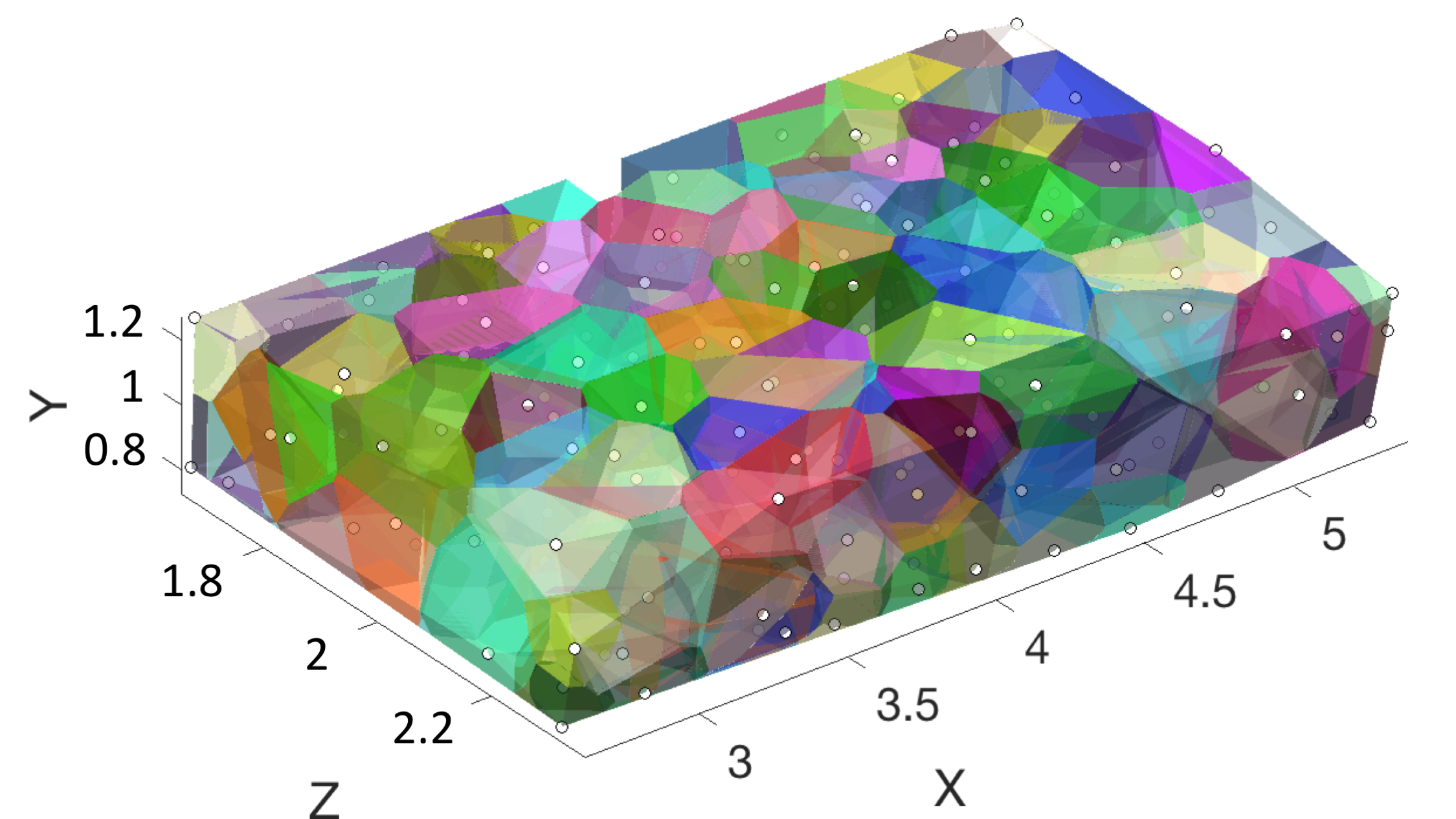}
\end{center}
\end{minipage}}								 																
\caption{\Voro tessellation of (a) the $xz$ plane located at $y = 3d_p/4$ and (b) the volume $0.8 < y < 1.2$   corresponding to one snapshot of particle centers in W0P5D. In (b) only a subset of the domain is visualized for the sake of clarity.   }   \label{fig:vor_sample}		
\end{figure}

The \Voro diagram \citep{okabe1992spatial} has previously been used to examine clustering in particle-laden turbulent flows \citep{monchaux2010preferential,garcia2012dns}, and will be adopted herein for cluster analysis.  
The domain, either a plane or a volume,  is partitioned into $n$ cells marking the neighboring region of $n$ particles,  i.e.\,the $i$th particle center is the closest particle center to all of the points that belong to the $i$th cell.
Two examples of \Voro tessellations are shown in figure \ref{fig:vor_sample}: the first is in a plane based on particles centers near the wall, and the second is in a volume for particles near the channel center.  
Isolated particles are located in larger cells, while smaller cells indicate high particle concentrations. 
For the near-wall particles clustering is visually discernible in this particular instance of the flow.

\begin{figure}		
\centering
\subfigure[]{\label{fig:vor_volume_near_wall_W0P5}
\begin{minipage}[b]{0.2925\textwidth}
\begin{center}
\includegraphics[width =\textwidth,scale=1]{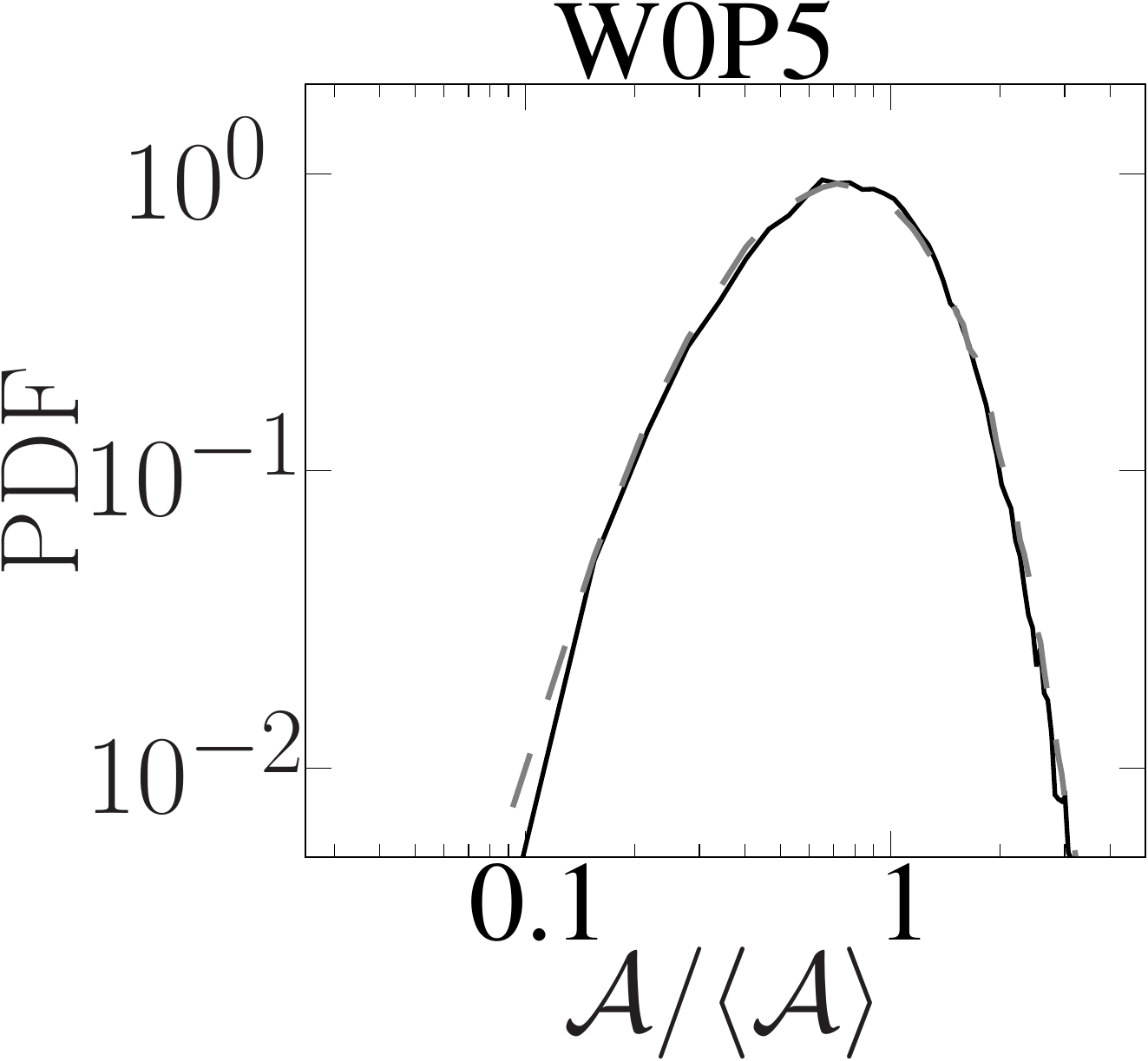}
\end{center}
\end{minipage}}
\subfigure[]{\label{fig:vor_volume_near_wall_W0P5D}
\begin{minipage}[b]{0.215\textwidth}
\begin{center}
\includegraphics[width =\textwidth,scale=1,trim={3.55cm 0 0 0},clip]{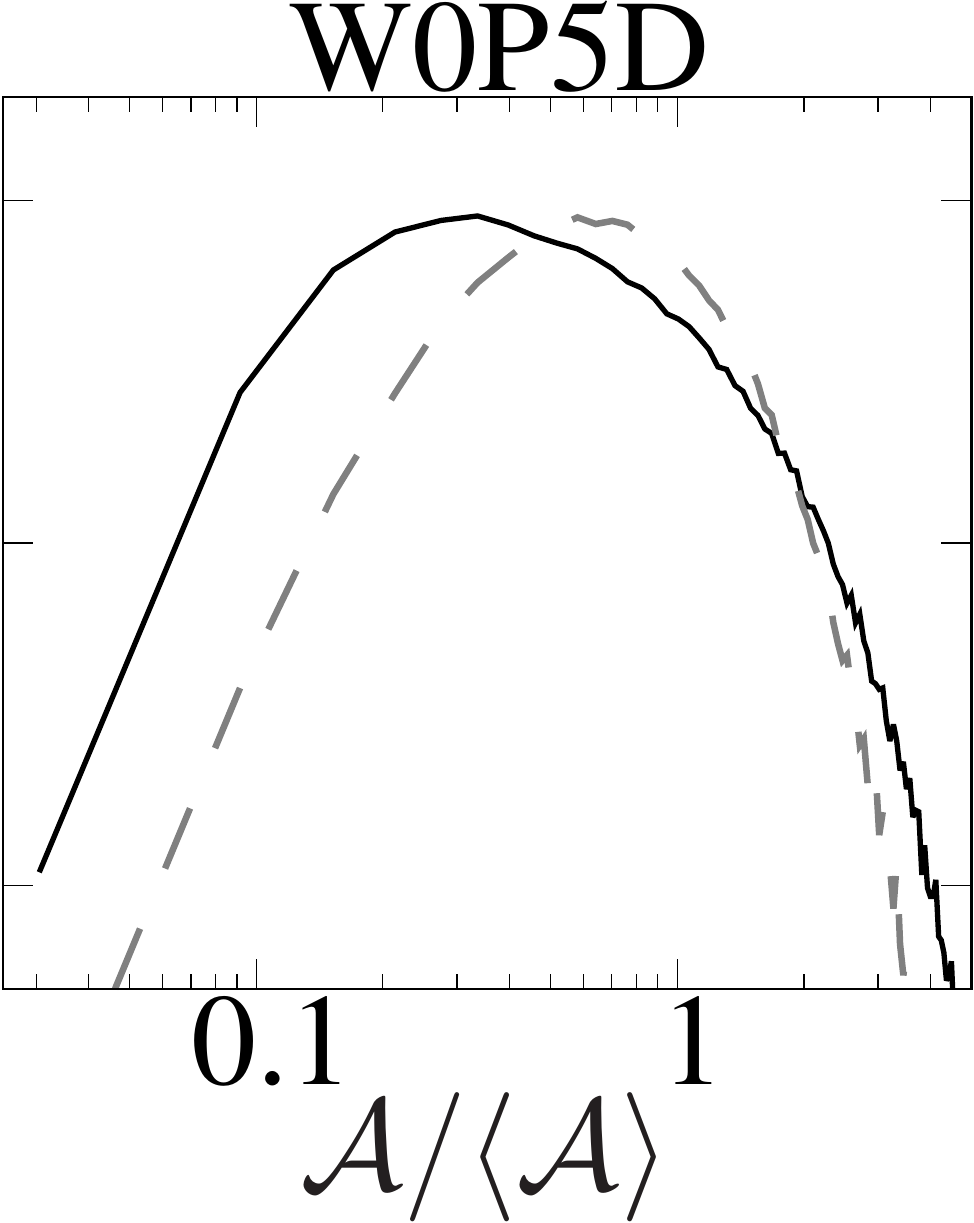}
\end{center}
\end{minipage}}	 
\subfigure[]{\label{fig:vor_volume_near_wall_W15P5}
\begin{minipage}[b]{0.215\textwidth}
\begin{center}
\includegraphics[width =\textwidth,scale=1,trim={3.55cm 0 0 0},clip]{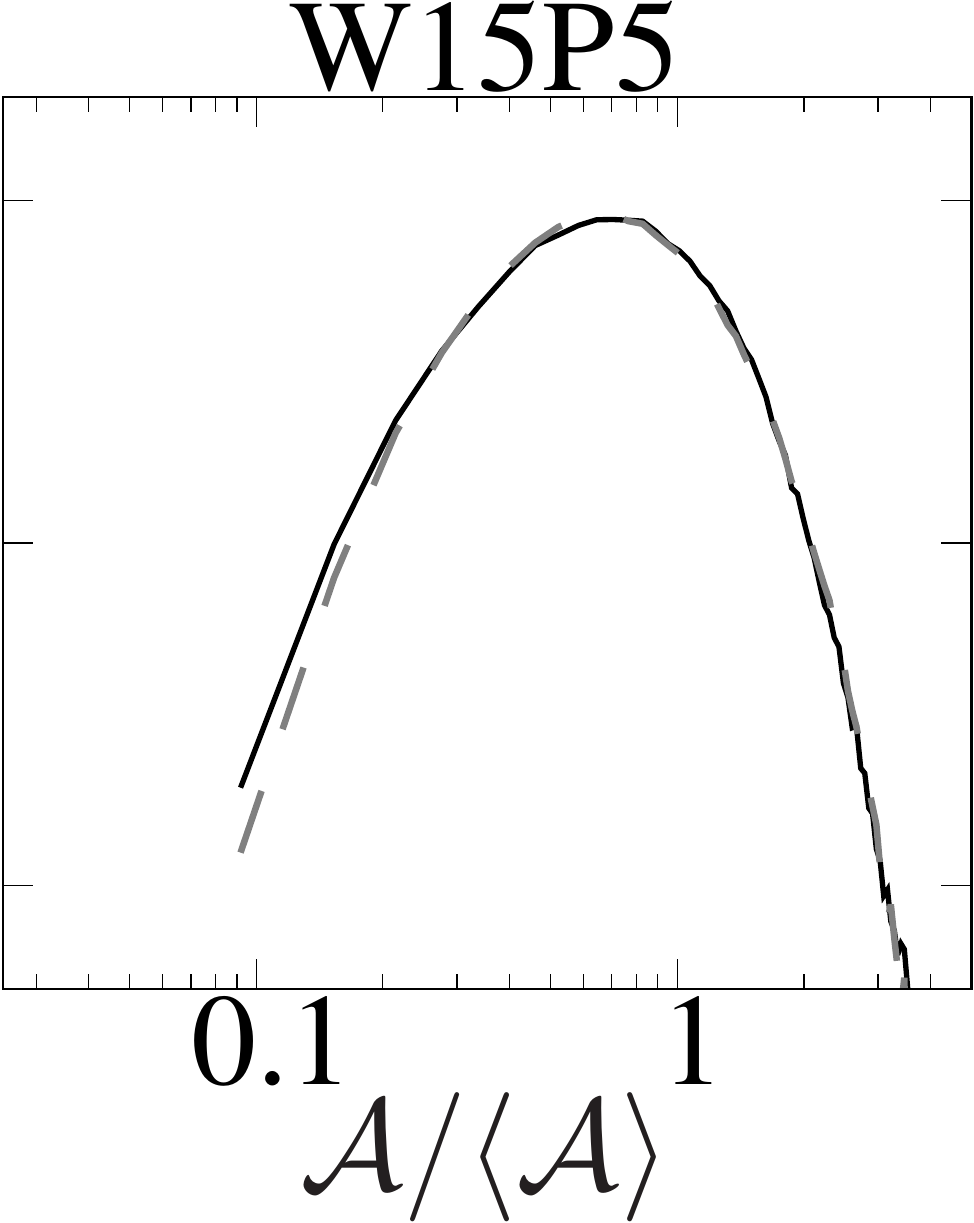}
\end{center}
\end{minipage}}				
\subfigure[]{\label{fig:vor_volume_near_wall_W15DP5}
\begin{minipage}[b]{0.215\textwidth}
\begin{center}
\includegraphics[width =\textwidth,scale=1,trim={3.55cm 0 0 0},clip]{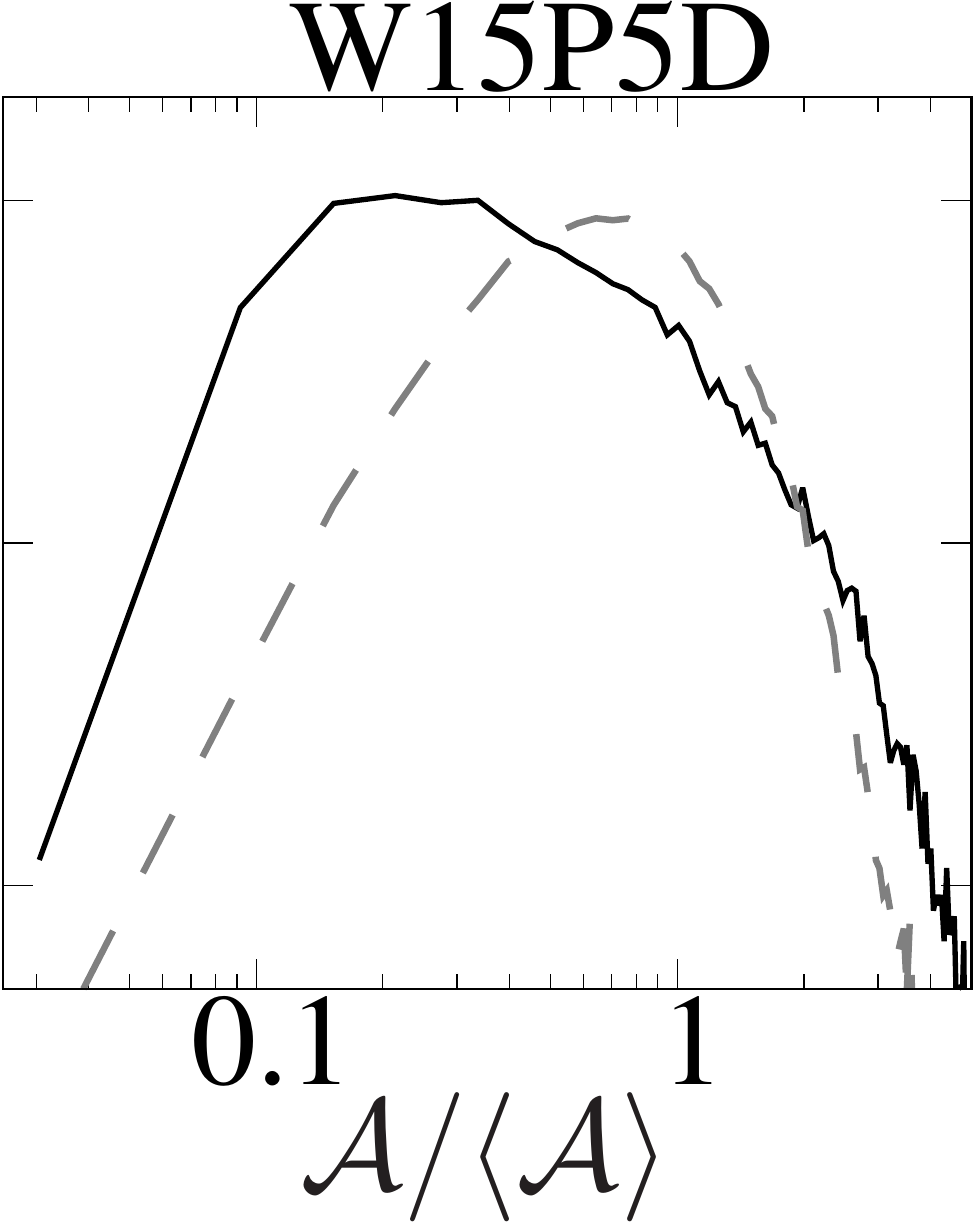}
\end{center}
\end{minipage}}
\caption{Probability density function (PDF) of normalized \Voro cell areas for near-wall particles; (a) W0P5, (b) W0P5D, (c) W15P5, (d) W15P5D.  For each case, PDF of the same quantity is also plotted for a random distribution of non-overlapping particles with a matching particle volume fraction (gray dashed lines ({\protect\greydashedlt})).   }   \label{fig:vor_volume}		
\end{figure}

A quantitative assessment of clustering is provided by evaluating the probability density function (PDF) of normalized surface areas of \Voro cells (figure \ref{fig:vor_volume}).
The position of particles located in a wall-parallel bin are projected onto the $xz$ plane. 
The extent of the bin in the wall-normal direction is $d_p/2$, while in the $x$ and $z$ directions it spans the full domain. 
The bin is placed at the closest location to the wall with a statistically significant number of particles, $y = 5d_p/4$ in W15P5D and $y=3d_p/4$ in all other cases. 
The width of the PDF highlights the heterogeneity of the particles distribution, and its tails indicate the likelihood of finding clustered and isolated particles.  
For example, for a uniformly structured layout of particles the area of \Voro cells are all equal, and therefore the PDF becomes a Dirac function.
For each case, the results for a random distribution of non-overlapping particles with a matching particle volume fraction is also presented as a reference.
For neutrally buoyant particles the differences between the PDFs of normalized \Voro areas for the present data and a random distribution are statistically insignificant.
In the case of dense particles in both Newtonian and viscoelastic fluids, however, the PDFs (a) have thicker tails and (b) their peaks are shifted towards smaller values of $\mathcal{A}/\langle \mathcal{A} \rangle$ in comparison to a random distribution.
The former observation highlights the high probability of finding significantly clustered and isolated particles, and the later point indicates that the particles are more likely to be in close proximity of one another. 
Both conclusions are in agreement with the aggregating patters visually seen in the sample snapshot (figure \ref{fig:vor_sample_near_wall}). 

\begin{figure}		
\centering
\subfigure[]{\label{fig:schematic}
\begin{minipage}[b]{0.35\textwidth}
\begin{center}
\includegraphics[width =\textwidth,scale=1]{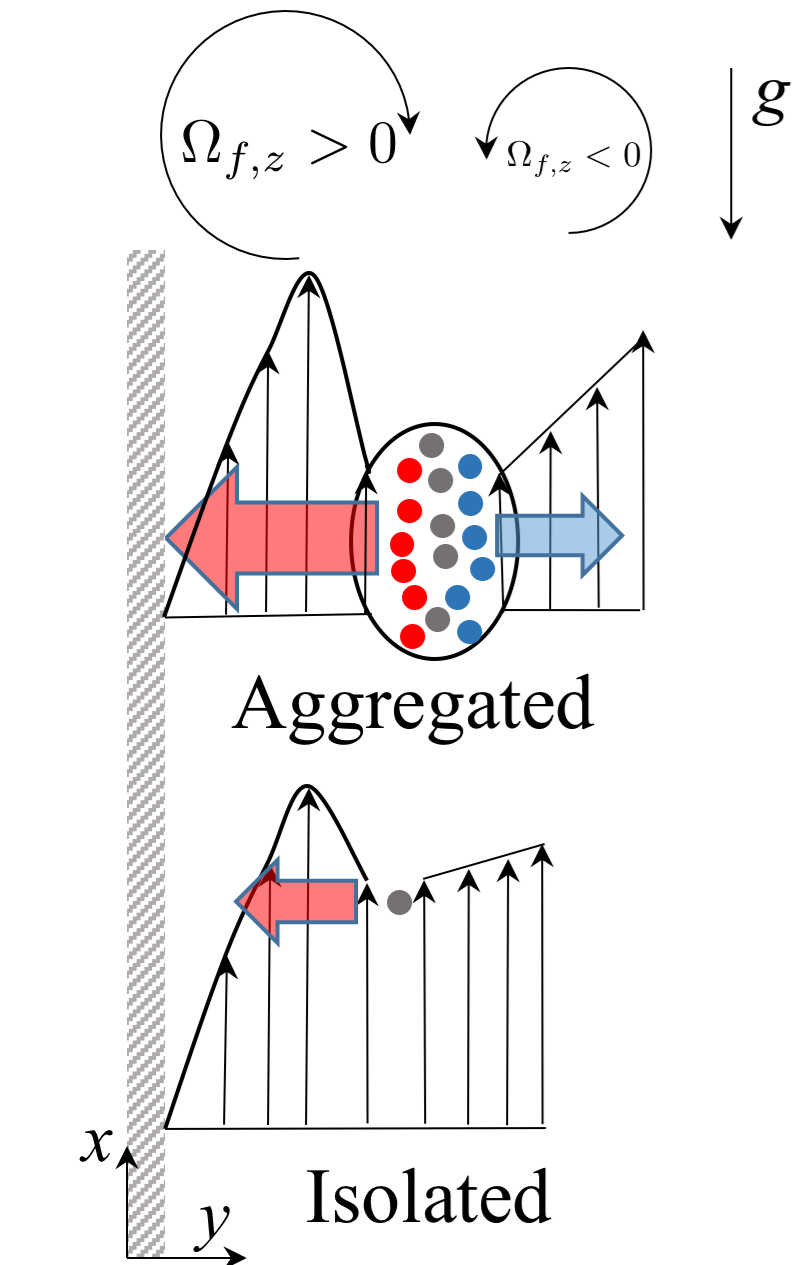}
\end{center}
\end{minipage}}	\hspace{10pt} 
\subfigure[]{\label{fig:pdf_crossing}
\begin{minipage}[b]{0.5\textwidth}
\begin{center}
\includegraphics[width =\textwidth,scale=1]{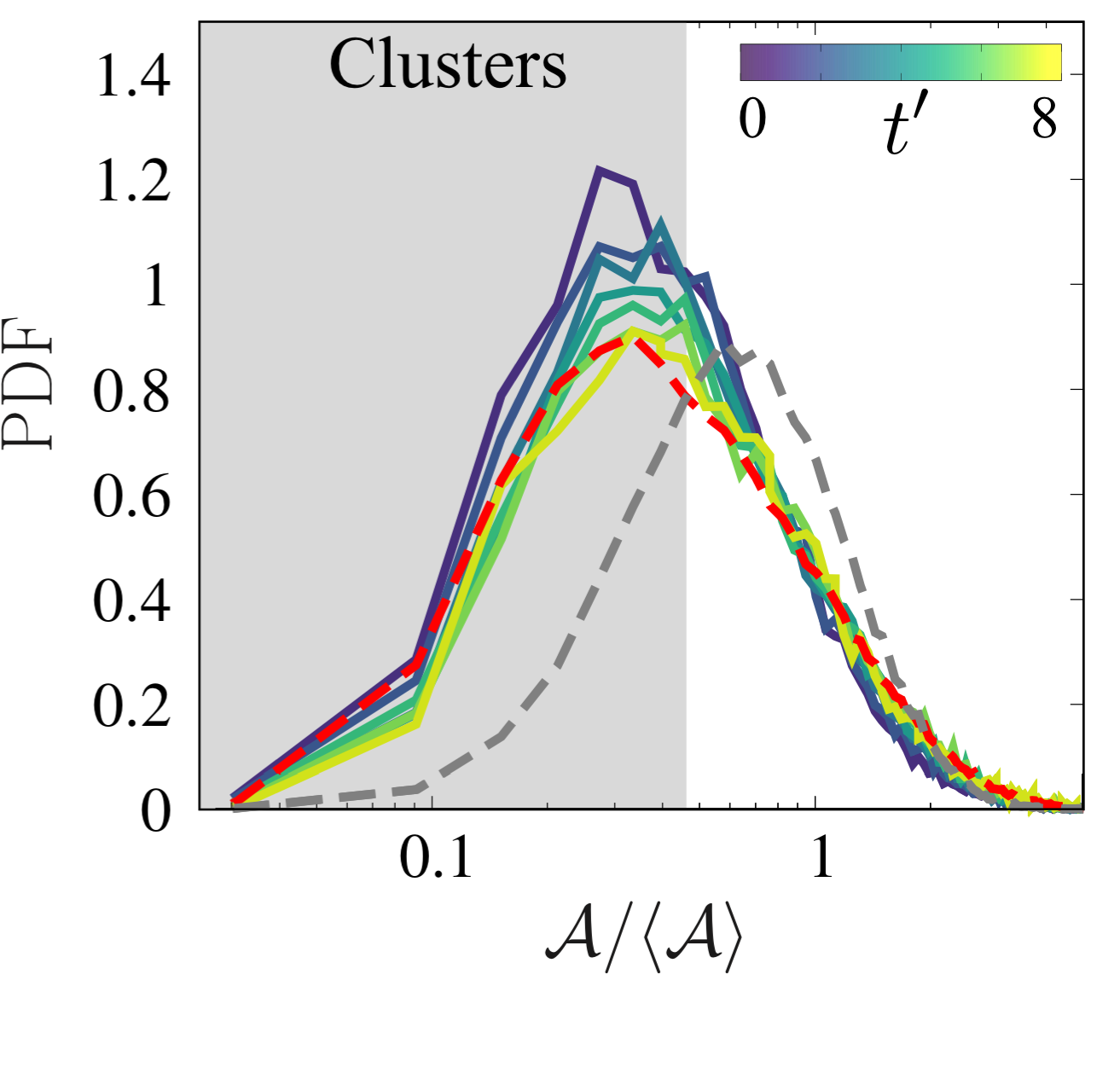}
\end{center}
\end{minipage}}
\caption{(a) Schematic of the rotation-induced lift forces acting on the aggregated and isolated particles within $d_p<y<2d_p$. 
(b) PDF of normalized \Voro cell areas $\mathcal{A}/\langle \mathcal{A} \rangle$ for case W0P5D and particles located at $y<d_p$. 
The data are conditioned based on $t'\equiv t-t_0$, where $t_0$ denotes the time at which a particle entered $y<d_p$ and $t$ is the time of data collection. 
The red dashed line ({\protect\reddashedlt}) is the unconditional PDF and the gray dashed line ({\protect\greydashedlt}) shows the PDF for a random distribution of non-overlapping particles.
Note that the average \Voro cell area $\langle \mathcal{A}\rangle$ is not conditioned. 
The shaded area shows the range of $\mathcal{A}/\langle \mathcal{A} \rangle$ for which the particles are clustered.   }   \label{fig:schematic_pdf_crossing}		
\end{figure}

Clustering in turbulent flows is often due to the preferential accumulation of particles in coherent turbulent structures \citep{kaftori1995particle}.
Nevertheless, clustering by turbulence is only observed for small particles ($d^+ < 10$) \citep{Hetsroni1994} and is therefore not relevant in our cases where $d^+ \approx \{28, 32\}$ for W0P5D and W15P5D. 
We argue that the observed clustering is due to the preferential transport of aggregated particles towards the wall. 
In \S\ref{sec:mean} we showed that the particles experience positive spanwise angular velocities when located within $y<0.2$  (figures \ref{fig:omegap_W0} and \ref{fig:omegap_W15}), and therefore are pushed towards the wall by a rotation-induced lift force.  Why this effect favors aggregated particles is shown schematically in figure \ref{fig:schematic}.  The slip velocity of the particles results in shear layers, with positive and negative vorticities on either side, and the same effect applies to the aggregate.  
For an isolated particles, the net effect is a weak rotation-induced lift force towards the wall. 
On the other hand, for particles within the aggregate, the effect is most substantial:  Particles in the wall-facing edge of the aggregate experience strong positive vorticity due to the near-wall accelerated fluid, while being shielded from the negative vorticity by neighboring particles. 
As a result, they attain a large positive angular velocity, and in turn, experience a strong rotation-induced lift force towards the wall. 

Additionally, as will be discussed shortly, the aggregated particles have smaller streamwise velocities, and in turn larger slip velocities compared to isolated ones\textemdash a fact that also contributes to a stronger rotation-induced lift force on clustered particles.
All together, the large rotation-induced lift forces on wall-facing particles within a cluster lead to higher likelihood of penetrating towards the wall and, as a result, strong clustering effect is observed in that region. 

For further evidence of the preferential wall-ward migration of aggregated particles, we revisit the PDF of $\mathcal{A}/\langle \mathcal{A}\rangle$ for particles within $y<d_p$.  This time, however, we condition the data with respect to $t'\equiv t-t_0$, where $t_0$ denotes the time at which the particle enters $y<d_p$ region and $t$ is the time of data collection.
Results are shown only for W0P5D and similar trends are also observed in W15P5D. 
Two additional curves are plotted (dashed curves), the unconditional PDF for the present data and of a random distribution of non-overlapping particles, and the first intersection of these two curves is a measure that identifies clustering \citep{monchaux2010preferential}.
The PDF for $t'=0$ represent the particles that entered $y<d_p$ region from the adjacent sub-volume, i.e.\,$d_p<y<2d_p$, at the time of data collection.
For those particles, the probability of being in a cluster is higher than the rest of the particles.
As $t'$ increases, the PDFs of conditional data tends to that of the unconditional curve, except in the extremely clustered range of the distribution ($\mathcal{A}/\langle \mathcal{A} \rangle < 0.1$). 
Interestingly, in that range the conditional PDFs move away from that of the unconditional data with increasing $t'$, implying that only the particles which recently penetrated the near-wall region can experience extreme clustering.
Overall, it is concluded that the clusters are transported from the adjacent sub-volume $d_p<y<2d_p$ rather than being formed in the near-wall region itself. 

\begin{figure}		
\centering
\subfigure[]{\label{fig:vor_JPDF_W0P5}
\begin{minipage}[b]{0.29\textwidth}
\begin{center}
\includegraphics[width =\textwidth,scale=1]{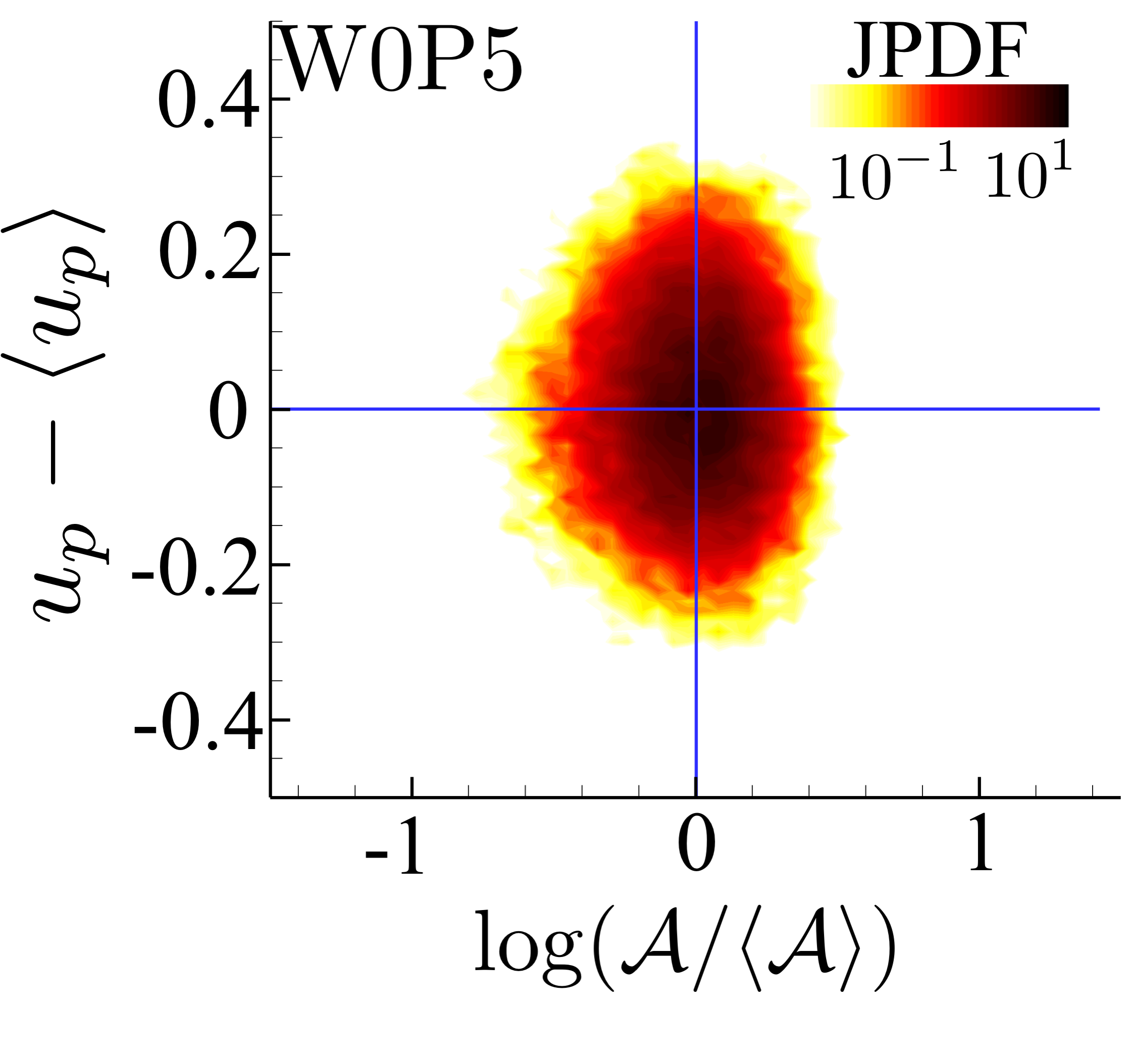}
\end{center}
\end{minipage}}		
\subfigure[]{\label{fig:vor_JPDF_W0P5D}
\begin{minipage}[b]{0.215\textwidth}
\begin{center}
\includegraphics[width =\textwidth,scale=1]{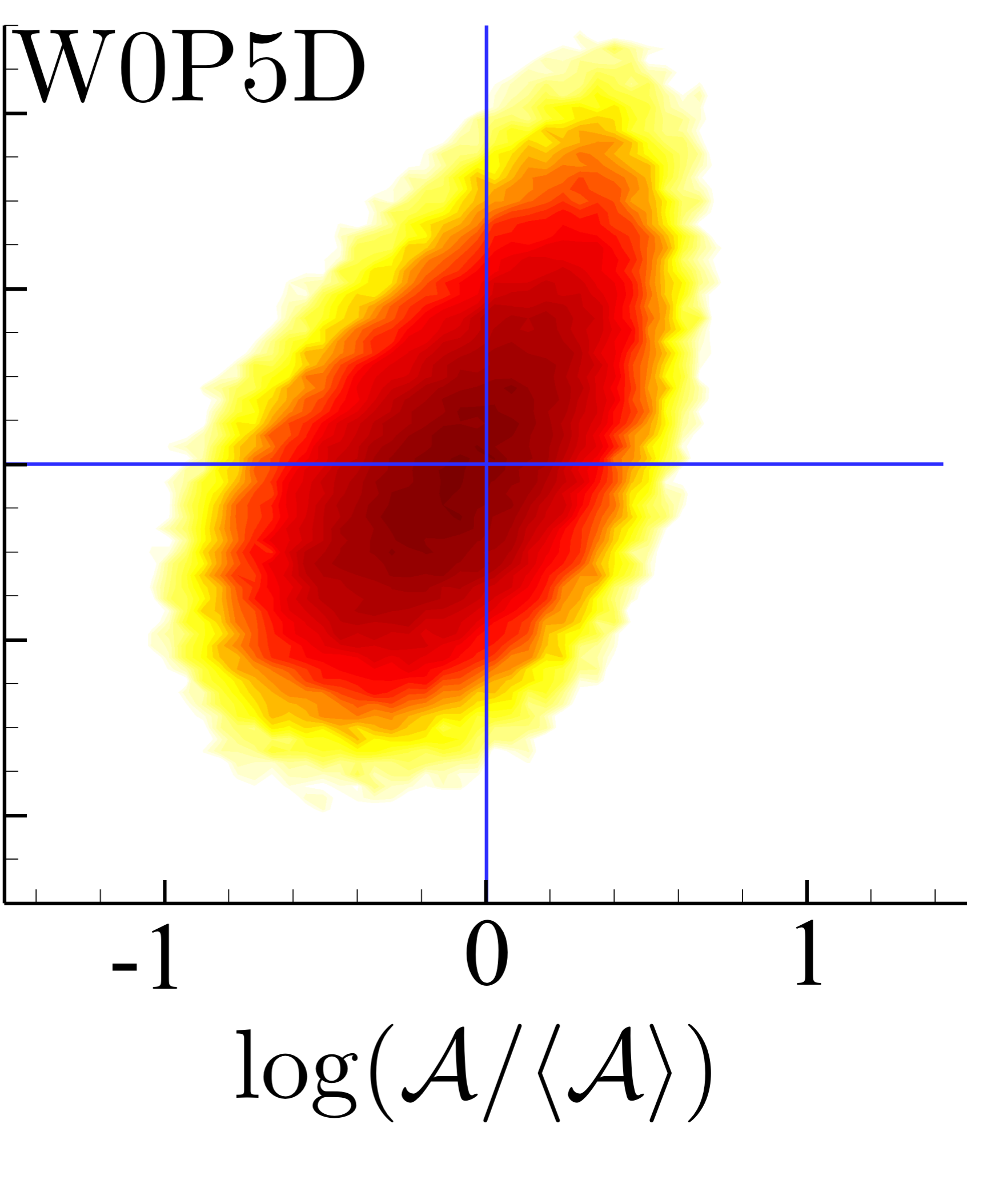}
\end{center}
\end{minipage}}										
\subfigure[]{\label{fig:vor_JPDF_W15P5}
\begin{minipage}[b]{0.215\textwidth}
\begin{center}
\includegraphics[width =\textwidth,scale=1]{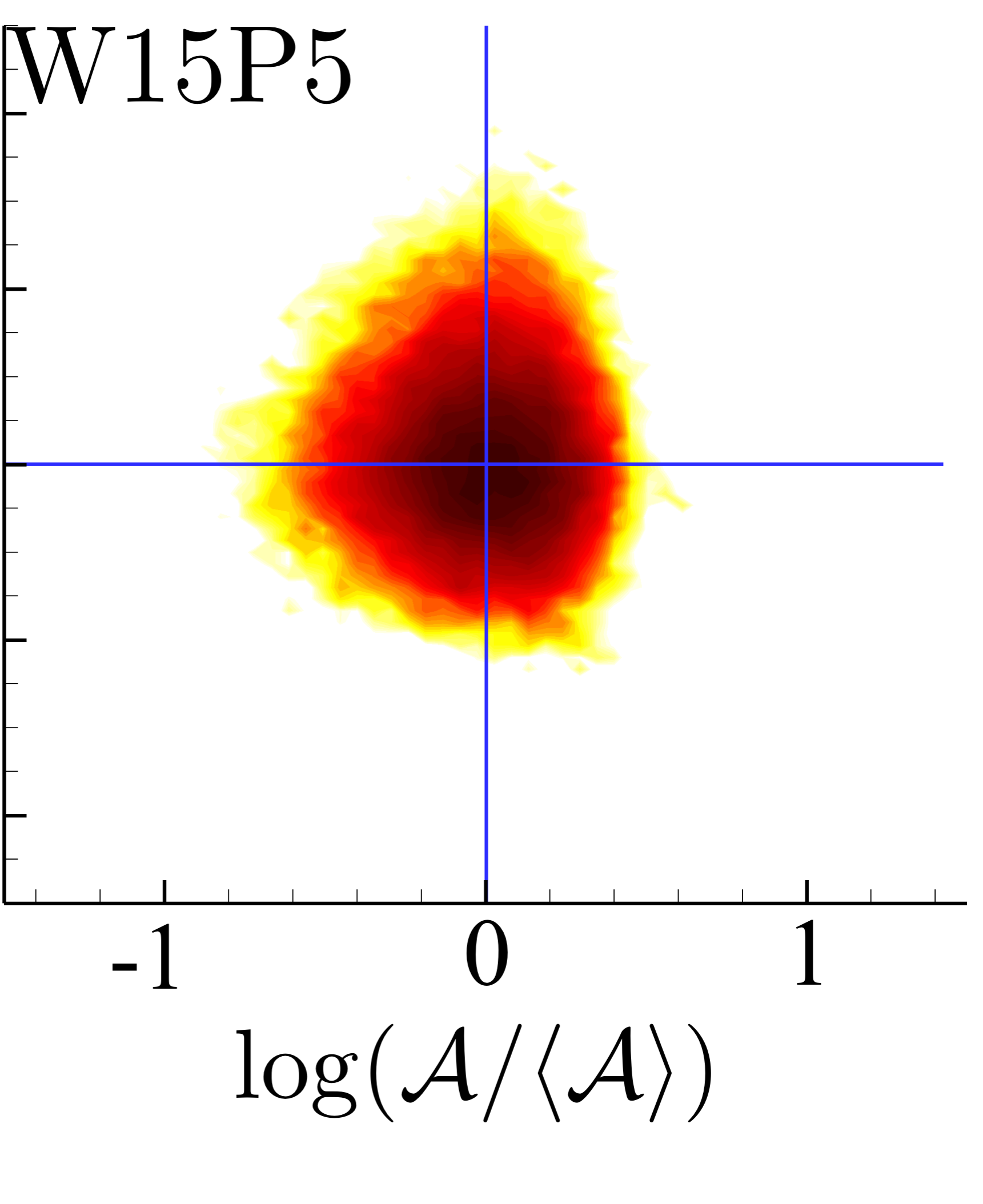}
\end{center}
\end{minipage}}	
\subfigure[]{\label{fig:vor_JPDF_W15P5D}
\begin{minipage}[b]{0.215\textwidth}
\begin{center}
\includegraphics[width =\textwidth,scale=1]{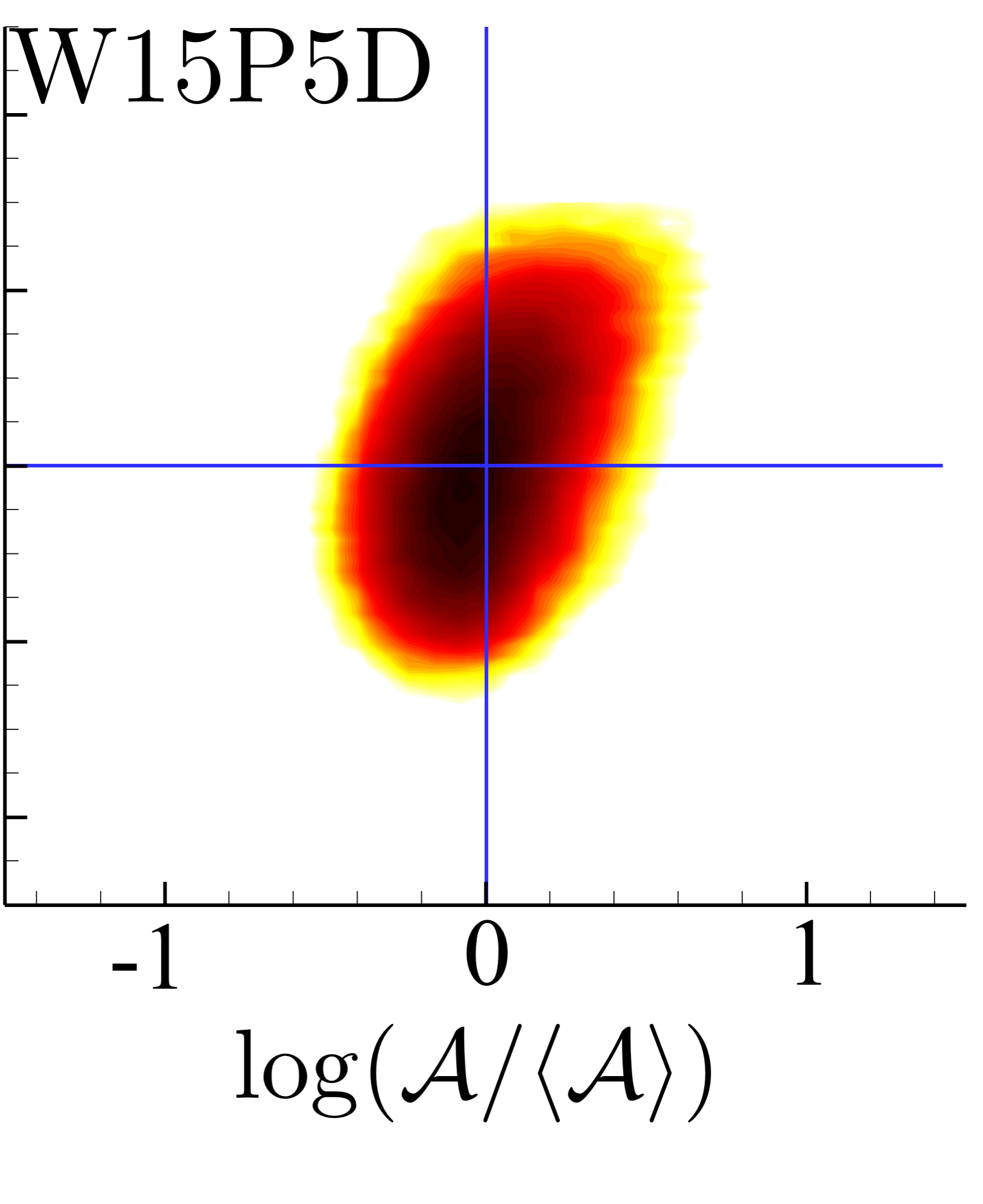}
\end{center}
\end{minipage}}										
\caption{Joint probability density functions of normalized \Voro areas $\log(\mathcal{A}/\langle \mathcal{A}\rangle)$ and mean particle relative velocity $(u_p -\langle u_f\rangle ) / \langle u_p \rangle$, (a,b) near the wall $y<0.1$ and (c,d) away from the wall $0.5<y<0.6$. 					
}   \label{fig:JPDF_vor_uprime}								
\end{figure}

Figure \ref{fig:JPDF_vor_uprime} demonstrates the impact of clustering on the particles streamwise velocities.
A correlation is observed between $u_p-\langle u_p \rangle$ and $\log(\mathcal{A}/\langle \mathcal{A} \rangle)$ for the dense particles only.
Particles in clusters have smaller streamwise velocities compared to the isolated ones and, as a result, larger slip velocities.
In all dense particle cases the slip velocity is inversely proportional to the drag coefficient, since the particles drag is balanced by the buoyancy force which is constant across the channel. 
Therefore, larger slip velocities in clusters imply that the aggregated particles experience a reduced drag coefficient due to the sheltering effect of neighboring particles. 
This phenomenon is observed in both Newtonian and viscoelastic conditions, while in the later the magnitudes of velocity fluctuations are smaller. 

\begin{figure}		
\centering
\subfigure[]{\label{fig:wakeW0P5D}
\begin{minipage}[b]{0.47\textwidth}
\begin{center}
\includegraphics[width =\textwidth,scale=1]{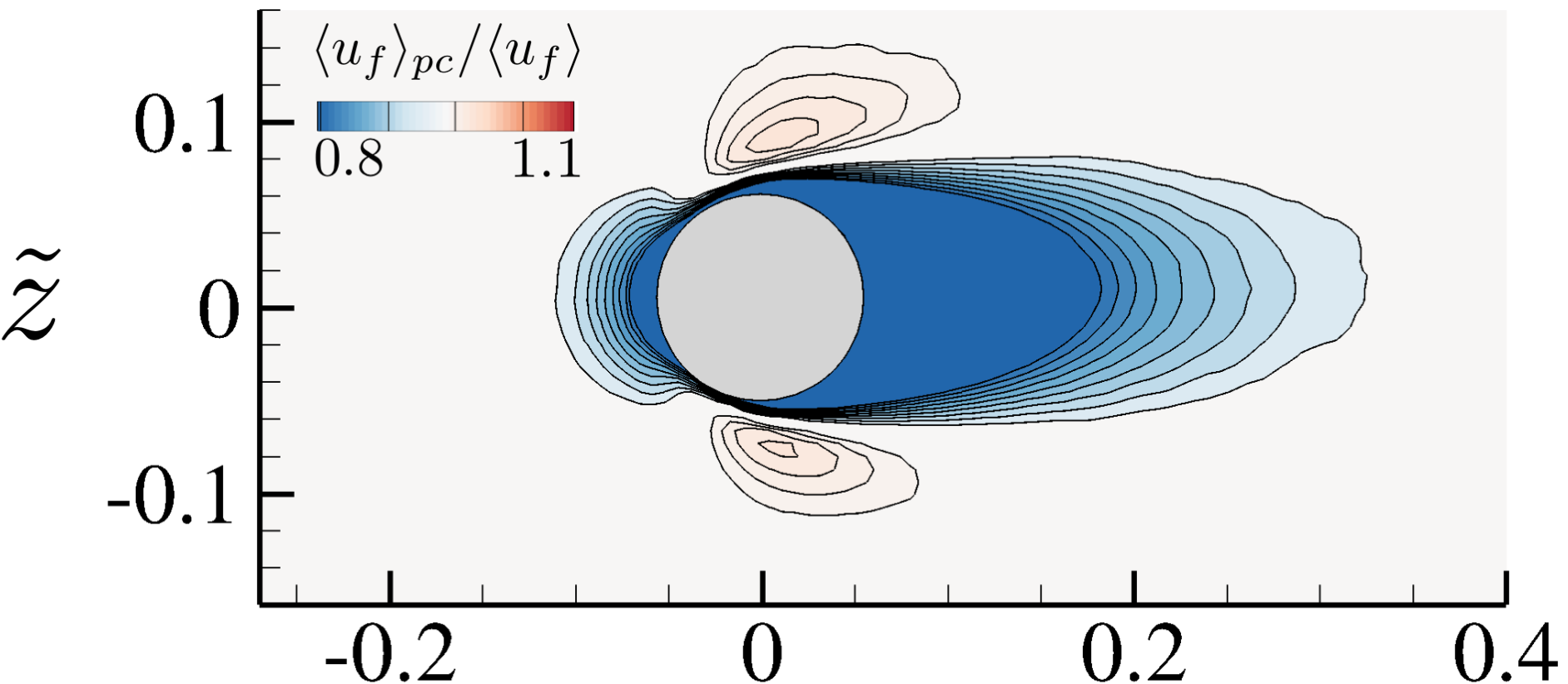}
\end{center}
\end{minipage}}
\subfigure[]{\label{fig:wakeW15P5D}
\begin{minipage}[b]{0.44\textwidth}
\begin{center}
\includegraphics[width =\textwidth,scale=1]{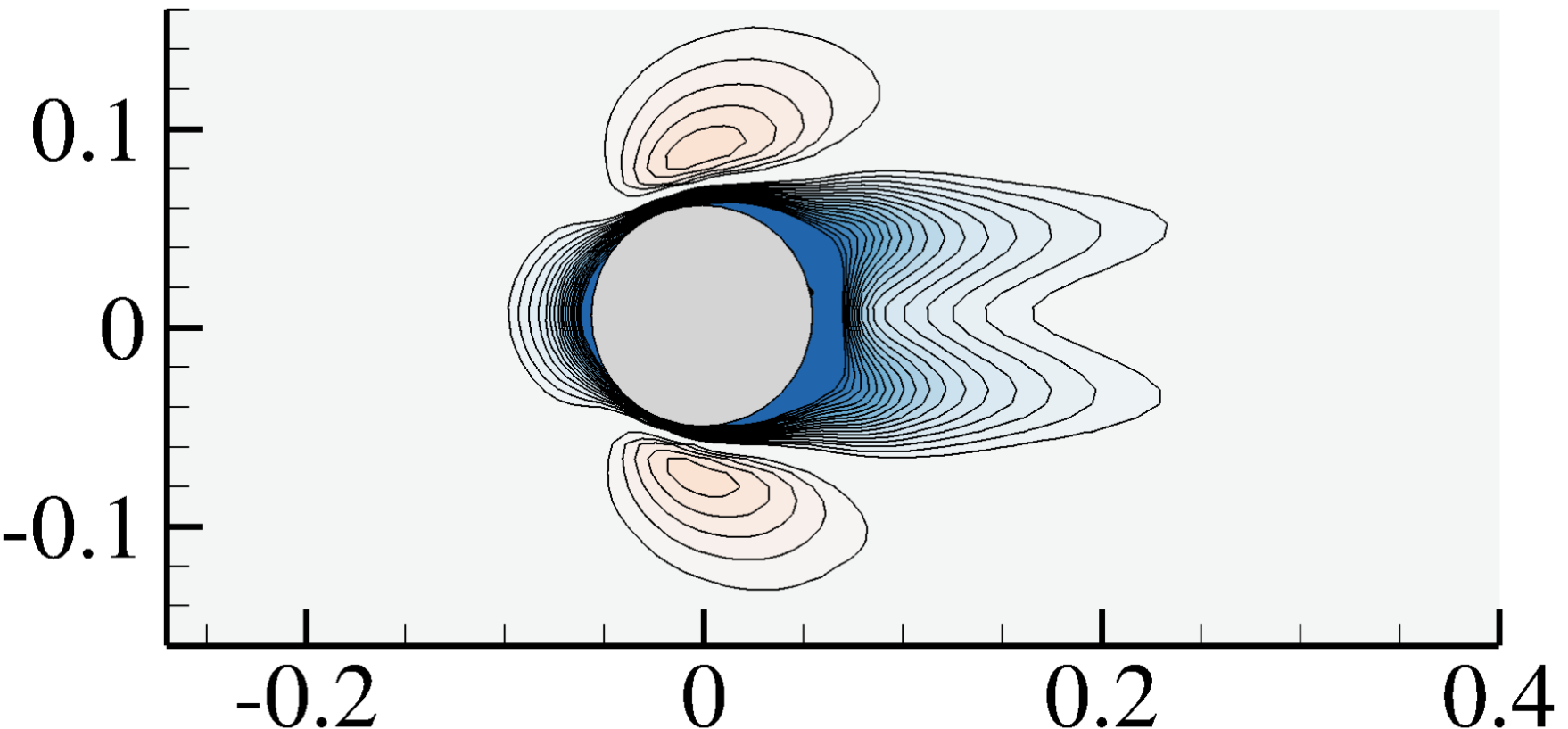}
\end{center}
\end{minipage}}	 \\
\subfigure[]{\label{fig:wakeW0P5D_Magnus}
\begin{minipage}[b]{0.47\textwidth}
\begin{center}
\includegraphics[width =\textwidth,scale=1]{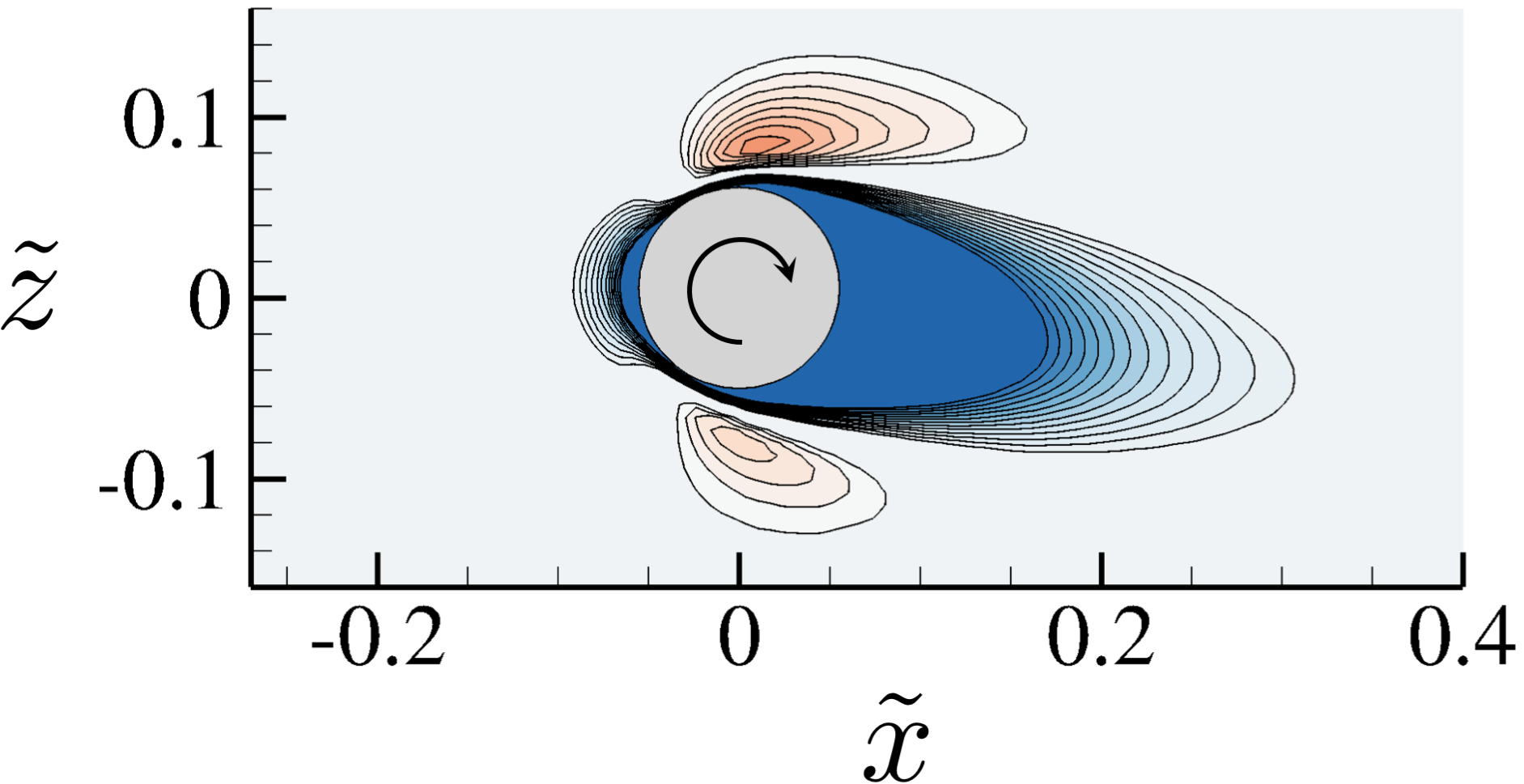}
\end{center}
\end{minipage}}				
\subfigure[]{\label{fig:wakeW15P5D_Magnus}
\begin{minipage}[b]{0.44\textwidth}
\begin{center}
\includegraphics[width =\textwidth,scale=1]{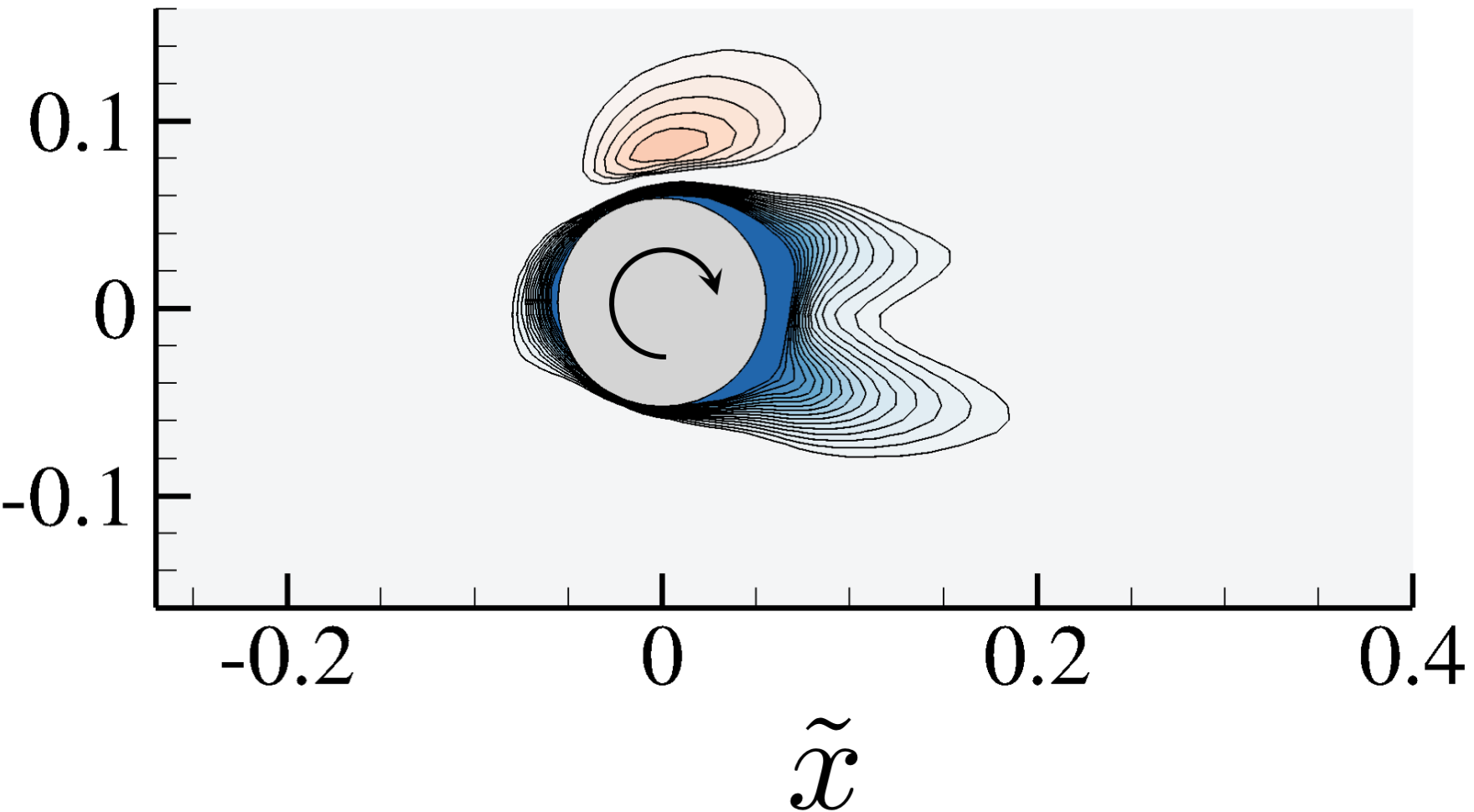}
\end{center}
\end{minipage}}				
\caption{ Particle conditioned mean flow field normalized by the unconditional fluid velocity $\langle u_f \rangle_{pc}/\langle u_f \rangle$ and plotted in the particle coordinates ($\tilde{\bm{x}} = \bm{x} - \bm{P}_\text{ref}$);
(a,c) W0P5D, (b,d) W15P5D. 
In (c) and (d) particles are further conditioned to have $\Omega_{p,y}>0$.
The conditional averages are performed for particles located at $y<0.4$. }   \label{fig:wake}		
\end{figure}

We now direct our attention to the influence of gravity and rotation-induced lift force on the micro-scale features of the flow.
Figure \ref{fig:wake} shows particle-conditioned mean flow fields normalized by the unconditional fluid velocity $\langle u_f \rangle_{pc}/\langle u_f \rangle$.
The conditional average is only performed for particles within $y<0.4$; the average wake is qualitatively similar for particles away from the wall. 
The flow decelerates downstream of the particle due to the negative slip velocity and, as a result, a strong shear layer is formed (figures \ref{fig:wakeW0P5D} and \ref{fig:wakeW15P5D}).
Viscoelasticity appreciably changes the wake structure, in particular leading to higher momentum in the core region of the wake.
Similar qualitative changes have previously been observed downstream of particles and bubbles in both experiments \citep{Kemiha2006} and numerical simulations \citep{esteghamatian2019dilute}, and are attributed to the competition between elastic and viscous stresses \citep{Frank2006}.
As a result of this increase in momentum in the core of the wake, the velocity profile recovers over a shorter distance downstream in the viscoelastic condition.
To examine the impact of particles rotation on the local flow field, averaging is further conditioned on the particles having a positive wall-normal angular velocity $\Omega_{p,y}>0$ (figures \ref{fig:wakeW0P5D_Magnus} and \ref{fig:wakeW15P5D_Magnus}).
As anticipated, particles rotation results in flow asymmetry: the fluid velocity increases on one side and decreases on the other. 
The net effect is a rotation-induced lift force in the positive $\tilde{z}$, and the incurred motion gives rise to a tilt in the downstream wake.

\begin{figure}		
\centering			
\subfigure[]{\label{fig:TPC_W0P5_NW}
\begin{minipage}[b]{0.3\textwidth}
\begin{center}
\includegraphics[width =\textwidth,scale=1]{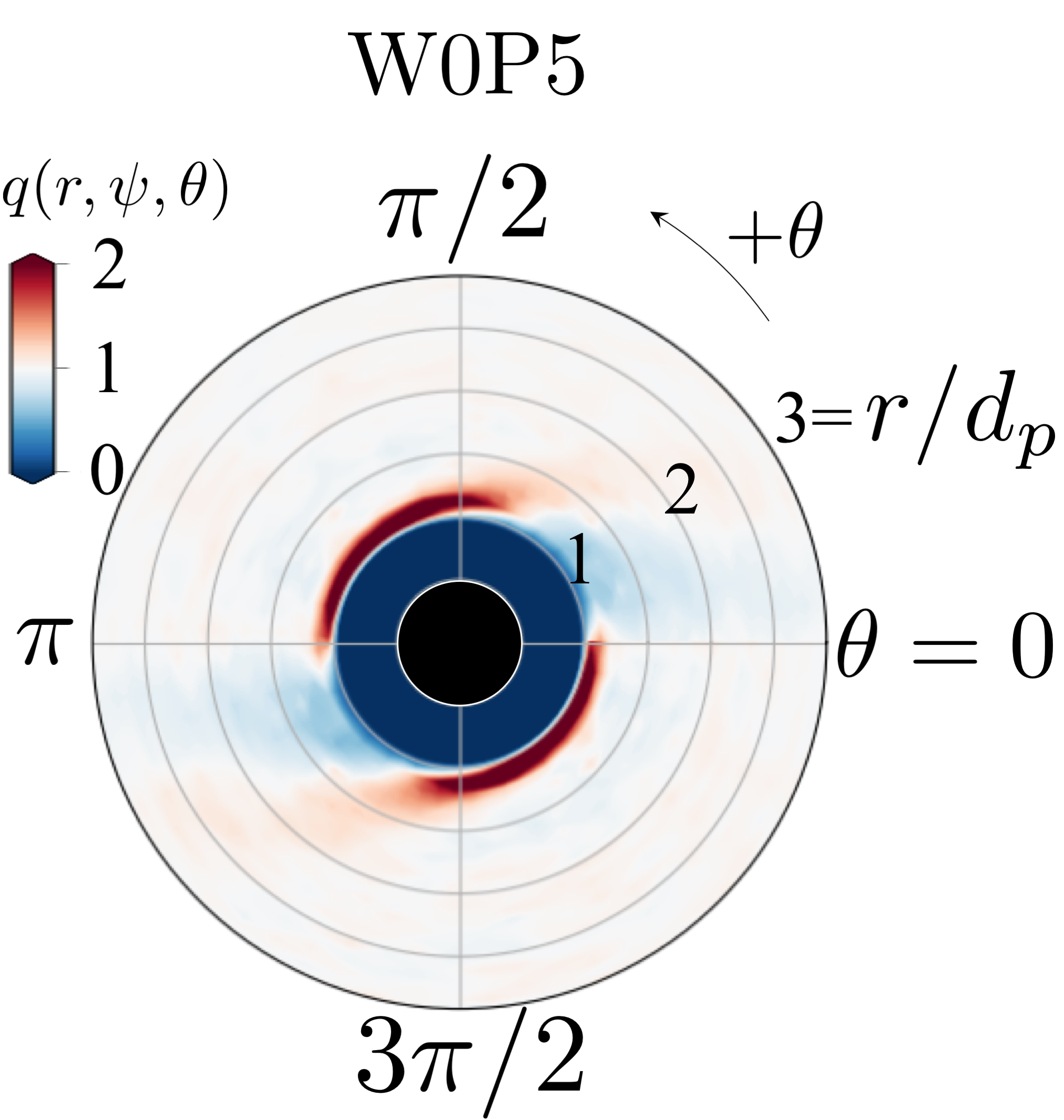}
\end{center}
\end{minipage}}			
\subfigure[]{\label{fig:TPC_W15P5_NW}
\begin{minipage}[b]{0.212\textwidth}
\begin{center}
\includegraphics[width =\textwidth,scale=1]{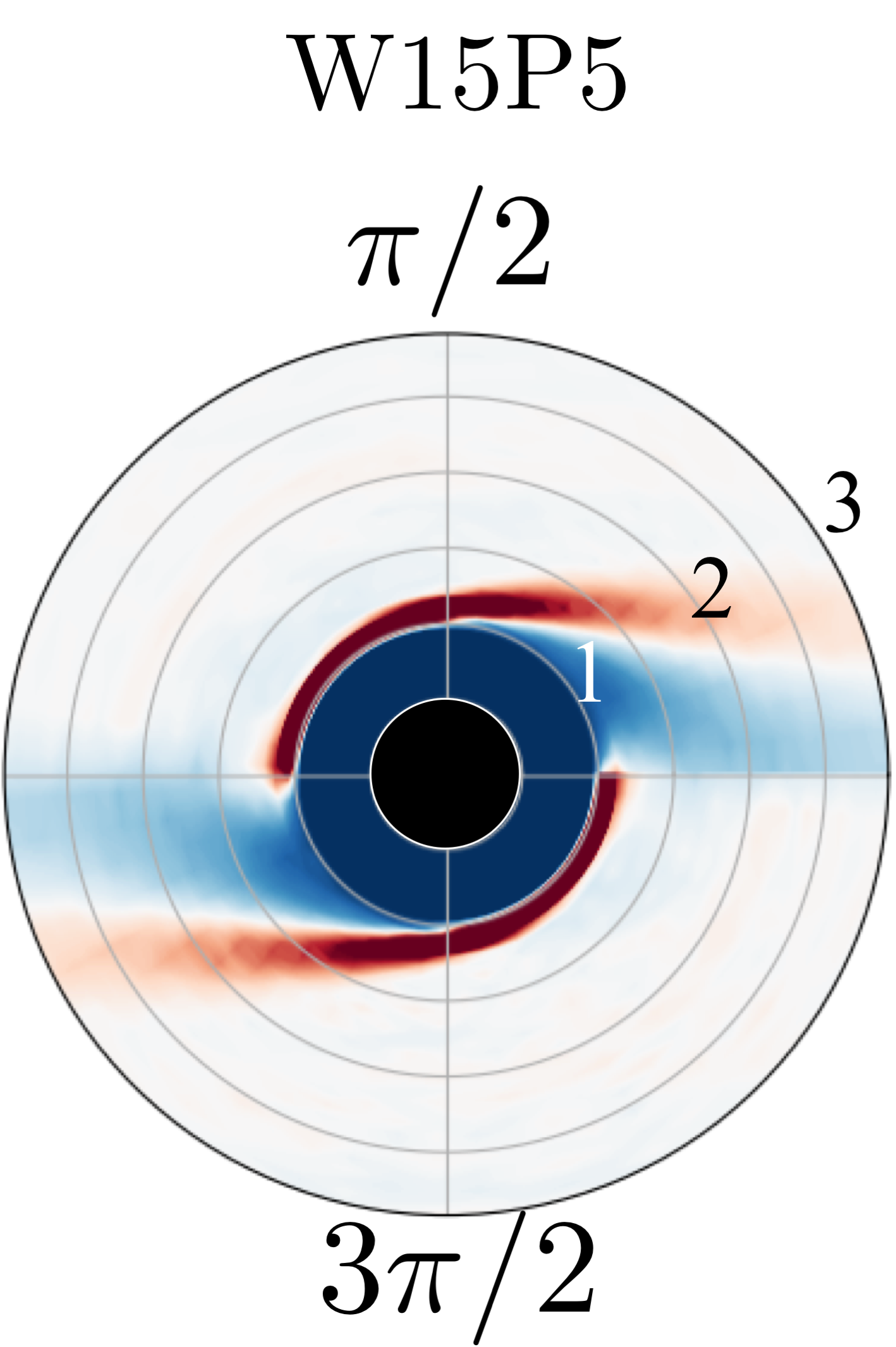}
\end{center}
\end{minipage}}			
\subfigure[]{\label{fig:TPC_W0P5D_NW}
\begin{minipage}[b]{0.212\textwidth}
\begin{center}
\includegraphics[width =\textwidth,scale=1]{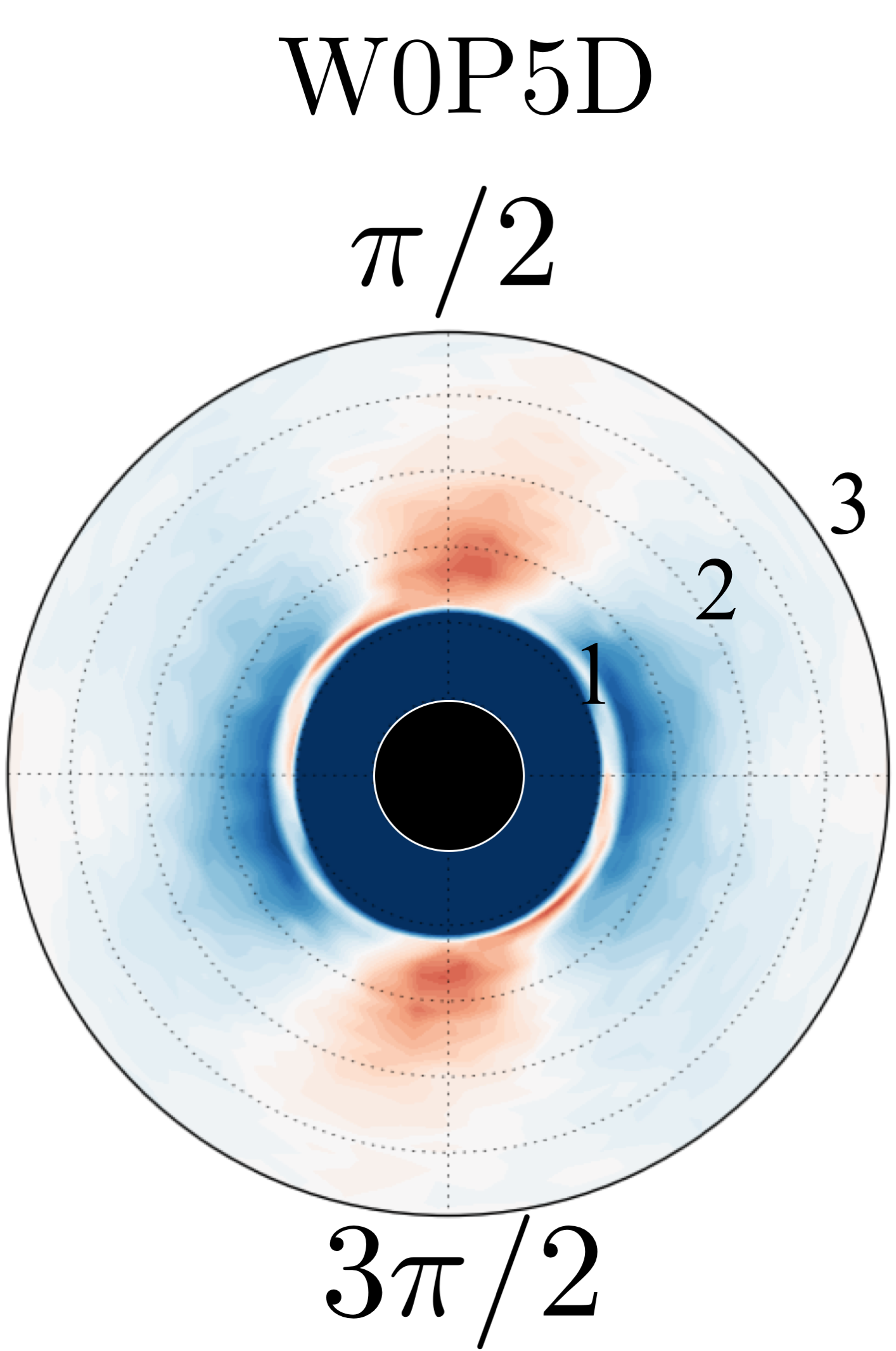}
\end{center}
\end{minipage}}					
\subfigure[]{\label{fig:TPC_W15P5D_NW}
\begin{minipage}[b]{0.212\textwidth}
\begin{center}
\includegraphics[width =\textwidth,scale=1]{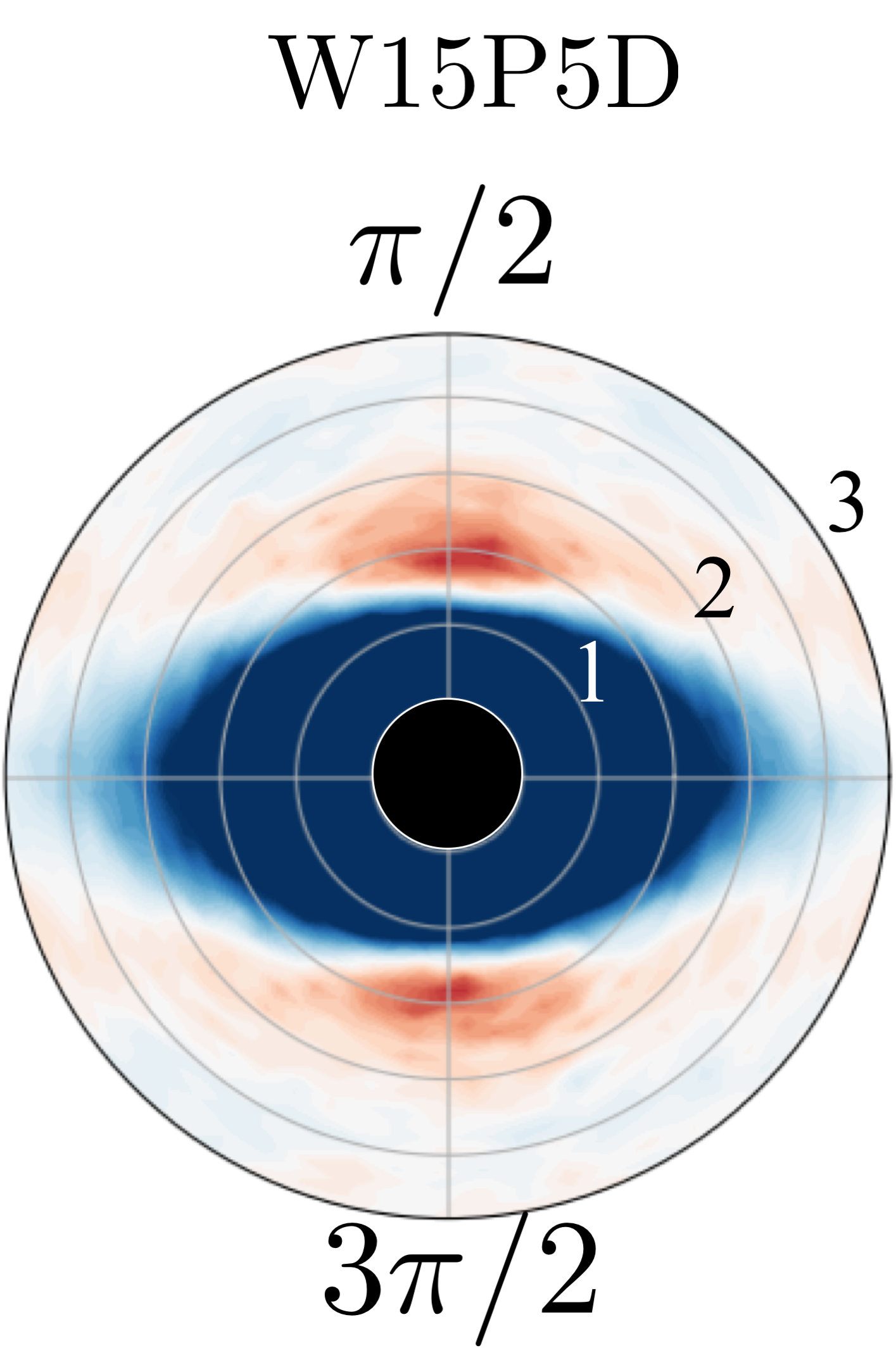}
\end{center}
\end{minipage}}																 		
\caption{Particle-pair distribution function $q(r,\psi,\theta) $ defined in \eqref{eq:ppd}, averaged in the range $-\pi/8<\psi - \pi/2 < \pi/8$. Reference particle is located at $y<0.4$. 				
}   \label{fig:TPC}								
\end{figure}

Figure \ref{fig:TPC} shows the particle-pair distribution function, $q(\bm{r},\psi,\theta)$, which depends on the pair separation $\bm{r}$, the polar angle $\psi$ relative to the positive $z$ axis, and the azimuthal angle $\theta$ measured anticlockwise from the positive $x$ axis, 
\begin{align}
& q(r,\psi,\theta) = \langle \text{d}N(r,\psi,\theta) \rangle / \langle \text{d} N(r,\psi,\theta)^\text{R} \rangle, \label{eq:ppd} \\
&\text{d}N(r,\psi,\theta) = \sum_{m=1}^{N_p} \delta(r-|\bm{r}_{m}|)\delta(\psi - \psi_{m})\delta(\theta - \theta_{m}), \\
& \bm{r}_{m} = \bm{P}_\text{m} - \bm{P}_\text{ref}, \\ 
& \psi_{m} = \cos^{-1}(\dfrac{\bm{r}_m\cdot \mathbf{e}_z}{|\bm{r}_{m}|}); \ \  \theta_{m} = \arctan(\dfrac{\bm{r}_m\cdot \mathbf{e}_y}{\bm{r}_m\cdot \mathbf{e}_x}),
\end{align}
where $\text{d}N$ denotes the number of neighboring particles within each bin and $\bm{P}$ is the particle position vector.
To avoid the bias due to the non-homogeneity in the wall-normal direction, $q$ is normalized by $dN^R$ which corresponds to a random distribution of particles with a non-overlapping condition and the case-specific wall-normal profile of the mean particle volume fraction \citep{esteghamatian2019dilute}.
The reference particle is located within $y<0.4$ (microstructure and clustering of particles away from the wall are reported in Appendix \ref{app:clust}).
For neutrally buoyant particles (figures \ref{fig:TPC_W0P5_NW}, \ref{fig:TPC_W15P5_NW}), $q$ is maximum at a separation distance of approximately one particle diameter due to collision forces. 
Also, the microstructure for neutrally-buoyant particles is dominated by the mean shear, with regions of particle accumulation in the second and forth quadrants.
For dense particles, microstructure is dominated by the particle wake.
In W0P5D and W15P5D, particle-pairs are preferentially aligned in the cross-stream direction (figures \ref{fig:TPC_W0P5D_C}, \ref{fig:TPC_W15P5D_C}), in agreement with previous observations in sedimenting suspensions \citep{yin2007hindered} and fluidization \citep{fortes1987nonlinear}. 
This preferential arrangement is a consequence of the wake interactions, in particular the drafting-kissing-tumbling mechanism.
One particle is trapped in the wake of another particle due to the reduced drag (drafting); it approaches the leading particle (kissing), rotates due to the lift force induced by the shear-layer near the edge of the wake (c.f. figures \ref{fig:wakeW0P5D}), and forms a cross-stream alignment (tumbling) \citep{fortes1987nonlinear}. 
The deficit of $q$ in the streamwise direction, however, is stronger in W15P5D, which is attributed to the viscoelastic wake structure. 
The enhanced momentum in the core region of the wake, as seen in figure \ref{fig:wakeW15P5D}, inhibits the kissing phase, while the shear-layer near the edge of the wake promotes the tumbling phase. 
Both of these effects reduce the formation of streamwise alignments and yield a stronger deficit of $q$ in the streamwise direction.

\subsection{Mean stress balance and velocity fluctuations  \label{sec:turb}}
\begin{figure}		
\subfigure[]{\label{fig:stress_barchart}
\begin{minipage}[b]{0.36\textwidth}
\begin{center}
\includegraphics[width =\textwidth,scale=1]{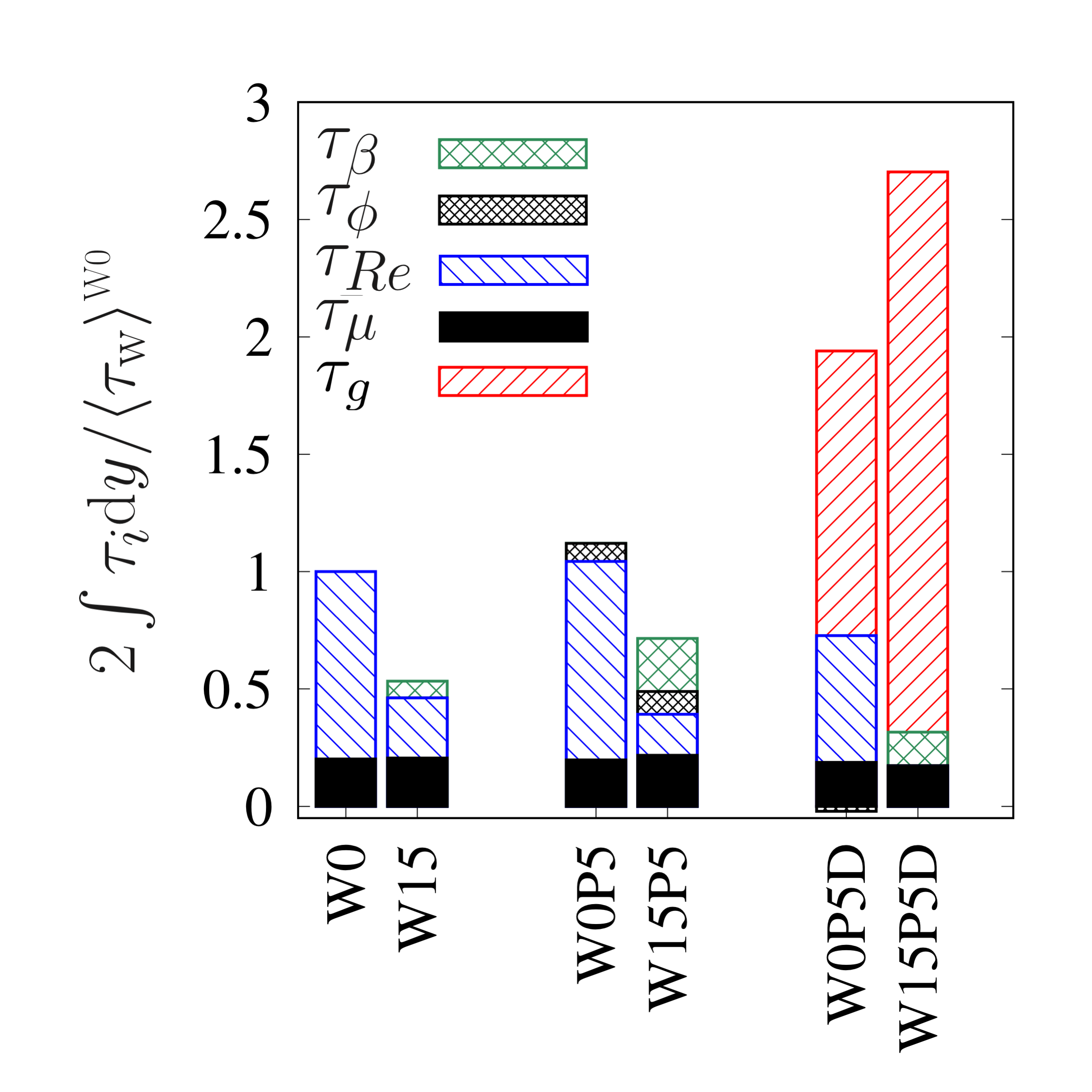}
\end{center}
\end{minipage}}		
\subfigure[]{\label{fig:stress_W0P5D}
\begin{minipage}[b]{0.32\textwidth}
\begin{center}
\includegraphics[width =\textwidth,scale=1,trim={0 -1.6cm 0cm 0},clip]{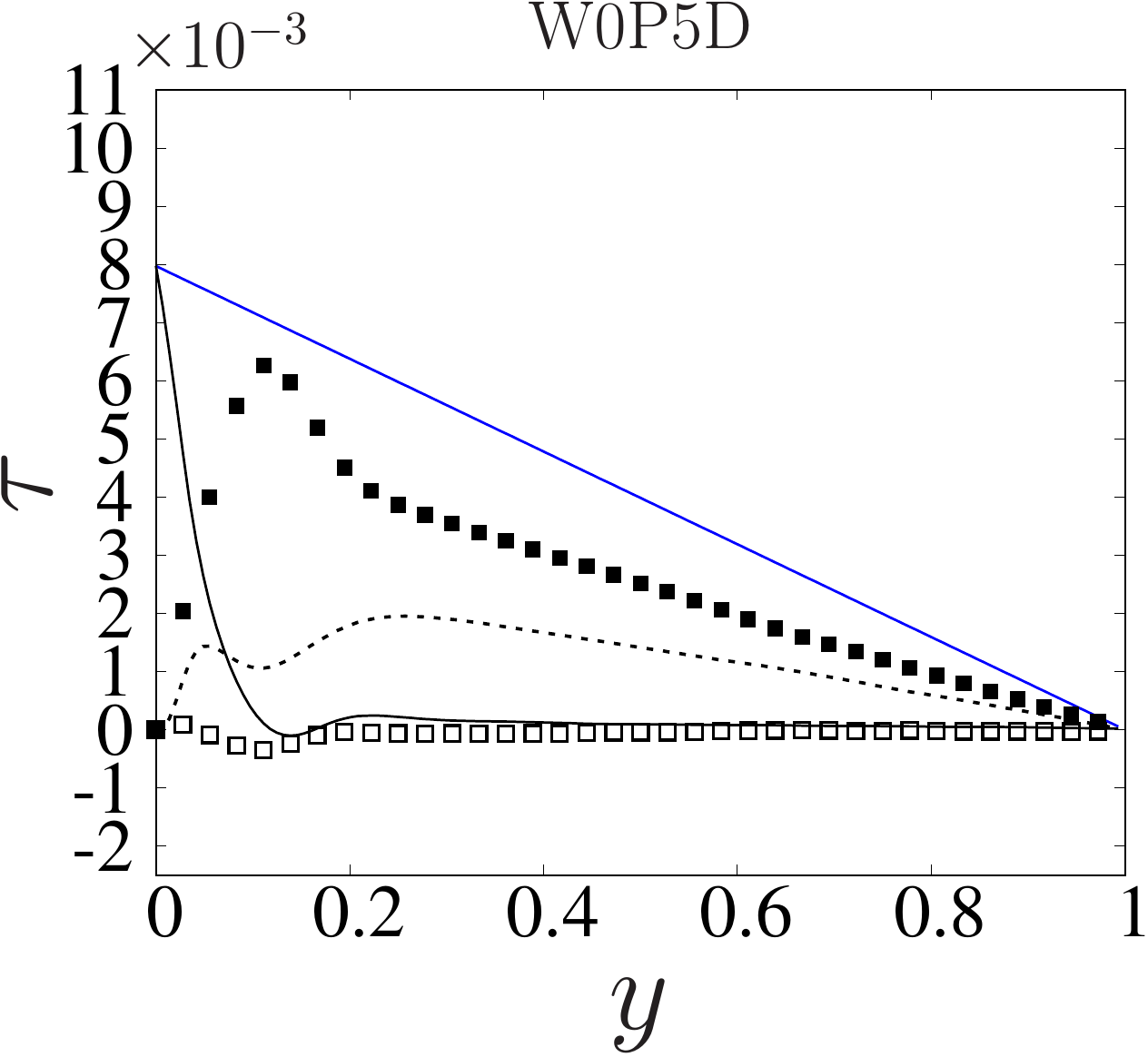}
\end{center}
\end{minipage}}		
\subfigure[]{\label{fig:stress_W15P5D}
\begin{minipage}[b]{0.285\textwidth}
\begin{center}
\includegraphics[width =\textwidth,scale=1,trim={1.55cm -1.6cm 0 0},clip]{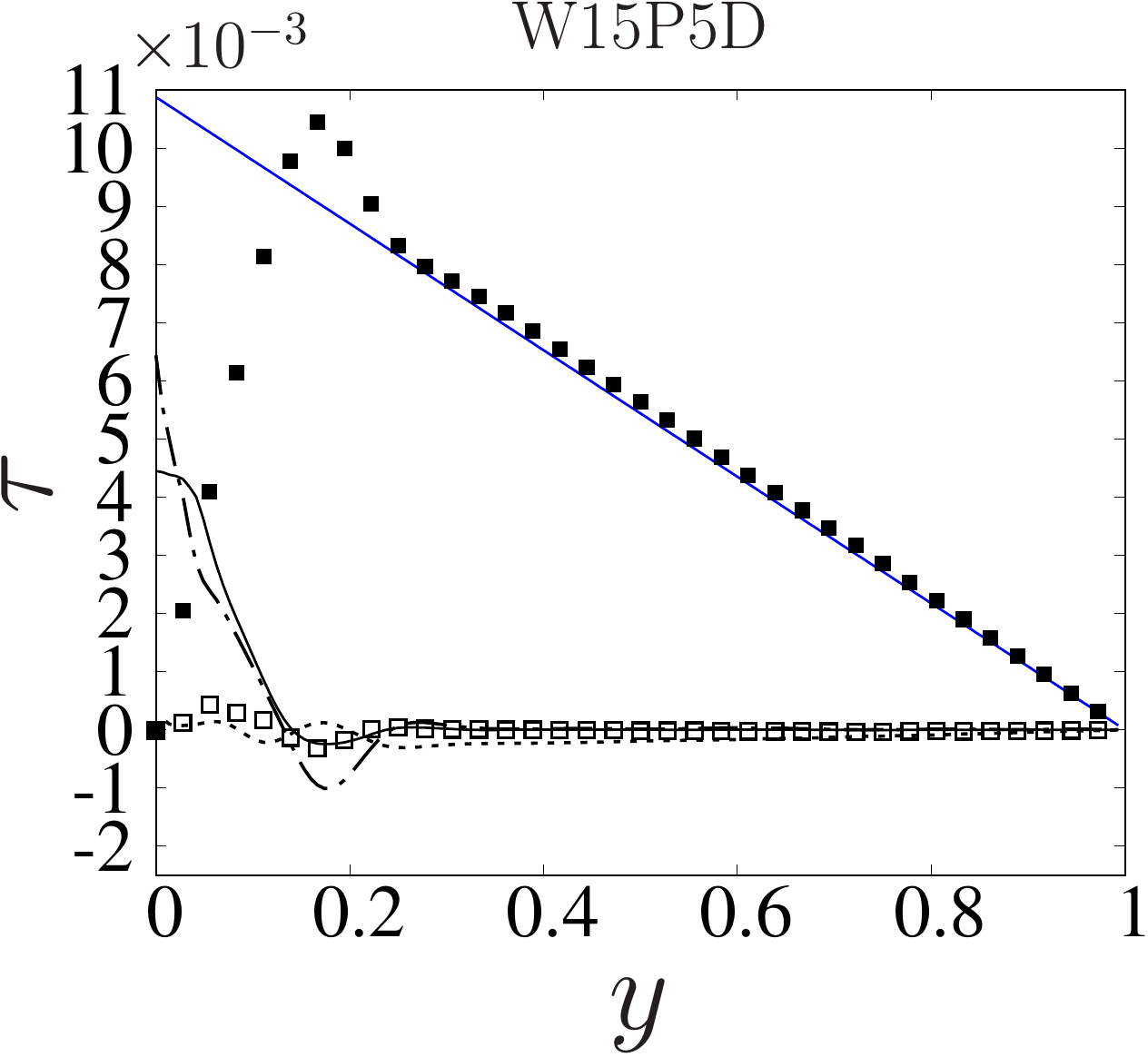}
\end{center}
\end{minipage}}
\caption{(a) Contribution of different stress components to the wall drag, integrated between $0\le y \le 1$.
Wall-normal profiles of the stresses defined in \eqref{eq:stressbudget} for cases (a) W0P5D and (b) W15P5D: 
({\protect\solidlt}) viscous stress $\tau_\mu$;
({\protect\dottedlt}) turbulent Reynolds stress $\tau_{Re}$;
({\protect\dashdotlt}) polymer stress $\tau_\beta$;
({\protect\symbolsquarelt}) particle stress $\tau_\phi$;
({\protect\symbolfillsquarelt}) gravity stress $\tau_g$;
({\protect\bluelt}) right hand side of \eqref{eq:stressbudget};
Case designations are listed in table \ref{tab:params}.
}   \label{fig:stress}		
\end{figure}

A derivation of the stress balance equation for neutrally-buoyant particles in a viscoelastic channel flow was previously provided \citep[see appendix A by][]{esteghamatian2019dilute} and is herein adapted to $\rho_r \ne 0$ condition following a similar derivation by \citet{uhlmann2008interface}. The various contributions to the wall stress $\tau_\text{w} $ at every $y$ position are given by,
\begin{align}
\overbrace{\dfrac{\beta}{Re} (1-\phi)  \DCP{\langle u_f \rangle}{y}}^{\Scale[1.2]{\tau_{\mu}}} ~
\overbrace{-~[(1-\phi) \langle u'_f v'_f \rangle  + \rho_r\phi \langle u'_p v'_p \rangle]}^{\Scale[1.2]{\tau_{Re}}}   \notag  \hspace{3cm}\\  +
\underbrace{\dfrac{(1-\beta)}{Re}(1-\phi) \langle \mathsf{T}_{xy} \rangle}_{\Scale[1.2]{\tau_{\beta}}} + \underbrace{\phi \langle \sigma_{p,xy} \rangle}_{\Scale[1.2]{\tau_{\phi}}}  \hspace{2pt}\underbrace{-g_x(\rho_r-1)\int_0^y(\varphi - \phi)\text{d}y}_{\Scale[1.2]{\tau_{g}}}= \langle\tau_\text{w}\rangle(1-y).
\label{eq:stressbudget}
\end{align}
From left to right, the components are the viscous stress $\tau_\mu$, the turbulent Reynolds stress $\tau_{Re}$, the polymer stress $\tau_\beta$, $\tau_{\phi}$ the particle stress, and $\tau_g$ the stress due to gravity. 
When integrated over $0 \le y \le 1$, the left hand side of \eqref{eq:stressbudget} expresses different contributions to the mean drag at the wall.
In figure \ref{fig:stress_barchart}, the components of the stress are normalized by $0.5\langle \tau_\text{w}  \rangle^{\Scale[0.55]{\text{W0}}}$ from the single-phase Newtonian case as a reference. 
In dense cases, the wall drag is significantly increased due to an increase in $\tau_g$.
This extra stress arises due to the wall-normal non-uniformity of the reaction forces of the particle drag on the fluid, which is larger when the particles accumulate away from the wall and vanishes when they are homogeneously distributed.
In viscoelastic cases, the particles are displaced further away from the wall due to the imbalance of elastic normal stresses \citep{d2015particle}, which increases the contribution of $\tau_g$. 
Interestingly, although the turbulent stress $\tau_{Re}$ vanishes in W15P5D, the larger contribution of $\tau_g$ in W15P5D compared with W0P5D results in an overall drag increase.

Different constituents of the drag in the dense cases are further examined in figures \ref{fig:stress_W0P5D} and  \ref{fig:stress_W0P5D}.  
Except for the near-wall region which is void of particles, $\tau_g$ is the dominant stress component across the channel.
The viscous and polymer contributions are a function of the mean shear rate $\langle \dot{\gamma} \rangle $ and follow similar trends to one another across the channel.
Thus, $\tau_\mu$ and $\tau_\beta$ are positive in the immediate vicinity of the wall, are slightly negative near $y\approx 0.2$, and vanish away from the wall (compare with profiles of $\langle \Omega_{f,z} \rangle= -\dot{\gamma}$ in figures \ref{fig:omegaf_W0} and \ref{fig:omegaf_W15}).

\begin{figure}		
\subfigure[]{\label{fig:TKE_W0}
\begin{minipage}[b]{0.31\textwidth}
\begin{center}
\includegraphics[width =\textwidth,scale=1]{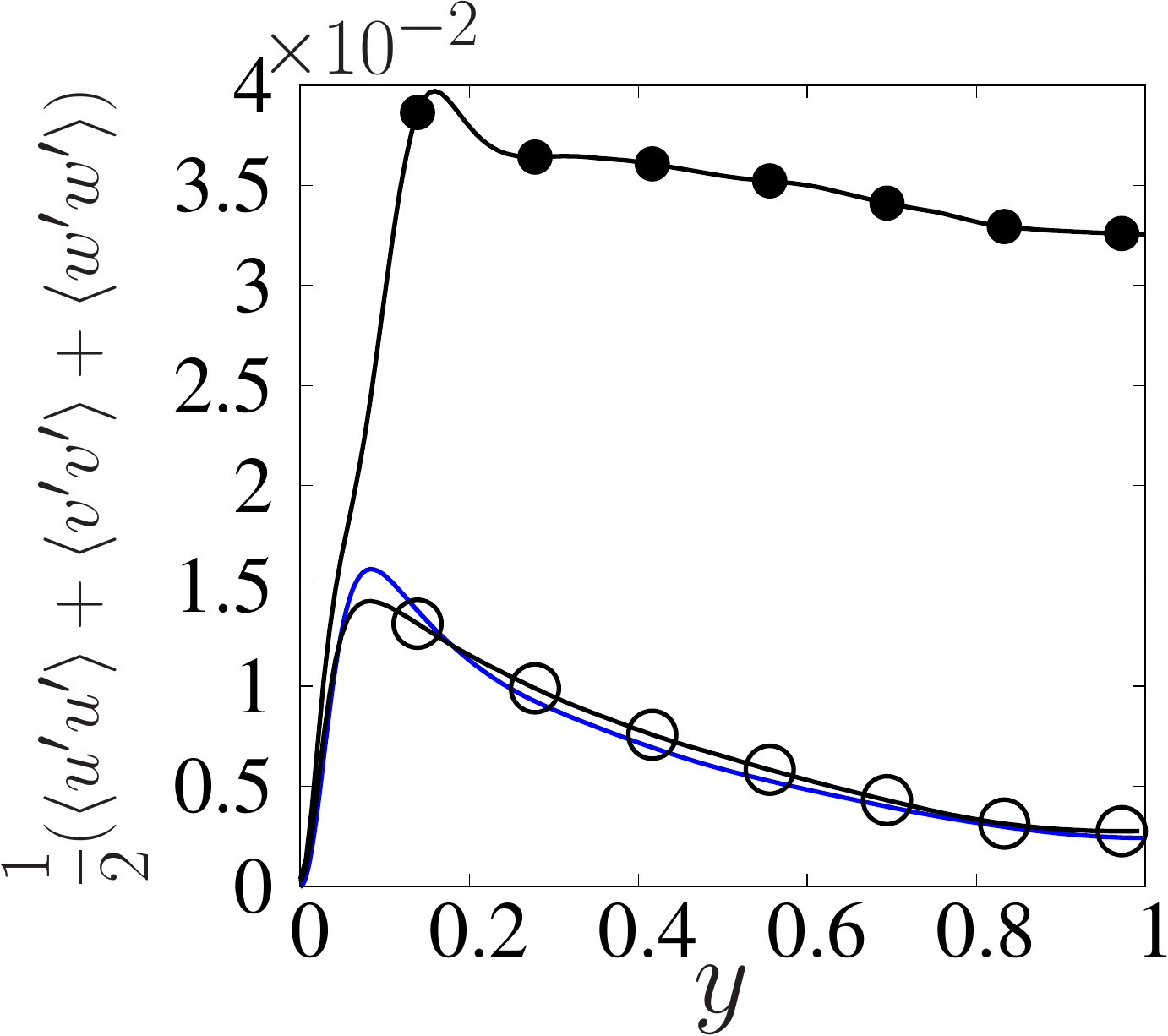}
\end{center}
\end{minipage}}		
\subfigure[]{\label{fig:uu_W0}
\begin{minipage}[b]{0.3\textwidth}
\begin{center}
\includegraphics[width =\textwidth,scale=1]{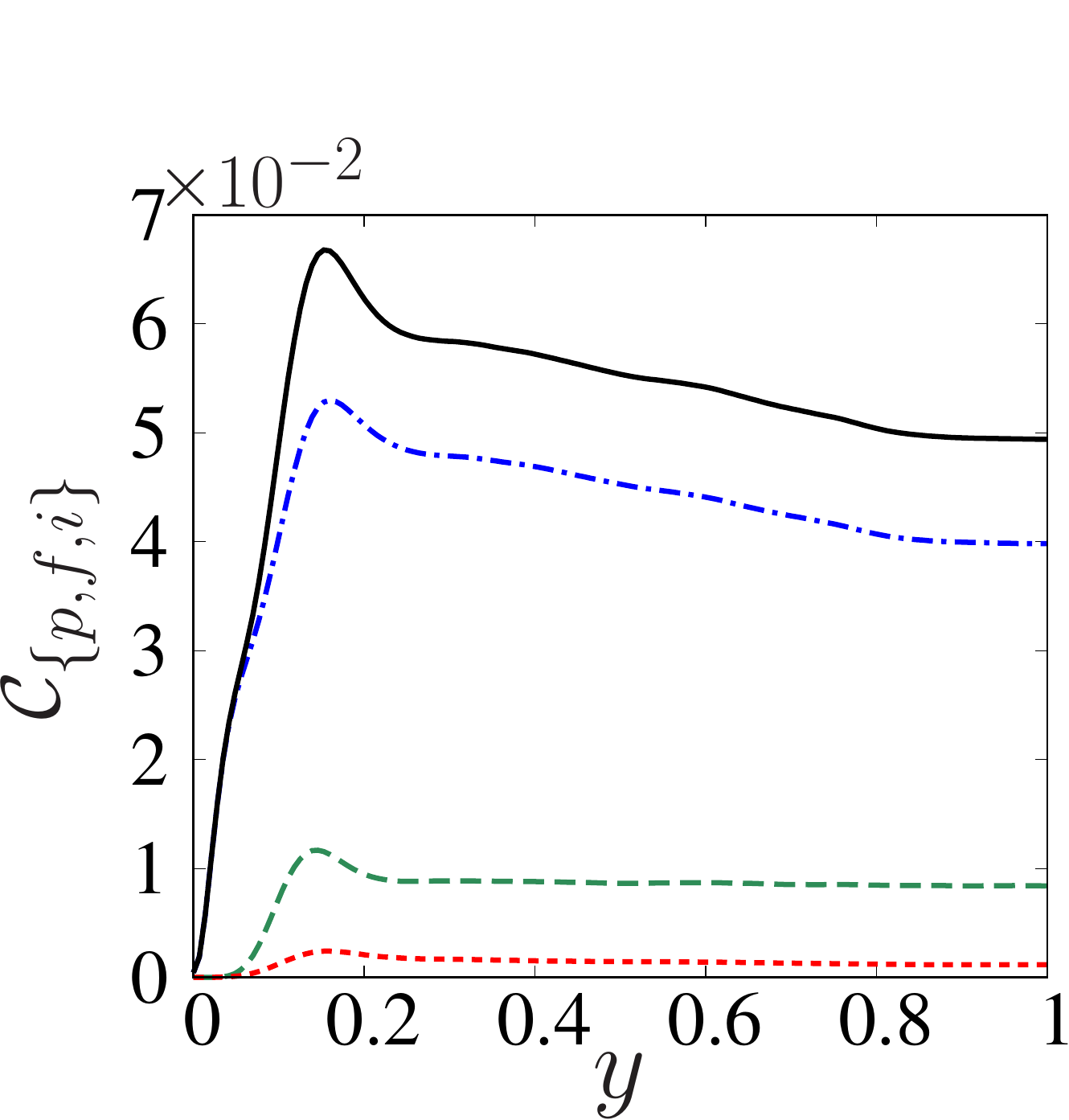}
\end{center}
\end{minipage}}			
\subfigure[]{\label{fig:ufuf_W0}
\begin{minipage}[b]{0.3\textwidth}
\begin{center}
\includegraphics[width =\textwidth,scale=1]{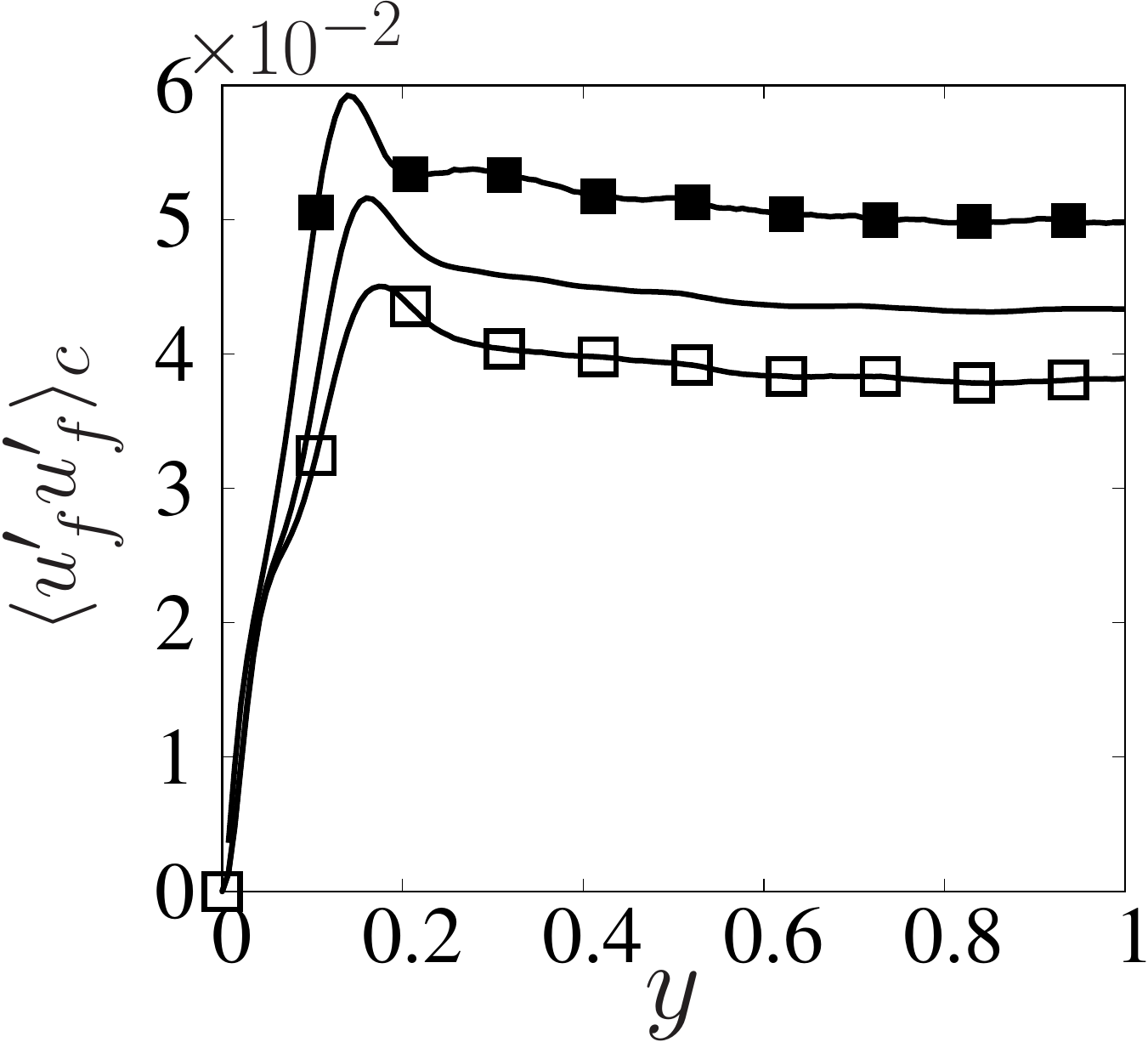}
\end{center}
\end{minipage}}	\\
\subfigure[]{\label{fig:TKE_W15}
\begin{minipage}[b]{0.31\textwidth}
\begin{center}
\includegraphics[width =\textwidth,scale=1]{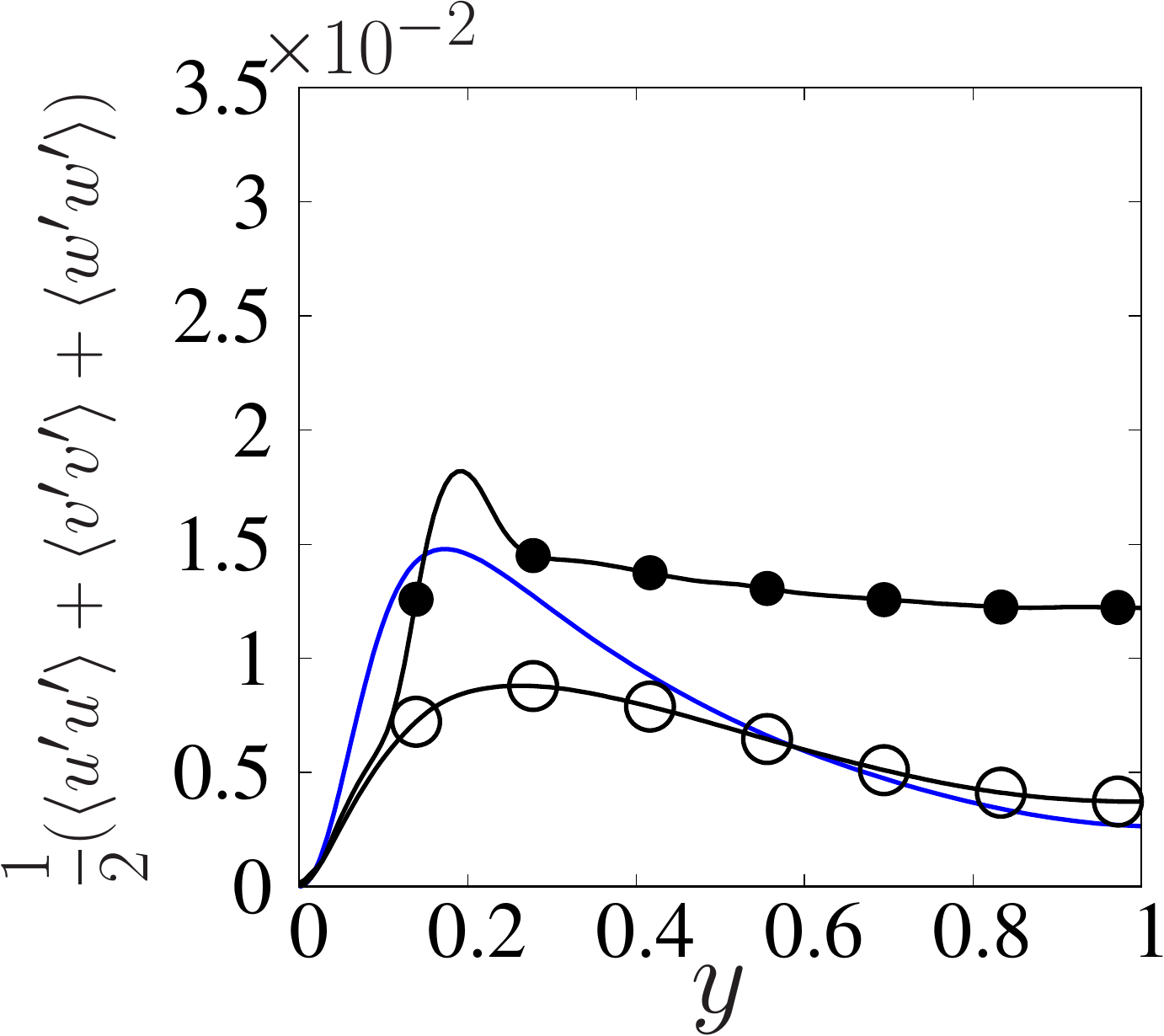}
\end{center}
\end{minipage}}					
\subfigure[]{\label{fig:uu_W15}
\begin{minipage}[b]{0.3\textwidth}
\begin{center}
\includegraphics[width =\textwidth,scale=1]{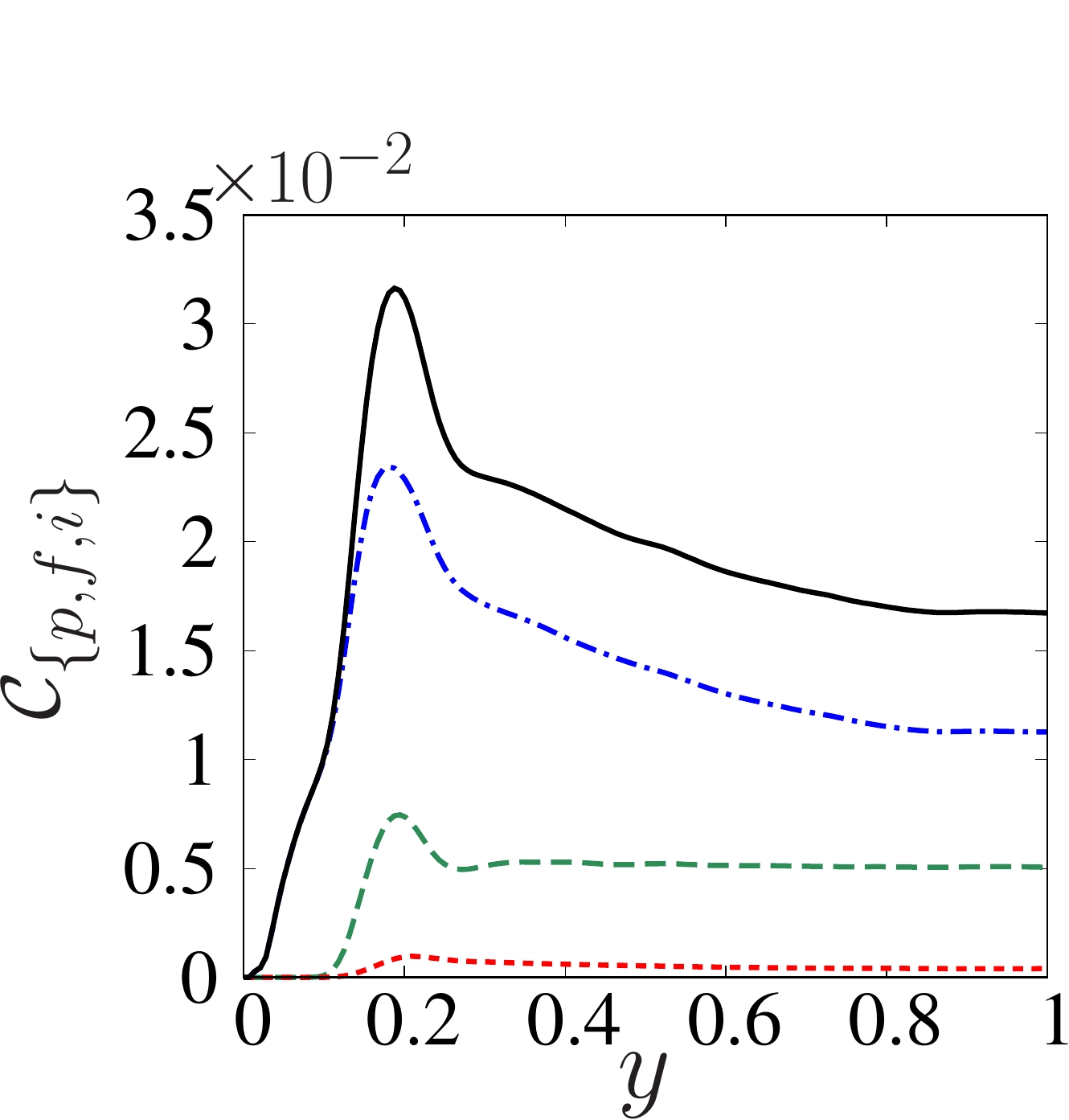}
\end{center}
\end{minipage}}			
\subfigure[]{\label{fig:ufuf_W15}
\begin{minipage}[b]{0.3\textwidth}
\begin{center}
\includegraphics[width =\textwidth,scale=1]{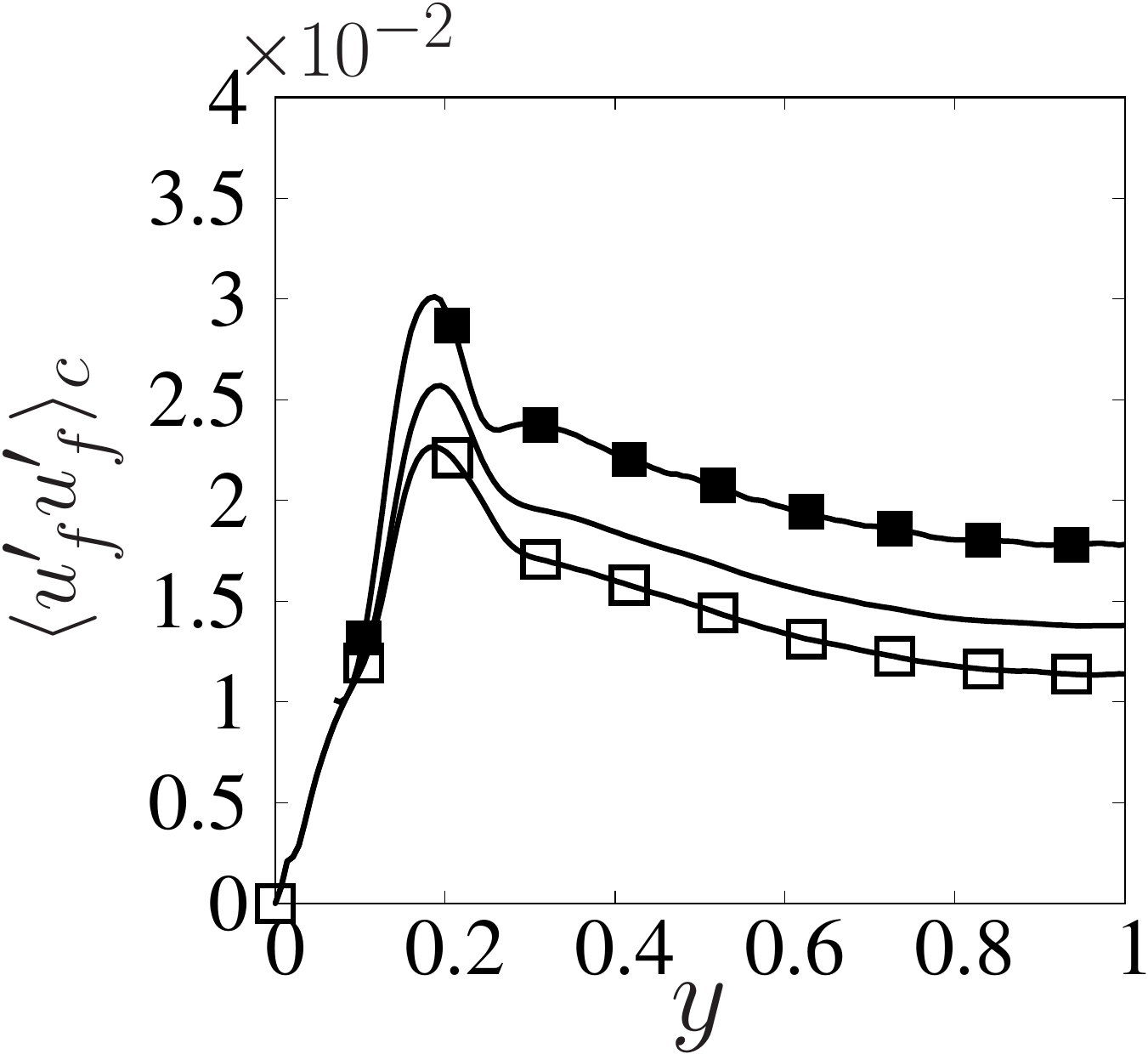}
\end{center}
\end{minipage}}																							
\caption{Mean profiles of (a,d) turbulent kinetic energy of the unconditional velocity field.
Suspensions with ({\protect\circlelineblack}) neutrally buoyant and ({\protect\filledcirclelineblack}) dense particles; ({\protect\bluelt}) single-phase condition.
(b,e) Mean squared streamwise velocity fluctuations and its constituents as defined in \eqref{eq:uu_cond},
({\protect\solidlt}) $\langle u'u' \rangle$,
({\protect\reddotlt}) $\mathcal{C}_p$,
({\protect\bluedashdotlt}) $\mathcal{C}_f$,
({\protect\greendashlt}) $\mathcal{C}_i$.
(c,f) ({\protect\solidlt}) Fluid fluctuations further conditioned to ({\protect\symbolfillsquaresolidline}) near-particle and ({\protect\symbolsquaresolidline}) far-field regions.
The fluid in the top row (a,b,c) is Newtonian and in the bottom row (d,e,f) is viscoelastic. }   \label{fig:fluid_stat}
\end{figure}

Although the turbulent shear stress is hindered in dense particle conditions, the turbulent kinetic energy of the unconditional velocity field increases (figures \ref{fig:TKE_W0} and \ref{fig:TKE_W15}).
We investigate the origins of this increase by conditional sampling, placing our focus on the streamwise velocity fluctuations which are the dominant contributors to the turbulent kinetic energy.
Unconditional velocity fluctuations include contributions from both the fluid and particle phases, as well as from the difference between the mean velocity of each phase \citep{you2019conditional}. In the streamwise direction this relation is given by,
\begin{align}
\langle u'u' \rangle = \underbrace{\phi \langle u'_pu'_p \rangle}_{\mathcal{C}_p} + \underbrace{(1-\phi) \langle u'_fu'_f \rangle}_{\mathcal{C}_f} + \underbrace{\phi  (1-\phi) \langle u_p - u_f \rangle^2}_{\mathcal{C}_i}, \label{eq:uu_cond}
\end{align} 
where fluctuations in each phase are defined with respect to the phase-specific mean, i.e.\,$u'_{\{f,p\}}\equiv u_{\{f,p\}}-\langle u_{\{f,p\}} \rangle$. The three terms on the right hand side correspond to contributions from the particle phase $\mathcal{C}_p$, fluid phase $\mathcal{C}_f$, and the difference of the means $\mathcal{C}_i$.
Figures \ref{fig:uu_W0} and \ref{fig:uu_W15} show the profiles of each term in Newtonian and viscoelastic conditions. 
While $\langle u'u' \rangle$ is dominated by the fluid phase fluctuations, the contribution of $\mathcal{C}_i$ is not negligible in both Newtonian and viscoelastic conditions; while the particle phase contribution is relatively insignificant.
In viscoelastic conditions, particle slip velocities are smaller due to the larger drag coefficient, and therefore the $\mathcal{C}_i$ is smaller compared to the Newtonian counterpart. 
The contribution of fluid phase fluctuations is also smaller in viscoelastic conditions.

We elaborate on $\mathcal{C}_f$ by further conditioning the fluid phase to near-particle and far-field regions. 
Based on the particle wake visualizations (figure \ref{fig:wake}), the near-particle region is defined as the fluid points inside spheres that are concentric with the particles and have a diameter of $2d_p$.
The fluctuations in near-particle regions are larger than those in the far field, implying that the particle wakes are an important contributor to $\langle u'_fu'_f \rangle$.
Since the particles wakes are weakened by viscoelasticity (compare figures \ref{fig:wakeW0P5D} and \ref{fig:wakeW15P5D}), the resulting $\langle u'_fu'_f \rangle$ is smaller in W15P5D compared to W0P5D. 

\begin{figure}		
\centering
\subfigure[]{\label{fig:struct_vis}
\begin{minipage}[b]{0.5\textwidth}
\begin{center}
\includegraphics[width =\textwidth,scale=1]{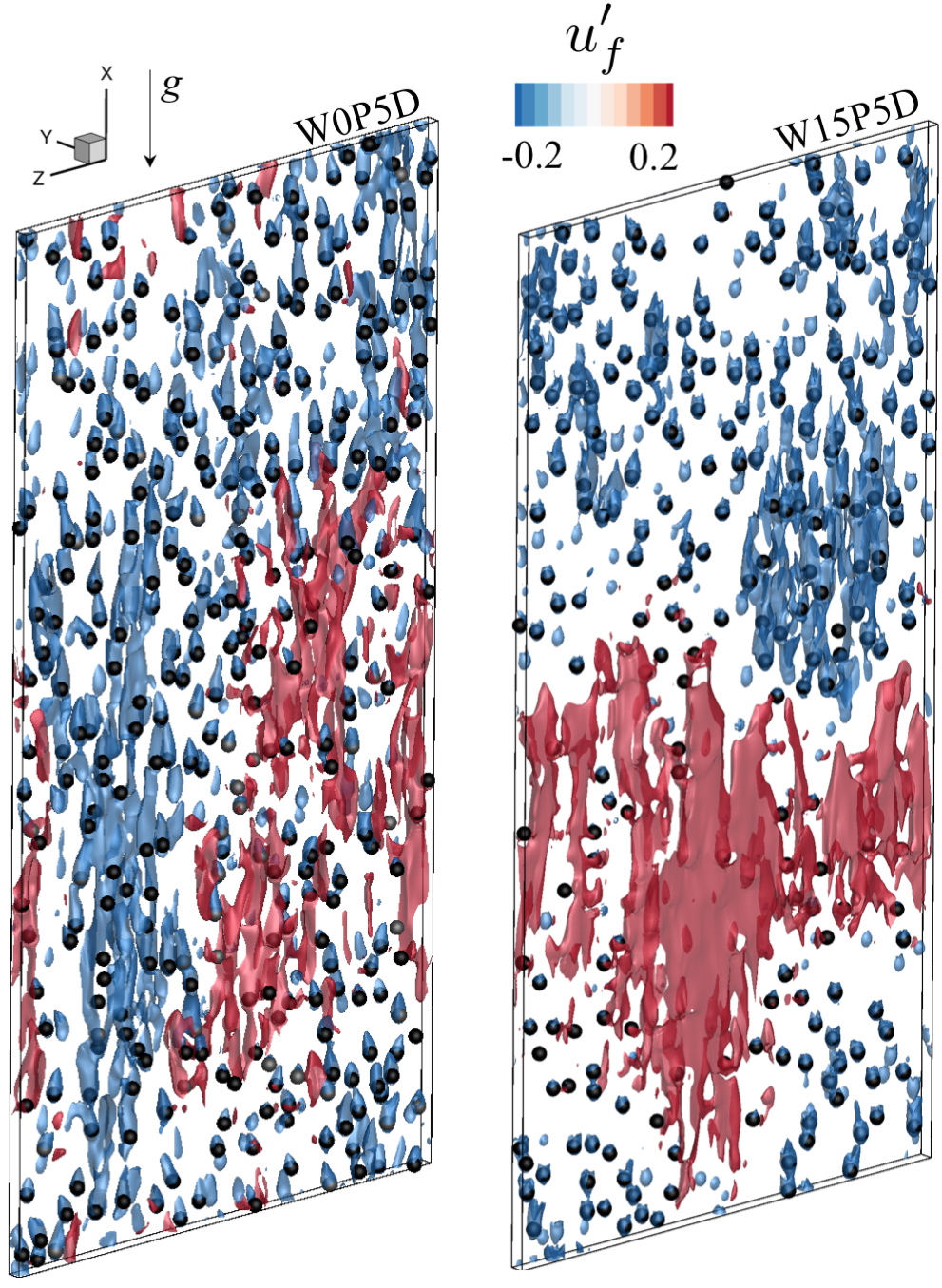}
\end{center}
\end{minipage}}		
\subfigure[]{\label{fig:Ruu}
\begin{minipage}[b]{0.35\textwidth}
\begin{center}
\includegraphics[width =\textwidth,scale=1]{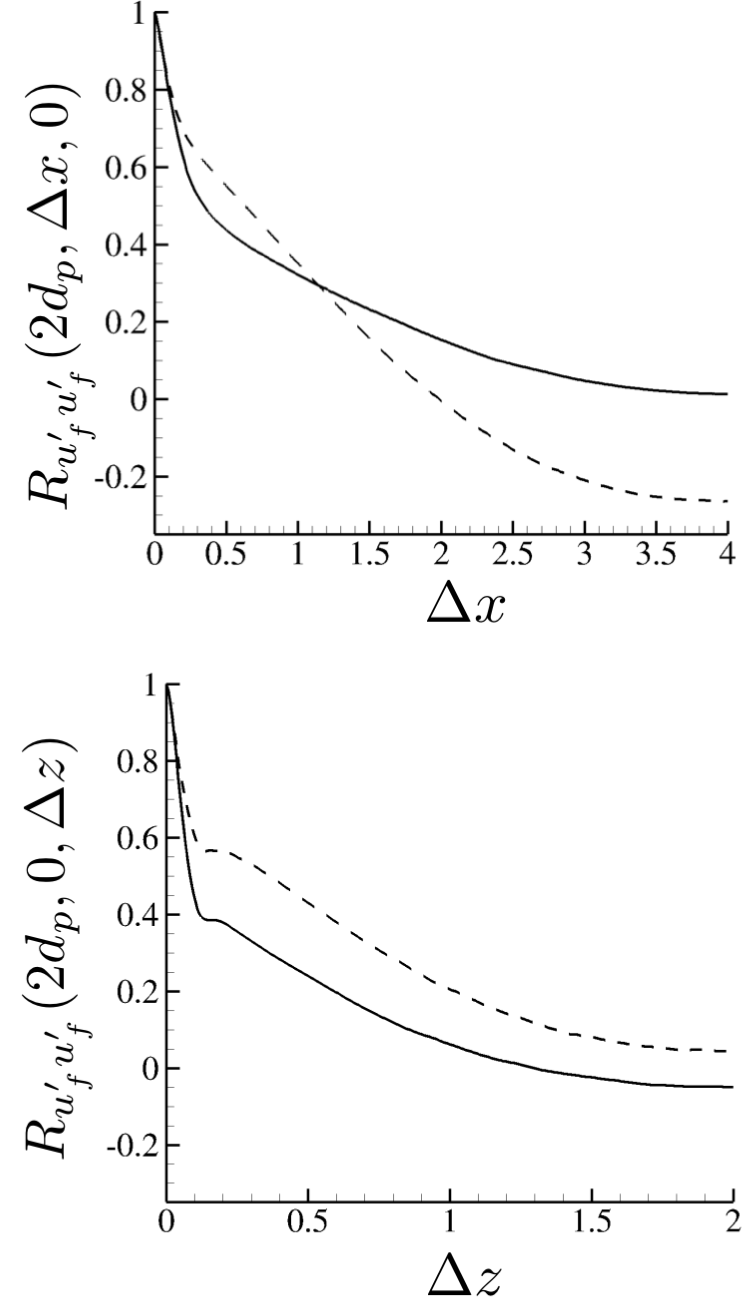}
\end{center}
\end{minipage}}																			
\caption{(a) Isosurfaces of streamwise fluid velocity fluctuations $u'_f$ and particle positions located at $y=2d_p$. (b) Two-point correlation of $u'_f$ defined in \eqref{eq:2pnts} as a function of (top) streamwise $\Delta x$ and (bottom) spanwise $\Delta z$ spacing for cases ({\protect\solidlt}) W0P5D and ({\protect\dashlt}) W15P5D.
$R_{u'_fu'_f}$ is evaluated at $y=2d_p$.
Case designations are listed in table \ref{tab:params}.	
}   \label{fig:velocity_structure}								
\end{figure}	

Finally we examine the interaction between dense particles and the streamwise velocity structures (figure \ref{fig:velocity_structure}).
Elongated flow structure are discernible in both W0P5D and W15P5D cases. The flow decelerates near the particles, and low-speed velocity regions are formed in the vicinity of particle aggregates. 
In contrast, the flow accelerates in interstitial regions void of particles. 
These observations suggest that $u'_f$ structures are generated by particles collective motion rather than the classical lift-up mechanisms of wall turbulence.
The $u'_f $ structures are visually wider in W15P5D, and more elongated along the streamwise direction in W0P5D.
In figure \ref{fig:Ruu} the extent of these structures are quantified by two point correlation of $u'_f$ as a function of streamwise $\Delta x$ and spanwise $\Delta z$ spacing, 
\begin{align}
R_{u'_fu'_f}(y, \Delta x, \Delta z) = \dfrac{\langle u'_f(x,y,z,t)u'_f(x+\Delta x,y,z+\Delta z,t)\rangle}{\langle u'_f(x,y,z,t)u'_f(x,y,z,t)\rangle}. \label{eq:2pnts}
\end{align} 
In both directions a sudden drop in $R_{u'_fu'_f}$  at $\{\Delta x, \Delta z\}=\{d_p,2d_p\}$ is induced by particles wakes, and is more pronounced in W0P5D in which these wakes are stronger.
Beyond this drop, $R_{u'_fu'_f}$ gradually decreases in both directions. 
In the streamwise direction, $u'_f$ structures decorrelate at almost half of the domain in Newtonian conditions, while in the viscoelastic case $R_{u'_fu'_f}$ becomes negative at $\Delta x \approx 2$. 
In the spanwise direction, $u'_f$ structures decorrelate at $\Delta x \approx 1$ in W0P5D, while they remain correlated across the span in W15P5D. 
The explanation is founded on the particles microstructure.
In viscoelastic conditions, we have previously seen that streamwise alignment is suppressed and particles are laterally attracted due to the wake structure (figure \ref{fig:TPC_W15P5D_NW}). 
Both effects promote formation of $u'_f$ structures that are wide in the spanwise direction.

\section{Conclusions \label{sec:conclusion}}
The streamwise settling effect of spherical particles was examined in Newtonian and viscoelastic vertical channel flows. 
Direct numerical simulations that resolve the flow at the scale of particles were performed with an immersed boundary method.
Viscoelastic effects were incorporated by solving the evolution equations of polymer conformation tensor with a FENE-P fluid model.
The vertical channel was laden with $\%5$ concentration of dense particles, and the gravity force was directed in the opposite direction of the flow.  The results were contrasted to additional reference simulations of neutrally-buoyant particles and single-phase flows.

In the dense cases, particles move slower than the background fluid, and experience a sustained slip velocity and a drag force which counterbalances their negative buoyancy.
The sustained slip was shown to have various implications in the dynamics of particles and fluid, from changes in mean quantities at a global scale to particle clustering and lateral forces acting on individual particles.

A primary consequence of sustained slip is the generation of lift forces that modify particle concentration profile and as a result the mean velocity. 
While the shear-induced lift forces repel the particles from the walls, the rotation-induced lift forces push the particles towards the wall\textemdash a competition that results in a remarkable peak in particle volume fraction. 
The same effect persists in the viscoelastic case where the elastic stress promotes migration away from the wall. 
Due to the non-uniform distribution of particles, the mean flow accelerates near the walls and is flattened in the bulk region. 

The lift forces were shown to dramatically alter the particles motion, and were further investigated by tracking the particle properties along their trajectories. 
The effect of rotation-induced lift was isolated by first examining the particle dynamics in the spanwise direction. 
The ensemble-averages along particle trajectories show that the particles wall-normal angular velocity determines the direction of their spanwise translation.
In the wall-normal direction, ensemble-averages confirmed that only particles with a positive angular velocity can penetrate the immediate vicinity of the wall, where they are eventually repelled by an intensive shear-induced lift force.

Lift forces in a vertical channel have been previously investigated for point particles with smaller diameters $d^+ < 1$ and a wide range of Stokes numbers \citep{marchioli2007influence,wang2019inertial}, highlighting the importance of the shear-induced lift force for a quantitative prediction of particle concentration profile. 
In the present study we showed that for particles with $d^+\approx 30$ and $St^+ \approx 25$, an account for the shear-induced lift force is not sufficient, and the rotation-induced lift force also has a significant influence on the particles motion.  
Unlike the shear-induced lift, the significance of the rotation-induced lift is not limited to the near-wall region, and has important implications in particle mixing and clustering throughout the channel.

Another consequence of sustained slip velocities is the formation of clusters in the near-wall region, which was quantified using a \Voro analysis. 
This phenomenon, which is absent away from the wall, is also a result of rotation-induced lift forces which preferentially transport aggregated particles towards the wall. 
The particle conditioned average flow field shows the striking differences in the wakes for Newtonian and viscoelastic conditions.
These differences in turn affect the particles microstructures as shown by two particle correlations. 

Despite the drag reducing effects of viscoelasticity in the single-phase and neutrally-buoyant particle suspensions, a drag increase by viscoelasticity was observed for dense particle suspensions of the same volume fraction.
We showed that the drag at the wall increases due to the non-uniformity of particles distribution.
Since the particle migration is increased by elasticity, the stress induced by gravity is larger in viscoelastic conditions and gives rise to a drag increase, thus entirely negating the favorable impact of viscoelasticity on turbulence suppression.

The present study showed that even a slight settling effect can alter the global dynamics of the flow.
All these qualitative changes originate from the particles sustained slip velocities: particle wakes increase the local velocity fluctuations; the balance of lift forces establishes the particle distribution; and aggregation at an offset from the wall significantly alters the velocity profile.
The implications of viscoelasticity were also understood through the lens of particle dynamics, rather than the commonly-anticipated viscoelastic turbulence suppression.
Viscoelasticity weakens the particles wakes, and therefore suppresses the wake-induced fluctuations, while enhancing the non-uniformity of particle concentration by migration and thus increasing the overall drag.

\appendix
\section{Clustering and microstructure away from the walls \label{app:clust}}
\begin{figure}		
\centering
\subfigure[]{\label{fig:vor_volume_inbulk_W0P5}
\begin{minipage}[b]{0.2925\textwidth}
\begin{center}
\includegraphics[width =\textwidth,scale=1]{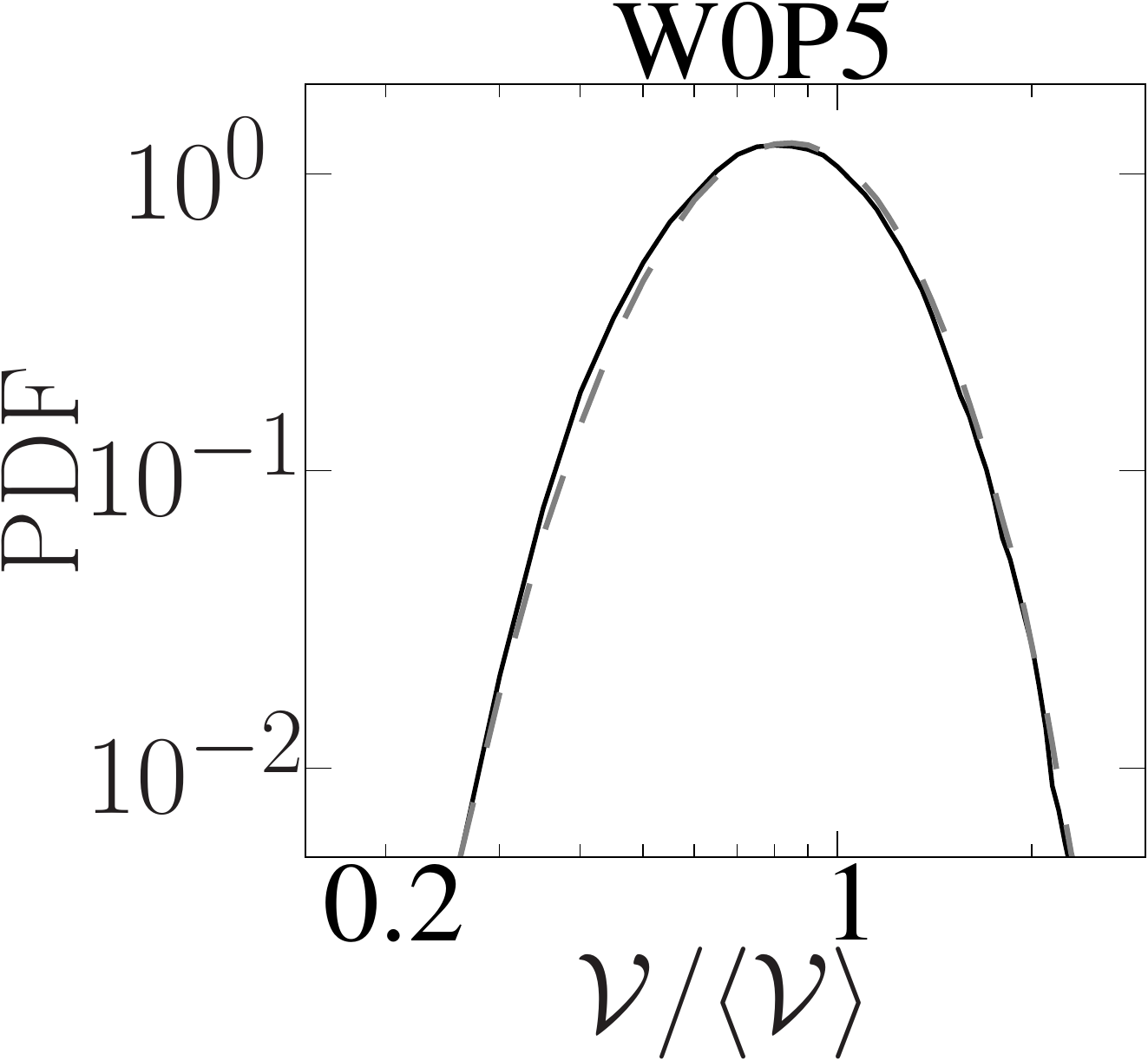}
\end{center}
\end{minipage}}
\subfigure[]{\label{fig:vor_volume_inbulk_W0P5D}
\begin{minipage}[b]{0.215\textwidth}
\begin{center}
\includegraphics[width =\textwidth,scale=1,trim={3.55cm 0 0 0},clip]{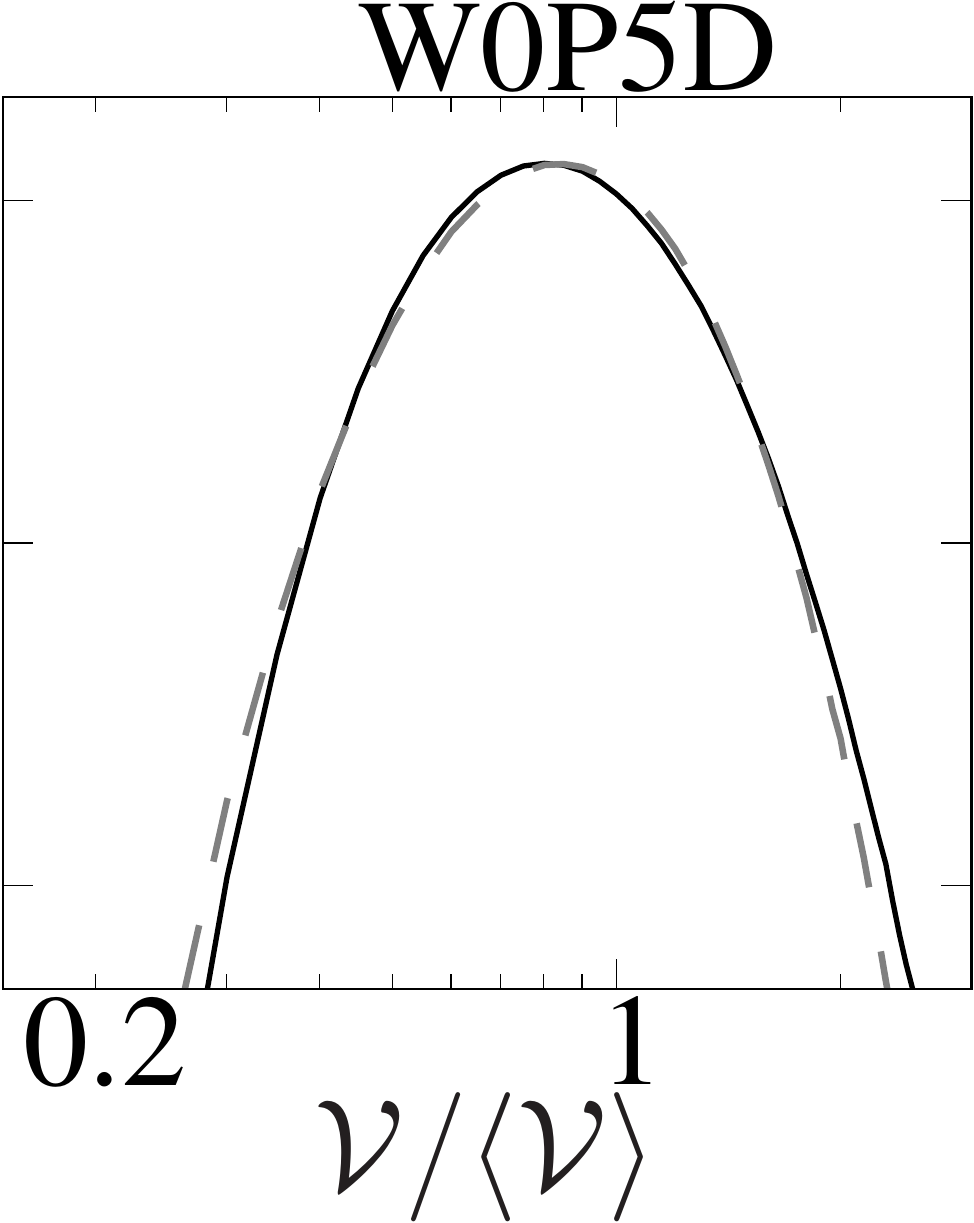}
\end{center}
\end{minipage}}	 
\subfigure[]{\label{fig:vor_volume_inbulk_W15P5}
\begin{minipage}[b]{0.215\textwidth}
\begin{center}
\includegraphics[width =\textwidth,scale=1,trim={3.55cm 0 0 0},clip]{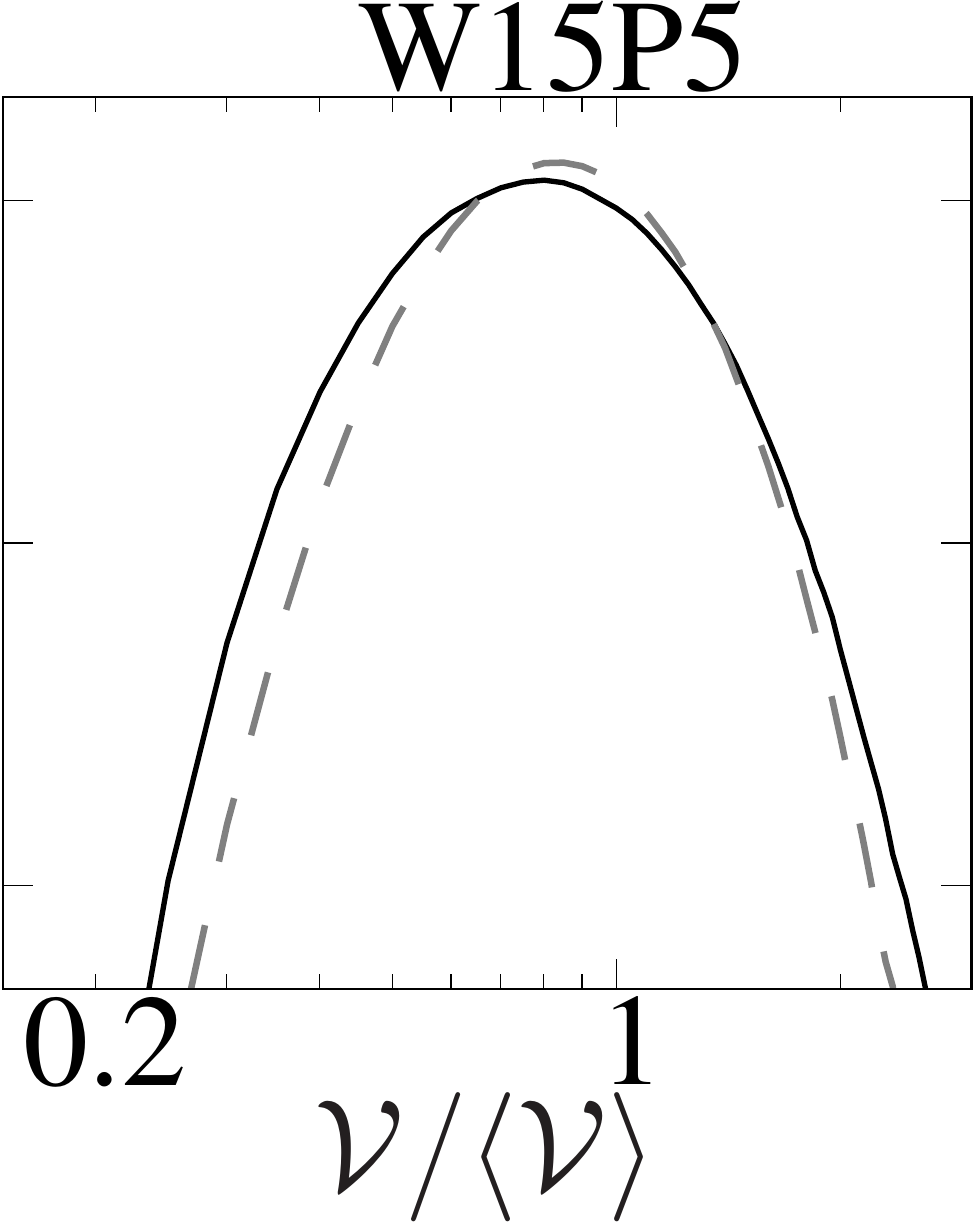}
\end{center}
\end{minipage}}				
\subfigure[]{\label{fig:vor_volume_inbulk_W15DP5}
\begin{minipage}[b]{0.215\textwidth}
\begin{center}
\includegraphics[width =\textwidth,scale=1,trim={3.55cm 0 0 0},clip]{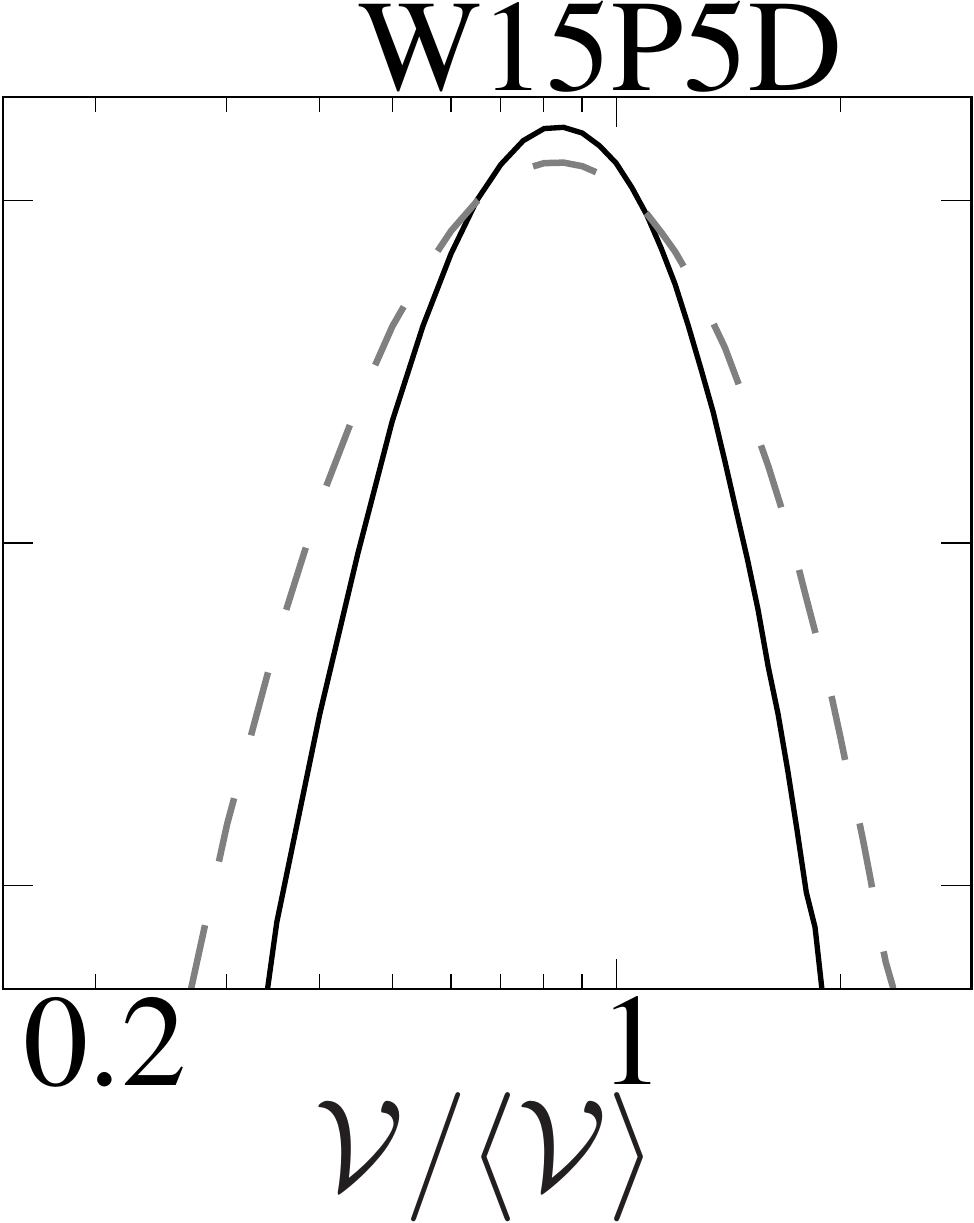}
\end{center}
\end{minipage}}		\\					 		
\subfigure[]{\label{fig:TPC_W0P5_C}
\begin{minipage}[b]{0.2925\textwidth}
\begin{center}
\includegraphics[width =\textwidth,scale=1]{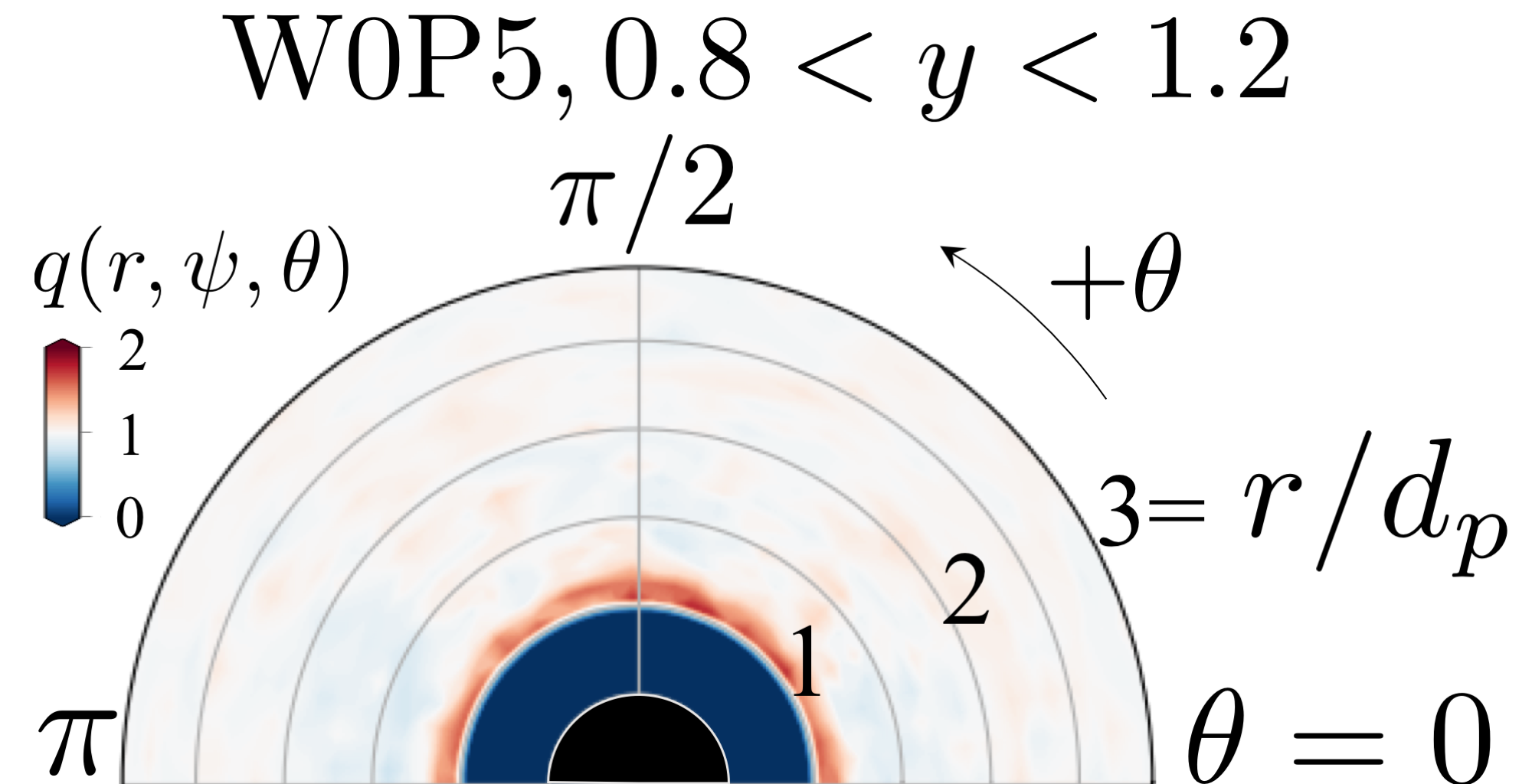}
\end{center}
\end{minipage}}		
\subfigure[]{\label{fig:TPC_W0P5D_C}
\begin{minipage}[b]{0.215\textwidth}
\begin{center}
\includegraphics[width =\textwidth,scale=1]{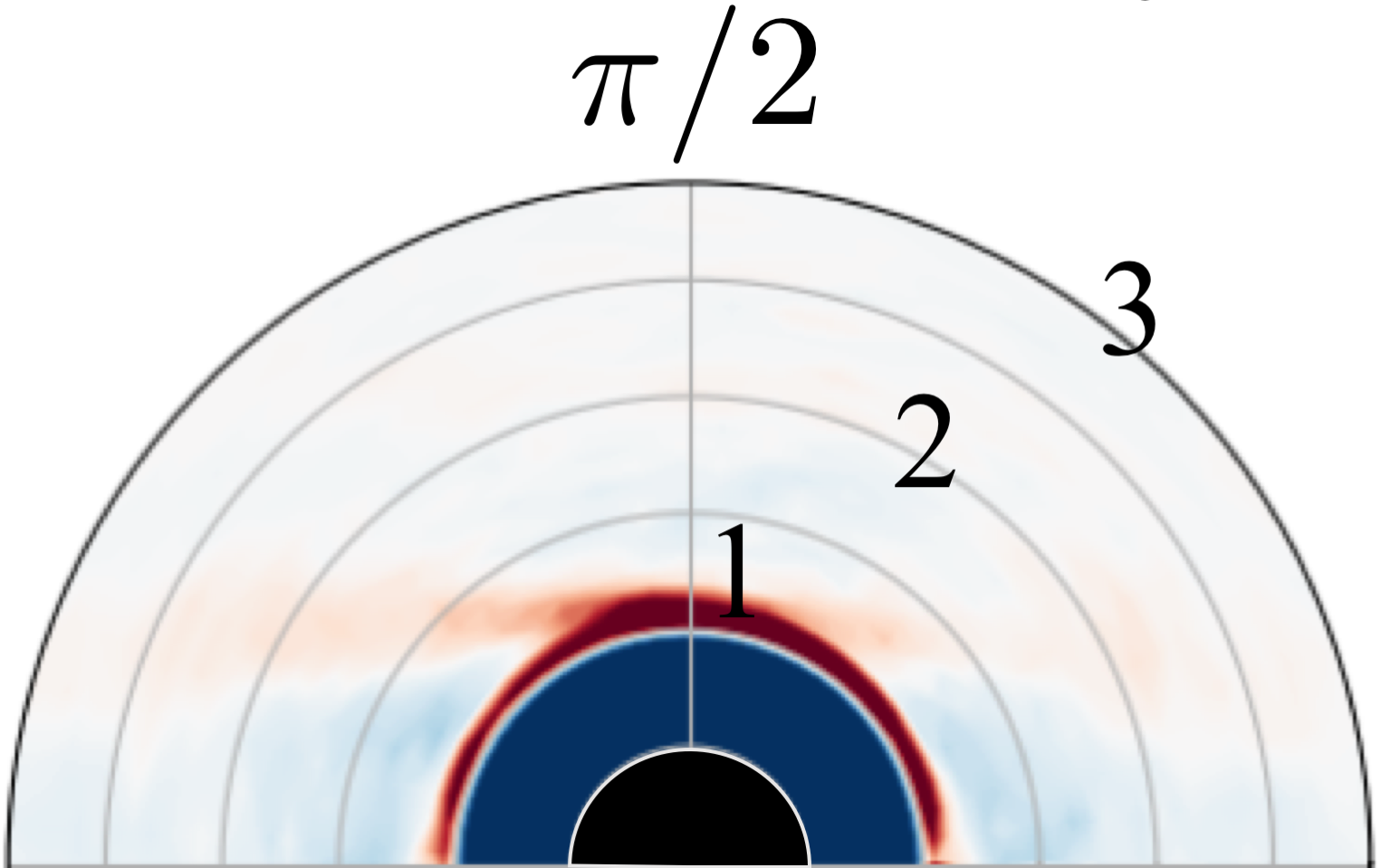}
\end{center}
\end{minipage}}											
\subfigure[]{\label{fig:TPC_W15P5_C}
\begin{minipage}[b]{0.215\textwidth}
\begin{center}
\includegraphics[width =\textwidth,scale=1]{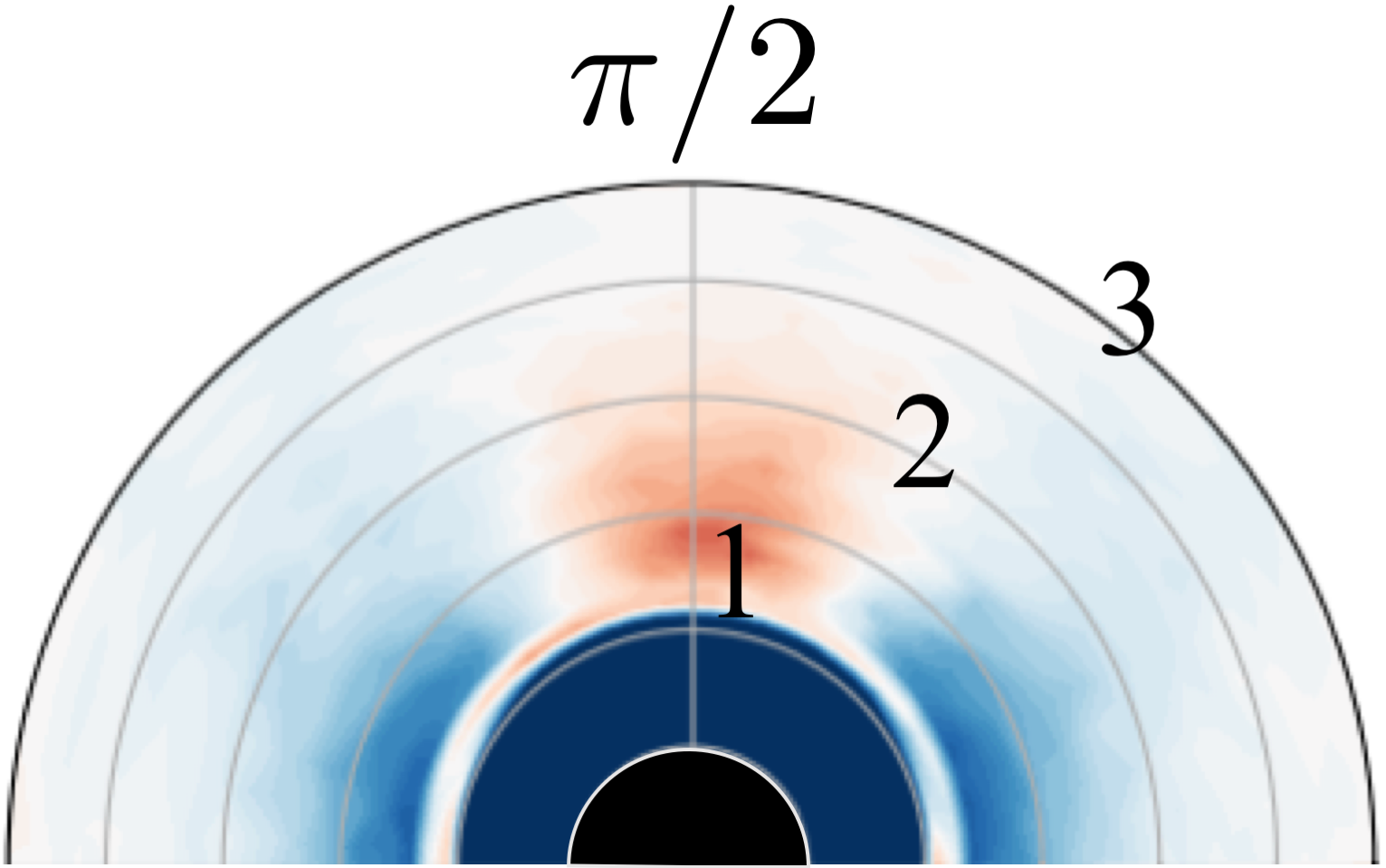}
\end{center}
\end{minipage}}					
\subfigure[]{\label{fig:TPC_W15P5D_C}
\begin{minipage}[b]{0.215\textwidth}
\begin{center}
\includegraphics[width =\textwidth,scale=1]{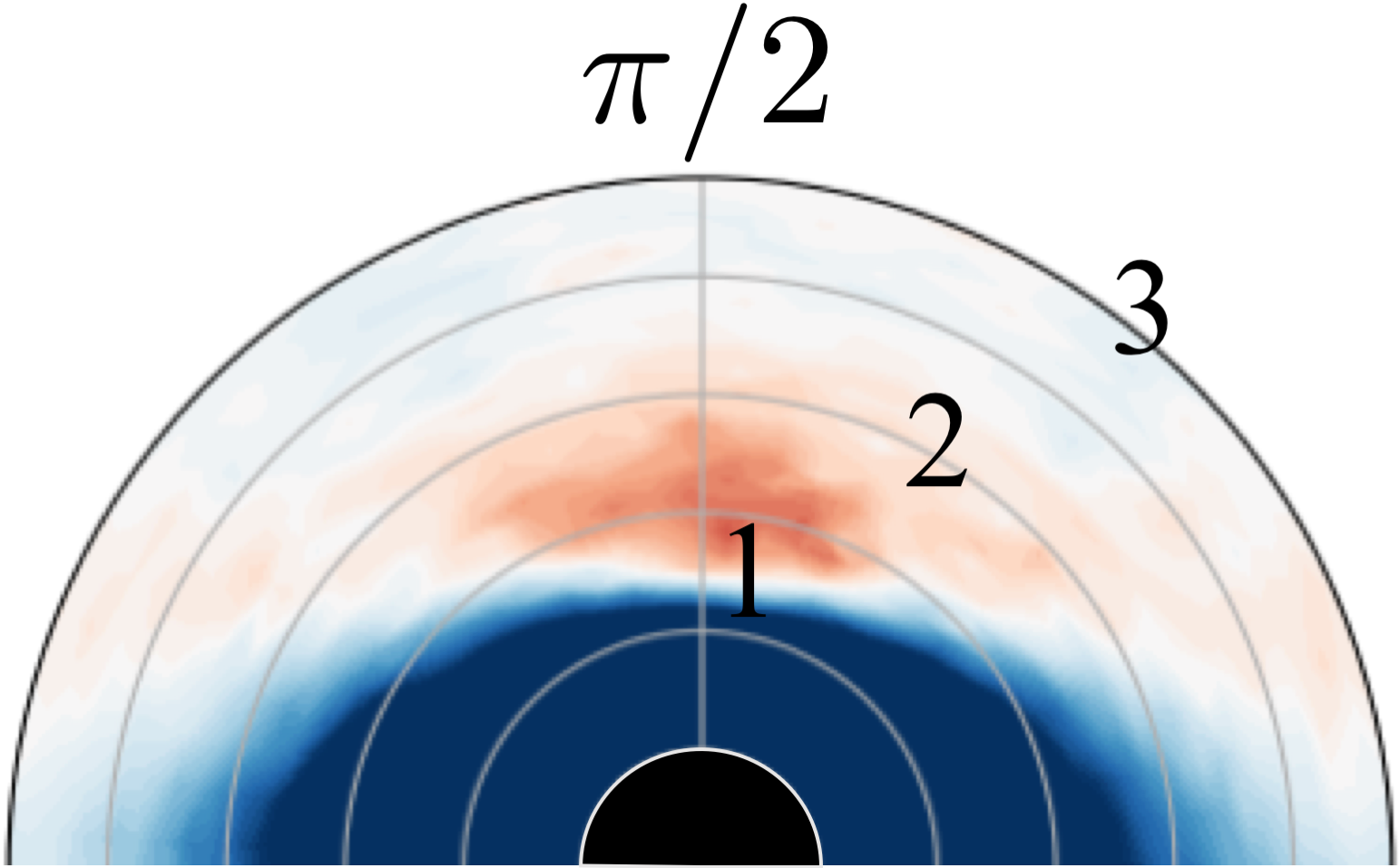}
\end{center}
\end{minipage}}																	
\caption{(a,b,c,d) Probability density functions (PDF) of normalized \Voro cell volumes $\mathcal{V}/\langle \mathcal{V}\rangle $ for particles in the bulk of the flow $0.8<y<1.2$; (a) W0P5, (b) W0P5D, (c) W15P5, (d) W15P5D.  
For each case, PDF of the same quantity is also plotted for a random distribution of non-overlapping particles with a matching particle volume fraction profile (gray dashed lines ({\protect\greydashedlt})).
(e,f,g,h) Particle-pair distribution function $q(r,\psi,\theta) $ defined in \eqref{eq:ppd}, averaged in the range $-\pi/8<\psi - \pi/2 < \pi/8$. 
Reference particle is located in $0.8<y<1.2$. 		   }   \label{fig:vor_volume_inbulk}		
\end{figure}	
A clustering analysis is performed for particles in the bulk of the flow with a \Voro tessellation of the sub-volume $0.8 < y <  1.2$.
Since the selected sub-volume is away from the boundaries in the wall-normal direction, a three-dimensional \Voro analysis is possible.

Figure \ref{fig:vor_volume_inbulk} shows the PDFs of normalized \Voro cell volumes $\mathcal{V}/\langle \mathcal{V}\rangle $. 
In the Newtonian cases (both dense and neutrally buoyant) the PDFs of $\mathcal{V}/\langle \mathcal{V}\rangle $ are very similar to a random distribution with a matching particle volume fraction profile. 
This observation confirms that the clustering effect seen in the near-wall region of W0P5D is not directly attributed to the wake interactions, since it is not observed away from the walls. 
With viscoelasticity, particles tend to weakly cluster in the neutrally-buoyant conditions while dense particles are less aggregated than a random distribution; these effects are however inappreciable compared to the near-wall clustering discussed in the main text (c.f.\,figure \ref{fig:vor_volume}).
While viscoelasticity promotes the formation of particle pairs in the cross-stream direction ($\theta = \pi/2$ in figures \ref{fig:TPC_W15P5_C} and \ref{fig:vor_volume_inbulk_W15DP5}), case W15P5D shows significant repulsion in the streamwise direction.
The net effect is a more homogeneous distribution compared with a random one (c.f.\,figure \ref{fig:vor_volume_inbulk_W15DP5}).

\bibliographystyle{jfm}
\bibliography{JFM_AE2020} 

\end{document}